# Formation, Composition, and History of the Pluto System: A Post-*New-Horizons* Synthesis

**William B. McKinnon**
Washington University in St. Louis

**Christopher R. Glein**
Southwest Research Institute, San Antonio

**Tanguy Bertrand**
NASA Ames Research Center

**Alyssa R. Rhoden**
Southwest Research Institute, Boulder

The Pluto-Charon system provides a broad variety of constraints on planetary formation, composition, chemistry, and evolution. Pluto was the first body to be discovered in what is now known as the Kuiper belt, its orbit ultimately becoming a major clue that the giant planets underwent substantial orbital migration early in Solar System history. This migration has been linked to an early instability in the orbits of the giant planets and the formation of the Kuiper belt itself, from an ancestral trans-Neptunian planetesimal disk that included Pluto. Pluto-Charon is emblematic of what are now recognized as small or dwarf planets. Far from being a cold, dead, battered icy relic, Pluto displays evidence of a complex geological history, with ongoing processes including tectonism, cryovolcanism, solid-state convection, glacial flow, atmospheric circulation, surface-atmosphere volatile exchange, aeolian processes, and atmospheric photochemistry, microphysics, and haze formation. Despite Pluto's relatively modest scale, the combination of original accretional heat, long-term internal radiogenic heat release, and external solar forcing, when combined with sufficiently volatile (and thus mobile) materials, yields an active world. Pluto may have inherited a large organic mass fraction during accretion, which may responsible, in part, for its surface and atmospheric volatiles. Charon, Pluto's major moon, displays evidence of extensive early tectonism and cryovolcanism. Dwarf planets are thus truly planetary in terms of satellite systems and geological and atmospheric complexity (if not ongoing activity). What they may lack in mass is made up in number, and the majority of the Solar System's dwarf planets remain undiscovered.

# 1. INTRODUCTION

The *New Horizons* encounter with Pluto revealed not just a remarkable dwarf planet, but a complex, scientifically rich planetary system beyond Neptune, out in the Kuiper belt (*Stern et al.*, 2018). This chapter draws on these encounter results, later analyses, and other chapters in this volume. It has as its goal to synthesize and summarize what we have learned from *New Horizons*, highlighting some of the less understood or appreciated aspects of the Pluto system. The chapter is organized in the following manner. We first discuss how Pluto fits in with the emerging picture of planetesimal and planet formation in the Solar System, and how it contributes to this understanding (Sec. 2). This is followed by a discussion of the composition of the bodies in the Pluto system, focusing on volatiles and carbon (Sec. 3). How Pluto and Charon inform understanding of other major icy satellites and Kuiper Belt objects (KBOs) is also addressed. A section on the evolution of the Pluto system through time follows, considering orbital, thermal, tectonic, geomorphologic, and climactic changes (Sec. 4). Emphasis is placed on the controls and duration of planetary activity and the roles of contingent events. A summary is then provided (Sec. 5). Finally, some thoughts are offered on how exploration of the Pluto system has expanded our view of planethood and the richness of nature and how we might deepen this understanding through future work (Sec. 6), for there is much that we do not yet know.

No single paper or chapter can do justice to these topics. The interested reader is invited to consult other chapters in this volume, the initial papers that discuss encounter results (*Stern et al.*, 2015; *Moore et al.*, 2016; *Grundy et al.*, 2016; *Gladstone et al.*, 2016; *Weaver et al.*, 2016; *Bagenal et al.*, 2016), subsequent *Annual Reviews* articles (*Stern et al.*, 2018; *Gladstone and Young*, 2019; *Moore and McKinnon*, 2021), two special issues of the journal *Icarus* (*Binzel et al.*, 2017a; *Grundy et al.*, 2020b), as well as papers in the original Space Science Series volume *Pluto and Charon* (*Stern and Tholen*, 1997), not all of which are out of date (and far from it, actually).

# 2. FORMATION OF THE PLUTO SYSTEM

In this section we examine the formation of the Pluto-Charon system, the timing and conditions thereof, and both the implications for the initial states of Pluto and Charon and the constraints that Pluto and Charon provide on the accretion process. We begin by considering the accretion processes that created the bodies that eventually populated the Kuiper belt (Sec. 2.1), and follow with a summary of current thinking regarding the dynamic instability that created the Kuiper belt (Sec. 2.2). This is followed by a discussion of the Charon-forming giant impact specifically and when and where it occurred (Sec. 2.3). Sections 2.4 and 2.5 then address, respectively, the thermal and structural implications for the progenitor (pre-giant-impact) bodies of the system and Pluto and Charon after their formation, focusing on the thorny issue of differentiation. Finally, we synthesize the overall discussion and consider continuing conundrums (Sec. 2.6).

**2.1. Accretion Scenarios**

Understanding of planetary accretion has evolved substantially in the last 25 years (i.e., since the publication of *Pluto and Charon*). Both the supposed initial configuration of the Solar System and the physics of planetesimal and planet formation have undergone profound if not revolutionary advances (see, e.g., chapters in *Beuther et al.*, 2014). Historically, accretion of Pluto at its present orbital distance by binary, Safronov-style hierarchical coagulation was difficult to understand. As extensively discussed in *Stern et al.* (1997), the issues reduced to insufficient surface mass density of accreting solids and the long orbital time scales in the classical Kuiper belt, making the formation of Pluto a drawn-out affair, potentially taking billions of years unless the velocity dispersion among the planetesimals was highly damped (e.g., *Goldreich et al.*, 2004). Moving Pluto's accretion zone from near Pluto's present semimajor axis (*a*) of 39.5 AU to somewhere in the 15-to-30 AU range, as in the original Nice planetary migration/instability models (e.g., *Tsiganis et al.*, 2005; *Levison et al.*, 2008a), goes a long way towards ameliorating this time scale issue, both because Keplerian orbital periods scale as $a^{3/2}$ and because the minimum mass necessary to construct all the planets must now be distributed over a much more compact range of semimajor axes (*Desch*, 2007) (i.e., overall collision rates scale with orbital speed and number density squared).

There are other long-standing issues with hierarchical coagulation, however, having to do with (1) nebular-gas induced turbulence preventing the formation of even small planetesimals, and (2) even if planetesimals could form by sticking, the velocity dispersion among them would result in destructive collisions between the nominally small, fragile bodies — problems together referred to as the "meter barrier" (see the reviews of *Johansen et al.* [2014] and *Johansen and Lambrechts* [2017]). A potential solution to these problems has been found in collective aerodynamic interactions of small particles with nebular gas (*Youdin and Goodman*, 2005;

*Johansen et al.*, 2007, 2014, 2015; *Simon et al.*, 2016), which cause high-density filaments and streams of mm-to-dm sized particles (dubbed "pebbles") to form. Termed the streaming instability, if the volume density of solids in these pebble streams increases sufficiently, then the streams can fragment into gravitationally contracting pebble clouds (see Fig. 1 in *Nesvorný et al.* [2019] for a compelling visualization). These infalling pebble clouds coalesce to form substantial planetesimals (10s-to-100 km or more in size) on short time scales (<$10^3$ years), and in the words of *Morbidelli et al.* (2009), are "born big" (though to be clear, not as big as Pluto itself). Recent results from the *New Horizons* flyby of the cold classical KBO Arrokoth strongly support this general view of planetesimal formation (see *Stern et al.*, 2019; *Spencer et al.*, 2020; *Grundy et al.*, 2020a; *McKinnon et al.*, 2020).

Measurements of nucleosynthetic anomalies in meteorites strongly imply that those planetesimals that formed "closer" to the Sun (here meaning out to and somewhat beyond 5 AU) formed over a range of times and places while the gas component of the protoplanetary disk existed (up to at least a few Myr after the condensation of the first calcium-aluminum inclusions, or CAIs) (*Kruijer et al.*, 2017; *Scott et al.*, 2018; *Desch*, 2018). One might then reasonably expect that formation of large planetesimals by pebble cloud collapse took place over the lifetime of the gas nebula elsewhere in the Solar System, such as in the region of the protoplanetary disk beyond Neptune (when the latter was much closer to the Sun), a zone we designate as the ancestral Kuiper belt (aKB). The characteristic planetesimal mass formed by the streaming instability (SI) followed by gravitational instability (GI) has been identified with the turnover (or "knee") in the size-frequency distribution (SFD) of KBOs (e.g., *Morbidelli and Nesvorný*, 2020), near diameter $D = 100$ km (see chapter by Singer et al. in this volume). Above this size the power-law slope of the differential size-frequency distribution is quite steep, $dN/dD \sim D^{-5}$, and below it is shallower, with a power-law exponent closer to -3 (*Fraser et al.*, 2014; *Lawler et al.*, 2018; and see *Greenstreet et al.*, 2015). (Note that these exponents refer to the ensemble of dynamically "hot" KBO populations, of which Pluto is an exemplary member.) The shallow power-law segment below the knee may be a direct outcome or signature of the SI/GI process (*Abod et al.*, 2019) or it may be due to later collisional evolution, a matter which is debated (though for the hot population collisional evolution is arguably the cause; *Morbidelli and Nesvorný*, 2020).

Regardless of the details of the SFD at small sizes, once planetesimals formed in the aKB, the path to planethood was open. Growth to larger sizes can occur by direct (hierarchical) planetesimal accretion or by gas-assisted accretion of pebbles (termed *pebble accretion*). *Johansen et al.* (2015) argue that the latter was more important for the aKB. Figure 1 illustrates both the original planetesimal SFD derived from their numerical SI model, and two outcomes of pebble accretion calculations. Two different planetesimal densities are used with two corresponding dimensionless turbulence parameters for the nebular gas ($\alpha$). Both models display ordered growth up to 300-km radii over a couple-to-several Myr, by planetesimal and pebble accretion, with a steep size distribution beyond 100-km sizes. This is followed by a runaway growth, due to pebble accretion, of a single, massive (Mars- or Earth-sized) body. This begs the question of what, if anything, might have limited the growth of planets in the aKB to Pluto-scale or similar.

The former existence of massive Kuiper belt bodies has in fact been proposed (e.g., *Gladman and Chan*, 2006). However, we are unaware of any dynamical scenario that permits the growth of one or more major ($\sim 10^{24-25}$ kg) planets in the aKB that is also consistent with creation of a Kuiper belt dynamical structure similar to that seen today (that is, one that has been numerically tested). In this context, *Shannon and Dawson* (2018) examined aspects of the survival of wide-binaries, cold classicals and the resonant populations in the Kuiper belt, but did not come to very restrictive conclusions regarding the number of massive aKB objects. Such massive bodies may also be subject to rapid inward migration, towards the ice giant region (*Izidoro et al.*, 2015), and thus removal from the aKB. In this review, we will assume that massive, Mars-scale planetary embryos *did not* form in the aKB. We leave the question of what were the largest bodies that did form (e.g., Triton is 65% more massive than Pluto) for future research.

The various parameters used by *Johansen et al.* (2015) in their study (see Fig. 1) are in some sense tuned (though not unreasonably so) to give plausible results. For example, the effective value of the turbulent viscosity $\alpha$ in the 20-to-30 AU zone of the protoplanetary disk is not known. Their Fig. 11 illustrates the effect of varying $\alpha$ about a canonical value of $10^{-4}$, albeit for the case of accretion in the asteroid belt (assuming a dead zone stirred by active disk layers). It shows that the growth of large protoplanets can either be rapid, or prolonged beyond the estimated lifetime of the protoplanetary gas disk (~5 Myr, based on the lifetimes of dust and gas disks around young solar-type stars [*Haisch et al.*, 2001; *Williams and Cieza*, 2011] or the Sun's protoplanetary disk magnetic field [*Wang et al.*, 2017]). In contrast, *Johansen et al.* (2015) find that varying $\alpha$ at 25 AU mainly serves to change the timing of runaway pebble accretion (Fig. 1).

We are interested in Pluto-scale bodies, of course, and models of Neptune's outward migration through the aKB planetesimal disk require ~$10^3$ or possibly several × $10^3$ Pluto-mass (~$10^{23}$ kg) objects to make Neptune's migration sufficiently "granular" that the mean-motion resonances (MMR) in the Kuiper belt do not get overpopulated. Such an overpopulation is the outcome of any smooth migration model (*Nesvorný and Vokrouhlický*, 2016). So, this implies that if it occurred, pebble accretion onto planetesimals in the aKB was likely widespread, but slow enough that it stalled before true runaways to Mars-mass and beyond occurred. *Johansen et al.* (2015) in fact state the aKB is marginal for substantial pebble accretion generally, and then only for weak turbulence (see Fig. 1).

Once the nebular gas in the aKB dissipates (by photoevaporation due to XUV radiation from the proto-Sun; see the reviews by *Williams and Cieza* [2011] and *Alexander et al.* [2014]), planetesimal growth by hierarchical coagulation resumes. And certainly, the end game for the formation of the Pluto system involved the Charon-forming giant impact, the ultimate expression of hierarchical, two-body accretion (discussed in greater detail in Sec. 2.3 below). In this context, Fig. 2 illustrates a "traditional" numerical model of planetesimal accretion in the aKB, via hierarchical accretion, from *Kenyon and Bromley* (2012). It is not entirely gas-free (the assumed nebular gas dissipation time scale is 10 Myr), but pebble accretion is not included (collisional debris <1 m in size is simply removed from the computation). Starting with a planetesimal distribution and a maximum size ($r_0$) of 1 km, the growth of the largest planetesimal in two example semimajor axis ranges is shown. Growth is at first slow, but enhanced gravitational focusing with increasing size eventually initiates a runaway. Accretion time scales are well under $10^9$ yr. (Indeed, runaway accretion was one solution offered in *Stern et al.* [1997] to the Pluto formation time scale puzzle.) We stress, however, that initial km-scale planetesimal "seeds" are needed to achieve such rapid growth: while motivated by classic works on planetesimal formation (e.g., *Goldreich and Ward*, 1973), such an assumption cannot be physically justified a priori based on our current understanding of nebular conditions and processes, as discussed above. Perhaps more interesting, from the point of view of planetesimal birth sizes (the initial mass function) resulting from the SI, is that the hierarchical coagulation time scales in the model increase if $r_0$ is larger, reaching Gyr to get to Pluto mass for $r_0 \approx 100$ km (*Kenyon and Bromley*, 2012; their Fig. 10).

We can conclude from these hierarchical coagulation models that as long as sufficiently small seed planetesimals form by some process, growth to Pluto size (or beyond) by hierarchical accretion is in fact actually plausible, on a time scale of millions to 10s of millions of years. On the other hand, starting with predominantly large planetesimals (100 km scale) greatly lengthens the accretion time scale. But the "problem" with accretion by hierarchical coagulation is deeper. It is actually not so much a matter of forming Pluto on a plausible time scale. As *Morbidelli and Nesvorný* (2020) point out, the problem is *making enough Plutos*. That is, the process is inefficient, whereas SI followed by pebble accretion can be both efficient and relatively rapid.

**2.2. Kuiper Belt Formation**

As now understood, the Pluto system was emplaced into its 3:2 MMR with Neptune as part of the overall dynamical rearrangement of the outer solar system attendant upon a compact, but ultimately unstable, arrangement of 4 or more giant planets emerging from the protoplanetary gas nebula (e.g., *McKinnon et al.*, 2017). The Kuiper belt as a whole is thought to be almost entirely derived from a ~15–20 $M_\oplus$ (Earth mass) remnant planetesimal disk originally orbiting exterior to Neptune, a disk whose main mass extended not much further than 30 AU, Neptune's present semimajor axis (see *Nesvorný* [2018] and *Morbidelli and Nesvorný* [2020] for recent reviews). The most natural time scale for this instability is *early*, within a few 10s of Myr of dissipation of the gaseous protoplanetary nebula, and not 100s of Myr later (*Nesvorný*, 2018; *Nesvorný et al.*, 2018; *Quarles and Kaib*, 2019; *Robeiro et al.*, 2020). Implantation into the Kuiper belt is not particularly efficient, of order $10^{-3}$ (*Morbidelli and Nesvorný*, 2020), meaning only about one out of a thousand bodies originally in the aKB planetesimal disk ends up in a stable hot classical or Neptune-resonant orbit. This implies many Pluto-scale dwarf planets were lost, ejected to the scattered/scattering disk, Oort cloud, or accreted by the giant planets, or possibly, in the case of Neptune's retrograde satellite Triton, captured (cf. *Stern*, 1991; *Nogueira et al.*, 2011).

Figure 3, from *Nesvorný and Morbidelli* (2012), illustrates a recent view of how the giant planet instability may have taken place. Notably, there is the ejection of a third ice giant. Nice-type models commonly feature ice giant ejections, and in this particular instance it is treated as a feature, not an issue that must be explained away (as it would be if the model started with only four giant planets). In Fig. 3 the three ice giants initially slowly migrate by scattering planetesimals. At 6 Myr into the simulation, the instability is triggered when the inner ice giant crosses an orbital resonance with Saturn, and the ice giant's eccentricity is pumped up. Following that, the ice giant has encounters with all other planets and is ultimately ejected from the Solar System by Jupiter. Orbital eccentricities of the remaining giant planets are then subsequently damped by dynamical friction from the planetesimal disk, and Uranus and Neptune, propelled by the planetesimal-driven migration, reach their current orbits some 100 Myr after the instability.

Neptune's slow, long-range migration in models of this type is able capture planetesimals from the aKB into mean-motion resonances, and pump up the eccentricities and excite the inclinations of the captured KBOs. In the real world, one of these was no doubt Pluto (or the Pluto-Charon system). Where specifically in the 20-30 AU aKB planetesimal disk Pluto accreted cannot be stated with any certainty. If captured into the 3:2 MMR by a *smoothly* migrating Neptune, as first proposed by *Malhotra* (1993), Pluto's eccentricity $e$ — though not its inclination — would have grown logarithmically from an initially small value (favorable for capture) as Neptune's semimajor axis ($a_N$) increased:

$$e_{\text{final}}^2 \approx e_{\text{initial}}^2 + \frac{1}{3}\ln\left(\frac{a_{N,\text{final}}}{a_{N,\text{initial}}}\right) \quad . \qquad (1)$$

Given Pluto's present large $e = 0.25$, the implication is that Pluto's orbit could have expanded by up to 20%. That would put Pluto's point of resonance capture somewhat beyond the outer limit of the aKB, implying some amount of planetesimal scattering beforehand. Neptune's likely grainy semimajor axis evolution (*Nesvorný and Vokrouhlický*, 2016; *Lawler et al.*, 2019) scrambles this picture, however. We note that the "catch and release" aspect of this grainy scenario (cf. *Murray-Clay and Chiang*, 2016) means that Pluto most likely (in terms of probability) entered into the 3:2 resonance with Neptune towards the end of the latter's planetesimal driven migration, and probably from an already non-circular and inclined orbit, after Pluto had in all likelihood spent 10s of millions of years or more dynamically interacting with Neptune, moving in and out of secular, Kozai and mean motion resonances with the ice giant (see *Gomes* [2003] and *Gomes et al.* [2005] for descriptions of the complex scattering dynamics during Neptune's migration). In other words, Pluto was likely first *scattered* by Neptune to a high-*e*, high-*i* but non-resonant orbit with a semimajor axis somewhat less than 39.5 AU, was then captured into the 3:2 resonance, and subsequently migrated some distance outward (a few × 0.1 AU?) in lock step with Neptune (*Nesvorný and Vokrouhlický*, 2016).

**2.3. Giant Impact**

The origin of the Pluto-Charon binary is widely regarded as due to a relatively giant (for the Kuiper belt) impact (chapter by Canup et al. in this volume). Other mechanisms for binary formation have been proposed for KBOs generally (*Nesvorný et al.*, 2010; *McKinnon et al.*, 2020), but the large masses of Pluto and Charon, the great specific angular momentum of the pair, and coplanar system of smaller satellites, all argue for an impact origin similar to that of the Earth's Moon (e.g., *Stern et al.*, 2018). The reader is directed to the chapter by Canup et al. in this volume for a general overview. Here we focus on the salient aspects or constraints from giant impact models.

Two are most important. The first is that to yield such a large satellite/primary mass ratio (0.122 for Pluto-Charon) implies comparably sized impactors and low encounter velocities, i.e., impact speeds close to the escape speed for the impacting pair, $v_{imp} \approx v_{esc}$. These inferences stretch back to earlier analytical estimates (*McKinnon*, 1989), but more modern numerical smoothed-particle hydrodynamic (SPH) calculations by *Canup* (2011) derive an upper limit $v_{imp}/v_{esc} \lesssim 1.2$, or a velocity at infinity $v_\infty \lesssim 0.7$ km s$^{-1}$. This result was confirmed by the subsequent SPH study of *Arakawa et al.* (2019).

This stringent velocity limit is not characteristic of KBO impact speeds onto Pluto today (*Greenstreet et al.*, 2015). In the aKB, it would imply an upper limit on characteristic eccentricities (*e*) and inclinations (*i*) such that $\sqrt{e^2 + \sin^2 i} \lesssim 0.1$. Such a dynamic limit would have been easily met during the pebble accretion phase before nebular gas dispersal (Fig. 1), but probably would have been met too well. Dynamical friction with the pebble swarm would have kept all the nascent proto-Plutos on near-circular, non-interacting orbits. Once the nebular gas dispersed, however, self-stirring of the proto-Pluto-rich planetesimal swarm (Fig. 2) plus long-range perturbations by Neptune should have triggered crossing orbits and ideal (i.e., low) velocity conditions for a Charon-forming impact (*Canup*, 2005). From calculations illustrated in *Morbidelli and Rickman* (2015), for 1000 embedded Pluto-mass bodies in the aKB disk, the orbital excitation limit above would have been easily met for planetesimal disk lifetimes $t_d$ <100 Myr, especially as self-excitation increases as $\sqrt{t_d}$ (see their Fig. 1).

The situation changes drastically once Neptune begins migrating through the disk, however. Collision velocities increase smartly (to ~3 km s$^{-1}$) and collision probabilities plunge as the disk is depleted by scattering (see Fig. 11 in *Nesvorný and Vokrouhlický* [2019]). Other things being equal, the earlier the better for the Charon-forming impact, because once the velocity dispersion between bodies in the aKB climbs above the *Canup* (2011) velocity threshold, the outcomes of oblique impacts between proto-Pluto class bodies (~2000 km in diameter) are restricted to smaller satellites at best (*Arakawa et al.*, 2019).

Were such Charon-forming impacts likely? This is a question distinct from their physical realism or plausibility (i.e., whether a Charon-like body can be formed in a giant impact). The chapter by *Canup et al.* in this volume estimates the mean free time between proto-Pluto collisions as ~few x 10$^7$ yr for 1000 Plutos embedded in the planetesimal disk, from a particle-in-a-box calculation without additional gravitational focusing. Considering that there may have been up to 4000 Plutos (*Nesvorný and Vokrouhlický*, 2016), or an even larger number of proto-Pluto (half-Pluto) mass or greater bodies, and strong gravitational focusing, the mean free time between collisions (for $v_\infty < 0.4$ km s$^{-1}$) could have been 10$^5$ yr or less. This is important because it is not enough to have a Charon-forming impact when the implantation efficiency into the Kuiper belt is of order 10$^{-3}$ (*Morbidelli and Nesvorný*, 2020). Unless Pluto is an exceptional or freak case, there had to be dozens if not hundreds of giant impacts in the aKB in order to make the capture of a dwarf-planet binary into the 3:2 mean motion resonance a likely event (*Stern*, 1991). Maybe most Pluto-sized objects in the aKB experienced at least one giant impact of one sort or another, if the Pluto-Charon binary is good indicator of their overall evolution. The results of all these giant impacts do not have to have resembled Charon of course; the outcomes are stochastic, and depend on $v_\infty$, impact angle (or parameter), initial mass ratio, and internal structural factors (*Canup* 2005, 2011; *Arakawa et al.*, 2019). That all the known dwarf planets of the Kuiper belt (the largest KBOs) have satellites or satellite systems is circumstantial evidence that giant impacts were the rule and not the exception (see *Barr and Schwamb* [2016] and Sec. 4.2 in *Stern et al.* [2018]).

The second constraint from giant impact modeling concerns the densities and structural state(s) of the impactors. Numerical simulations of the Charon-forming impact to date favor, for Pluto-like densities, partially differentiated precursors (*Canup*, 2011). Completely differentiated precursors (i.e., bodies with ice mantles and rock cores) yield, post-impact, a very icy Charon in orbit about Pluto, contrary to Charon's known mean density of 1700 kg m$^{-3}$, whereas totally undifferentiated precursors yield very rock- and organic-rich small satellites, in apparent contradiction to their extremely icy nature (*McKinnon et al.*, 2017). This preference for only partially differentiated precursors has recently been muddled somewhat by the results of *Arakawa et al.* (2019), who were able to produce intact moons with ice mass fractions between 0.4 and 0.6 from impacts between two fully differentiated precursors. The moon/primary mass ratios in these cases range between 10$^{-2}$ and in one instance a Charon-like 10$^{-1}$ (see their Supplementary Fig. 3). The SPH calculations of *Arakawa et al.* (2019) were carried out for a slightly smaller mass scale (about one-half Pluto mass total), but the physical inferences derived their calculations should still be applicable to the Pluto system.

## 2.4. Thermal State, Pre Giant Impact

Constraints on giant-impact-progenitor structural state are most valuable for the inferences they provide on the timing and mode of proto-Pluto accretion. Here we assume that the proto-Pluto impactors (1) matched the Pluto system's bulk composition, (2) were fully dense, and (3) were indeed only partially differentiated prior to the Charon-forming impact (*Canup*, 2011), following the general discussion in *McKinnon et al.* (2017) (though we relax this last constraint below). Surface ice layers in the simulations of *Canup* (2011) comprised 10–15% by mass of the precursor bodies, which can be compared with the ≈35 wt% water ice for the actual system as a whole (when considered in terms of a hydrated rock plus H$_2$O ice composition). The true range of pre-impact differentiation states that lead to a correct Charon bulk density has not been determined, but we will assume that differentiation must have proceeded to ice melting at least, and buoyancy driven separation of some 25-50% of the total available ice (the question of potential solid-state separation is addressed in Sec. 2.6).

Initial, post-accretion interior temperatures are determined by three things: *timing*, *time scale*, and *planetesimal size*. *Timing* refers to the time of accretion with respect to the first condensation of calcium-aluminum inclusions ($t = t_{acc} - t_{CAI}$), i.e., the beginning of the Solar System, and the possibility of radiogenic heating by the short-lived isotope $^{26}$Al. The integrated heat release subsequent to accretion is given by $H(0)\lambda^{-1}e^{-\lambda t}$, where $H(0)$ is the rate of radiogenic heating per unit mass of Pluto rock at $t = 0$ and $\lambda$ is the decay constant for $^{26}$Al. For a solar composition, carbonaceous chondrite-like Pluto rock (*McKinnon et al.*, 2017), $H(0)$ is 1.5 x 10$^{-7}$ W kg$^{-1}$ and $\lambda = 3.07 \times 10^{-14}$ s$^{-1}$ (*Castillo-Rogez et al.*, 2009; *Palme et al.*, 2014). The heat capacity of bulk KBO solid (assumed to be ~2/3 rock and ~1/3 water ice by mass) is approximated as $1150 \times (T/250\ \text{K})$ J kg$^{-1}$ K$^{-1}$, because both the heat capacities of rock and ice are temperature ($T$) dependent and because the carbonaceous fraction is poorly constrained (*McKinnon*, 2002; and see Section 3.2 below). Therefore, the heat necessary to increase the temperature of initial proto-Pluto solids at 40 K (a plausible background temperature in the protoplanetary disk at 25 AU) to 273 K (the low-pressure melting temperature of water ice) would have been provided radiogenically for $t \leq 3$ Myr (cf. *Sekine et al.*, 2017; their Supplementary Fig. 10). Given the latent heat of melting of water ice (335 kJ kg$^{-1}$), 25% (50%) ice melting would have been achieved for $t \leq 2.9$ (2.8) Myr.

These time limits assume instantaneous accretion; prolonged accretion with respect to the $^{26}$Al half-life of 720 kyr will change the estimates. But the important point is that *if* no more than 50% of the ice in the Pluto precursors melted prior to the Charon-forming impact, then the precursor bodies could not have *finished* accreting any earlier than 2.8 Myr after $t_{\text{CAI}}$. Similarly, *Bierson and Nimmo* (2019) find, in their study of KBO porosity evolution, that if multi-100-km KBOs are to retain their inferred high porosities (based on densities <1000 kg m$^{-3}$), they cannot have accreted any earlier than 4 Myr after $t_{\text{CAI}}$. These time scale limits are compatible, though the smaller KBOs are different bodies, and they may have formed later or more slowly than the Pluto progenitors (which may be why the former remained relatively modest in scale).

*Time scale* and *planetesimal size* refer to or control the magnitude and depth (within a body) of accretional heating. The temperature distribution $T(r)$ in a symmetrically accreting uniform sphere can be written (*Schubert et al.*, 1981) as

$$T(r) = \frac{h}{\bar{C}_\text{P}} \left( \frac{4}{3}\pi \rho G r^2 + \frac{\langle v \rangle^2}{2} \right) + T_0 \quad , \tag{2}$$

where $r$ is the instantaneous radius, $\langle v \rangle$ the mean encounter velocity at "infinity," $\rho$ and $T_0$ are the density and temperature of the incoming planetesimals, respectively, $h$ the fraction of the impact energy retained, and $\bar{C}_\text{P}$ the mean heat capacity averaged over the interval $T - T_0$. Figure 4 is an update of Fig. 8 from *McKinnon et al.* (1997) but with the temperature dependent $C_\text{P}(T)$ from above and an uncompressed density for the Pluto precursors of 1800 kg m$^{-3}$ (*McKinnon et al.*, 2017); as in that earlier calculation the kinetic energy term in equation (2) is ignored, in order to focus on the effect of gravitational potential energy and thus to provide *minimum* estimates of $T(r)$ (a point we return to later). In Fig. 4 the temperature dependence of $\bar{C}_\mathrm{P}$ makes $T(r)$ nearly linear with radius.

The significant unknown in Fig. 4 is the appropriate value of the empirical parameter $h$. Its value depends on 1) the burial depth of impact heat, 2) whether buried impact heat can be effectively conducted or advectively mixed by subsequent impacts toward the surface, and 3) whether surface heat can be effectively radiated to space or not (e.g., *Squyres et al.*, 1988). *McKinnon et al.* (2017) offered a simple scaling length to assessing whether heat can be conductively transported to the surface over the accretion time scale $\tau_\mathrm{acc}$:

$$\frac{\kappa}{u_\mathrm{acc}} \sim 10\ \mathrm{m} \times \left(\frac{1000\ \mathrm{km}}{R_\mathrm{final}}\right) \times \left(\frac{\tau_\mathrm{acc}}{10^6\ \mathrm{yr}}\right), \qquad (3)$$

where $\kappa$ is the thermal diffusivity (assumed to be that of porous ice-rock) and $u_\mathrm{acc}$ is the radial rate of growth of the body. For impacts much larger than this scale, impact heat is effectively buried ($h \sim 0.5\text{–}1$), whether directly by shock heating or by ejecta that is too thick to cool before it is buried by subsequent impact debris. Equation (3) implies, even for long accretion times ($\tau_\mathrm{acc} \sim$ several Myr), planetesimal sizes in hierarchical accretion scenarios (e.g., *Kenyon and Bromley*, 2012) are likely far too large for impact heat to be efficiently radiated away during accretion. From Fig. 4, $h \gtrsim 0.3$ implies some water ice melting during accretion of 1000-km-scale proto-Plutos, and for $h \gtrsim 0.7$ the total volume fraction of ice melted within the body becomes substantial (>50%).

In contrast, small-scale, pebble accretion appears ideal for depositing accretional energy right at the surface, where it can be radiated away efficiently. Following *Stevenson et al.* (1986), the radiative equilibrium temperature of an accreting surface, for energy deposited right at the surface, is

$$T(r) = \left[\frac{\rho}{\sigma_\mathrm{SB}}\left(\frac{GM(r)}{r} + \frac{\langle v \rangle^2}{2}\right)\frac{dr}{dt} + T_0^4\right]^{1/4}, \qquad (4)$$

where $M(r)$ the mass contained within a radius $r$, $G$ the gravitational constant, $\sigma_\mathrm{SB}$ the Stefan-Boltzmann constant, and $dr/dt$ the radial growth rate. For a proto-Pluto (~50% of Pluto's mass) accreting at a constant radial rate over $10^5$ ($10^6$) years, $T_0 = 40$ K, and ignoring any contribution from $\langle v \rangle$ (which is justifiable for pebble accretion), the surface temperature at the end of accretion could be as high as ~260 K (150 K). These temperatures are upper limits because no account is taken of heat capacity in equation (4), but they do serve to illustrate that pebble accretion alone is very unlikely to lead to wholesale ice melting, unless accretion occurs on a time scale much more rapid than considered here. On the other hand, even pebble accretion is unlikely to prevent accretional heating to 100 K or more, which implies bulk vaporization of "supervolatiles" such as $N_2$, CO, and $CH_4$, the rapid crystallization of any amorphous $H_2O$ ice (and expulsion of trapped supervolatiles; *Kouchi and Sirono*, 2001), and the early formation of atmospheres on protoplanets in the aKB (*Stern and Trafton*, 2008).

**2.5. Thermal State, Post Giant Impact**

Given one or both precursor bodies in a partially differentiated state, the Charon-forming impact may have pushed at least Pluto over the differentiation finish line. The release of gravitational potential energy plus any kinetic energy at infinity (which, unlike the case for pebble or other forms of runaway accretion, cannot justifiably be assumed to be negligible) can result in a global temperature rise of ~50-75 K for Pluto (*McKinnon*, 1989; *Canup*, 2005). The corresponding heat release would have been ~50-90 kJ kg$^{-1}$, sufficient to melt 50% of Pluto's ice complement if globally distributed. Impact heating can be highly localized of course, and numerical results in *Canup* (2005) and *Sekine et al.* (2017) show regional increases of ~150-200 K, so the implications for global differentiation are less clear.

There are additional sources of heat that could have driven differentiation in the early years after the giant impact: tidal heating owing to Charon's orbital evolution, heat of reaction due to serpentinization of ultramafic minerals, radiogenic heating by long-lived isotopes (U, Th, and $^{40}$K) and the gravitational potential energy released by differentiation itself (e.g., *McKinnon et al.*, 1997; *Robuchon and Nimmo*, 2011; cf. chapter by Nimmo and McKinnon in this volume). None of these heat sources is necessarily huge on a (geologically) short time scale, but over $10^7$ yr or more have could easily provided an additional, global temperature increase in excess of 50 K to a body that is already partially differentiated and post giant impact. We conclude that as long as the precursor bodies were partially differentiated, the giant impact and its aftermath likely resulted in a fully differentiated (ice from rock) Pluto. A further inference is that Pluto's ocean dates from this time (again see chapter by Nimmo and McKinnon in this volume; cf. *Bierson et al.* [2020]).

The outlook for Charon is less clear. It is likely that it too emerged from the giant impact in an at least partially differentiated state, if only because it would have accumulated icy impact disk debris in the first $10^2$-to-$10^3$ yrs as it orbited Pluto (*Canup*, 2011; *Arakawa et al.*, 2019). Direct giant-impact heating of Charon would nominally have been modest, because Charon would derive (in the "intact capture" mode) almost entirely from material distant from the impact point. Both *Canup* (2005) and *Sekine et al.* (2017) estimate very low temperature enhancements for Charon, ≲30 K. None of the impact calculations to date have incorporated material strength or frictional dissipation, however. This does not affect the shock heating or gravitational potential energy aspects of the giant impact, but for bodies that undergo substantial distortion or strain (such as Charon), such dissipation and heating may prove quite important to the internal energy balance (e.g., *Melosh and Ivanov*, 2018; *Ensenhuber et al.*, 2018). For further discussion of Charon's early thermal evolution, see the chapter by Spencer et al. in this volume.

**2.6 Synthesis and Unresolved Issues**

Figure 5 summarizes the steps involved in forming the Kuiper belt and Pluto-Charon within it, as outlined in this section. Starting with the gaseous and dusty protosolar disk, collective instabilities such as the streaming instability caused local concentrations of small particles (pebbles) to intermittently exceed the threshold for gravitational instability. These instabilities created the initial size distribution of planetesimals in the original trans-Neptunian region (from ~20 to 30 AU) or aKB, with characteristic sizes near 100 km (e.g., *Morbidelli and Nesvorný*, 2020). Further growth was first driven by hierarchical coagulation, but eventually pebble accretion became dominant (*Johansen and Lambrechts*, 2017). As long as the nebular gas persisted, aerodynamic gas damping drove continual, relatively efficient "pebble accretion" onto the growing bodies. How far this proceeded is unclear, but mass growth by an order of magnitude at least seems likely (e.g., Fig. 1), whereas growth to Mars mass and beyond does not appear to have occurred (though why not, and why hundreds if not thousands of Pluto-scale bodies formed instead, is not exactly clear). We emphasize that pebble accretion is a process distinct from the original gravitational instabilities involving pebbles (pebble cloud collapse). In any event, pebble accretion relies on the presence of gas, so once the gas component of the disk dissipates, accretion via hierarchical coagulation necessarily takes over. Such a gas-free planetesimal disk, on its own, was likely an ideal dynamical environment — in terms of sufficient number density and low encounter speeds — for the accretion of dwarf-planet mass bodies, and plausibly saw thousands of relatively "giant" impacts among them, including that which birthed Charon.

But this aKB Camelot was likely short-lived. According to current thinking, Neptune began migrating into the planetesimal disk within a few 10s of Myr, ultimately leading to the giant planet instability and rearrangement of the orbits of much of the Solar System. The bodies in the aKB were scattered, the population there plummeted and encounter velocities increased so that collisional grinding of smaller bodies ensued and giant impacts could no longer yield Charon-like outcomes (except rarely, as there is almost always a low-velocity tail to encounter velocity distributions among planetesimals), and the modern Kuiper belt was emplaced or installed (modern in terms of structure; the populations were larger than today of course, and have been decreasing in number ever since; *Greenstreet et al.*, 2015). Neptune's orbital migration during this time is thought to have been relatively slow, and grainy (jittery, in terms of orbital elements) due to scattering encounters with Pluto-mass bodies (Fig. 3; *Nesvorný*, 2018). Pluto-Charon was ultimately captured into the 3:2 MMR and other resonances with Neptune (*Malhotra and Williams*, 1997), and its outward migration in *a* did not end until Neptune's did, perhaps some 100 Myr after the beginning of the Solar System. The implications of these orbital changes for Charon and the small satellites are taken up at the end of Sec. 4.

Dynamical inferences for a relatively slow Charon-forming collision are consistent with late growth of the progenitor bodies in the ancestral planetesimal disk beyond Neptune. That the progenitor or precursor protoplanets were only partially differentiated is likely a signature of earlier growth by pebble accretion in the presence of nebular gas, i.e., by collisions so small that accretional heat was not deeply buried. Accretion of the progenitors could not have occurred too early, however, or full differentiation would have been driven by $^{26}$Al decay. Nor was pebble accretion the entire story, as accretion of larger planetesimals appears necessary to provide the (buried) impact heat required for partial differentiation. Once the giant planet instability initiated, however, impact speeds in the planetesimal disk would have necessarily climbed to several km s$^{-1}$, inconsistent with forming Charon according to our best simulations, though not with violent collisions generally (e.g., the formation of Sputnik basin, which we note is far too small to be the impact scar of the Charon-forming impact).

If a rock-rich, as opposed to an ice-rich, Charon can indeed be formed in a giant impact involving fully differentiated precursor protoplanets (*Arakawa et al.*, 2019), these inferences change. In this case the progenitor bodies would have had to either form early (to take advantage of $^{26}$Al heating) or after nebular dispersal (to accrete from substantial planetesimals). Only by forming in the latter, waning half of protoplanetary gas nebula's nominal lifetime of ~5 Myr can the combination of pebble and planetesimal accretion yield partially differentiated bodies of the proto-Pluto scale, especially as sizeable (~100 km scale) bodies may have dominated the planetesimal swarm in the emerging planetesimal disk. Further numerical work may shed light on the issue of precursor differentiation state, perhaps involving benchmarking between and higher resolution simulations of the Pluto-system-forming giant impact.

More work on the pathways to differentiation may also prove fruitful. The chapter by Canup et al. in this volume illustrates some possible initial conditions and outcomes. Those models adopt the perspective of *Desch et al.* (2009) in which differentiation is triggered when the ammonia-water peritectic temperature of 175 K is reached. Ammonia has been detected on Pluto and Charon (chapters by Cruikshank et al. and Protopapa et al. in this volume), and comets (which ultimately came from the same planetesimal disk as KBOs; *Morbidelli and Nesvorný*, 2020) contain $NH_3$ at the 1% level compared with $H_2O$ ice (*Mumma and Charnley*, 2011), so the formation of minor ammonia-water melt at low temperatures is a well-supported inference. *Desch et al.* (2009) argue that this melt should allow separation of rock from ice, that is, descent of rock through a weakened ice-rock matrix, but this obviously requires large "rock" masses or density concentrations (not pebbles) to be effective, masses or concentrations whose existence cannot be decided on the basis of theory alone. Our approach above is more conservative, requiring complete local melting of the icy component to insure rock from ice differentiation, but is not necessarily more correct. Further ice deformation experiments in the ammonia-water and ammonia-water-silicate systems at stresses and strain rates more appropriate to icy bodies (cf. *Durham et al.*, 1993) are needed.

In terms of observations, ever deeper, more complete, and better characterized surveys of the Kuiper belt and inner Oort Cloud, e.g., OSSOS–the Outer Solar System Origins Survey (*Bannister et al.*, 2018) and DEEP–the Deep Ecliptic Exploration Project (*Trujillo et al.*, 2019), and new large surveys such as Pan-STARRS and V. C. Rubin Observatory (LSST), in coming years will obviously improve enormously our picture of the structure (SFDs, dynamical classes) of the trans-Neptunian populations. The individual character of the KBOs, their colors, binarity, etc., will also come into better focus. All of these in turn will drive improvements in numerical models that account for these characteristics, thus fostering a deeper understanding of how the Solar System as a whole, and Pluto in particular, came to be. The perplexing puzzle of the most extreme trans-*Plutonian* objects (i.e., Sedna and brethren) and their possible relation to a distant massive planet (e.g., *Sheppard et al.*, 2019) promise to be a revelation in this regard.

## 3. COMPOSITION OF PLUTO AND ITS MOONS

In this section we examine compositional issues for the Pluto system, with a focus on the volatile ice budget of Pluto specifically. Section 3.1 introduces our understanding of the volatile ice reservoirs (volatile ice generally meaning ices other than water ice, which is assumed to be a major bulk component of the Pluto system [*McKinnon and Mueller*, 1988; *McKinnon et al.*, 2017]), and discusses at length the possible origin of Pluto's all-important $N_2$. This is then contrasted with the apparent paucity of CO ice on Pluto in Sec. 3.2. Tied to these discussions is the possible role of organic matter in Pluto's interior, and the possibility of bulk organic carbon and/or graphite within Pluto and Charon is addressed in Sec. 3.3. Comparisons with other KBOs or icy satellites that may or may not resemble Pluto or Charon in terms of volatile ice abundance form a brief Sec. 3.4, and we end with a synthesis and discussion of unresolved issues in Sec. 3.5. Implications of the apparent compositions of the small moons are taken up later, in Sec. 4.4.

### 3.1. Volatile Budgets and Cometary Provenance

The volatiles that comprise the atmosphere and surface ices of Pluto, Charon, and the small moons Nix, Styx, Kerberos, and Hydra are described in the chapters by Summers et al., Cruikshank et al., Protopapa et al., and Porter et al. in this volume. Regarding Pluto's neutral atmosphere during the *New Horizons* encounter, the primary gas is $N_2$ followed by minor $CH_4$ (~0.3%; *Young et al.*, 2018) and a trace of CO (~0.05%; *Lellouch et al.*, 2017). Minor though $CH_4$ may be, it is the feedstock for the UV photochemical production of hydrocarbons such as $C_2H_2$, $C_2H_4$, and $C_2H_6$ and nitriles such as HCN in the upper atmosphere. The principal atmospheric components $N_2>CH_4>CO$ (in order of abundance) are maintained in vapor pressure equilibrium with $N_2$, $CH_4$ and CO ices in various combinations at Pluto's surface. Regarding surfaces, $H_2O$ and $NH_3$-hydrate ices were detected on the surface of Pluto and Charon (*Grundy et al.*, 2016; *Dalle Ore et al.*, 2018, 2019; *Cook et al.*, 2019), confirming and extending earlier ground-based detections, and $H_2O$ ice and an ammoniated species dominate the surfaces of the small moons (*Cook et al.*, 2018). Molecular $O_2$, abundant in the coma of comet 67P/Churyumov-Gerasimenko (hereafter 67P) (*Bieler at al.*, 2015), was not detected in Pluto's atmosphere at the ~$10^{-5}$ level with respect to $N_2$ (*Kammer et al.*, 2017).

Less volatile methanol ($CH_3OH$) and hydrocarbon ices have been detected within and near to the dark Cthulhu terrain on Pluto (*Cook et al.*, 2018), and are thought to be derived ultimately from both UV photochemistry and ion irradiation at Pluto's surface, as are the dark, red, tholin-like materials that nominally coat Cthulhu and other dark regions ("Tenebrae") on Pluto (*Protopapa et al.*, 2017; *Grundy et al.*, 2018). What has not been detected on either Pluto or Charon is $CO_2$ ice, which is common on many icy satellites and abundant and well distributed on Triton in particular (*Grundy et al.*, 2010; *Merlin et al.*, 2018; and references therein).

In the rest of this subsection we focus on the implications of Pluto's $N_2$ abundance, both in bulk and in relation to other volatiles and chemical reservoirs, drawing from and expanding upon the analysis in *Glein and Waite* (2018).

***3.1.1. Pluto's $N_2$ inventory.*** On Pluto today, the arguably most abundant observed volatile is molecular nitrogen (see the chapter by Cruikshank et al. in this volume). The total amount of $N_2$ serves as a key constraint on Pluto's origin and evolution, as will be shown below. *Glein and Waite* (2018) attempted to estimate the inventory of $N_2$ that can be deduced from *New Horizons* data. They separated this inventory into four reservoirs, which they termed atmosphere, escape, photochemistry, and surface (Table 1). The amount of $N_2$ in the atmosphere ($1\times10^{15}$ mol) was calculated from the atmospheric pressure (~12 μbar) in 2015. The amount of $N_2$ that has escaped from Pluto's atmosphere was estimated to be a relatively modest $5\times10^{16}$ mol (compared with pre-encounter thinking), if the cold exobase in 2015 (65-70 K; *Young et al.*, 2018) is representative of Pluto's history. This is a conservative assumption, as Pluto's atmosphere has likely gone through major changes over geologic time (see Sec. 4.2). The photochemical inventory at the surface corresponds to the amount of $N_2$ that has been incorporated into the products of $CH_4$-$N_2$ photolysis (e.g., HCN) from photochemical modeling. This amount was taken to be $2\times10^{18}$ mol, again based on the notion that the past might have been like the present. The surface $N_2$ reservoir was assumed to be dominated by the glacial ice sheet called Sputnik Planitia (SP; see the chapter by White et al. in this volume). Based on estimated dimensions of this feature (e.g., a depth of 3-10 km to enable convection), it was found that Sputnik Planitia could contain $(0.4\text{-to-}3)\times10^{20}$ mol of $N_2$.

Because the other reservoirs above appear to be substantially smaller than that within SP, we take the range for SP to be a current best estimate for the inventory of $N_2$ on Pluto. If this estimate is normalized by *Glein and Waite*'s (2018) value for the water abundance on Pluto ($2\times10^{23}$ mol), then an $N_2/H_2O$ ratio of $(0.2-1)\times10^{-3}$ is obtained. However, one should keep in mind that this range could be a serious underestimate as it excludes any subsurface reservoirs (e.g., $N_2$ liquid trapped in crustal pore space [cf. Table 1]; $N_2$ trapped as clathrate in Pluto's ice shell, $N_2$ dissolved in a subsurface water ocean), and does not account for possible $N_2$ loss from the Charon-forming giant impact, or added by later cometary bombardment (the latter perhaps $2\times10^{17}$ mol; *Singer and Stern*, 2015).

***3.1.2. Accreted $N_2$ as nitrogen source?*** It is thought that there were three major reservoirs of nitrogen atoms in the early outer solar system (*Miller et al.*, 2019): $N_2$, ammonia, and organic matter. (Recently detected ammonium salts on the surface of comet 67P are plausibly the product of chemical reactions within the comet [*Poch et al.*, 2019].) Molecular nitrogen is considered to have been the most abundant form of nitrogen, because the solar ratio of $^{14}N/^{15}N$ (~440) is much different (isotopically lighter) from those in primordial ammonia (~135) and organic nitrogen (~230; see *Füri and Marti*, 2015; *Miller et al.*, 2019). The formation temperature of pebbles and planetesimals is a key factor in determining whether appreciable $N_2$ can be accreted by larger solid bodies such as Pluto. It is difficult to accrete $N_2$ in solid materials because of its great volatility. However, at sufficiently low temperatures in the solar nebula (e.g., <50 K), $N_2$ can be trapped in clathrate hydrates or amorphous ice (*Hersant et al.*, 2004; *Mousis et al.*, 2012). At even lower temperatures (e.g., ≲20 K), $N_2$ ice (or a solid solution of CO-$N_2$) can directly condense (*Hersant et al.*, 2004; *Mousis et al.*, 2012). To first-order, the accreted amount of $N_2$ should be inversely related to the formation temperature. The "formation temperature" may be thought of as the average temperature experienced by pebbles and planetesimals that formed across a potentially wide range of heliocentric distances and at different times (see Section 2.1). If primordial $N_2$ is present on Pluto, it could provide us with some insight into the thermal history of the environment where Pluto formed.

Until recently, it was not clear whether primordial $N_2$ is present in solids in the solar system. But then, the *Rosetta* mission discovered $N_2$ being outgassed from comet 67P in 2014. Subsequent analysis provided an estimate of the $N_2/H_2O$ ratio [$(8.9\pm2.4)\times10^{-4}$] in ices at the near-surface of the comet (*Rubin et al.*, 2015, 2019). This range is remarkably similar to that estimated above for Pluto, if Pluto is ~1/3 $H_2O$ by mass. This similarity serves as the foundation of *Glein and Waite*'s (2018) suggestion that Pluto's $N_2$ might be primordial. However, there are complications that could make a primordial origin either less or more likely. One is the apparently low CO/$N_2$ ratio at Pluto's surface compared with comets (see section 3.2). The second potential issue is that this scenario may impose implausible restrictions. If we assume that Pluto started with an $N_2/H_2O$ ratio similar to comet 67P, then at least ~20% of all accreted $N_2$ must already be accounted for (*Glein and Waite*, 2018). This may leave too little margin to accommodate other internal reservoirs or adjustments such as loss from giant impact.

On the other hand, a recent observation suggests that it is possible to obtain larger inventories of primordial $N_2$. This observation is the $N_2$-rich comet C/2016 R2 (PanSTARRS). It is a rare "blue comet" whose coma is dominated by CO. Water appears to be only a trace species at this comet ($H_2O/CO \approx 3 \times 10^{-3}$), and various other volatiles have been observed including $CO_2$, methanol, and methane (*Biver et al.*, 2018; *McKay et al.*, 2019). The $N_2/H_2O$ ratio is estimated to be ~15 (*McKay et al.*, 2019), which is roughly four orders of magnitude higher than that at comet 67P.

The provenance of comet R2 is unknown, but one possibility is that it formed at/near the CO/$N_2$ ice line in the protosolar nebula, where these volatiles could have been concentrated (e.g., *Öberg and Wordsworth*, 2019). Formation in such an ultracold region of the nebula does not in and of itself explain the very low water vapor or dust production of an Oort cloud comet observed between 2.6 and 2.8 AU, however, production which should be apparent at these distances (*McKay et al.*, 2019). Nor would an (apparently) water-ice- and dust-poor composition be consistent with supposing that *all* cometary bodies originally formed in the aKB were once R2-like but underwent minor $^{26}$Al heating and $N_2$ loss (cf. *Mousis et al.*, 2012). Accordingly, *Biver et al.* (2018) suggest that R2 may represent a collisional fragment from the volatile-ice-rich surface of a large (and at least somewhat differentiated) KBO. Setting aside this possibility for the sake of argument, the high $N_2$ abundance of R2 implies that *Glein and Waite*'s (2018) $N_2$ inventory for Pluto can be reproduced if the water fraction in Pluto that came from compositionally anomalous, R2-like comets is $(1-7) \times 10^{-5}$. This does not seem prohibitive. If Pluto accreted a larger fraction of R2-like comets, then it could have started with more $N_2$ than in the inventory of *Glein and Waite* (2018).

*3.1.3. Accreted $NH_3$ as nitrogen source?* Another candidate source of nitrogen atoms is ammonia. There are three reasons why primordial, condensed ammonia is a plausible source of Pluto's $N_2$. First, ammonia in some form has been detected on Pluto, Charon, and Nix and Hydra, so it is available (e.g., *Dalle Ore et al.*, 2018, 2019; *Cook et al.*, 2018). Second, Pluto could have accreted a large quantity of ammonia. The canonical cometary abundance of ~1% with respect to water (*Mumma and Charnley*, 2011) would translate to an $N_2/H_2O$ ratio of up to ~0.5%, above the current upper limit of ~0.1% for Pluto (Section 3.1.1, *Glein and Waite*, 2018). An ammonia source of $N_2$ may be ~5-25 times larger than what is needed. However, the amount of margin is likely to be less, as the current existence of ammonia means that not all of it has been converted to $N_2$. There could even be a shortfall if Pluto has an ammonia-rich subsurface ocean (*Nimmo et al.*, 2016). The third plausibility argument for ammonia as an $N_2$ source on Pluto is one of analogy. It is inferred that ammonia can be converted to $N_2$ on icy worlds. The nitrogen isotopic composition of atmospheric $N_2$ on Titan ($^{14}N/^{15}N \approx 168$) shows that isotopically heavy, primordial ammonia ($^{14}N/^{15}N \approx 136 \pm 6$; *Shinnaka et al.*, 2016) must have made a major contribution to Titan's $N_2$ (*Mandt et al.*, 2014). Ammonia could have been the sole contributor if sufficient photochemical fractionation has occurred to reconcile these two values (*Krasnopolsky*, 2016). The exact mechanisms that converted ammonia to $N_2$ on Titan (and hence Pluto) remain uncertain. The general requirement is a high-energy process to decompose ammonia. It has been proposed that atmospheric chemistry (*Atreya et al.*, 1978; *McKay et al.*, 1988), impact chemistry (*Sekine et al.*, 2011), or internal geochemistry (*Glein*, 2015) might be important in this respect for Titan (see *Atreya et al.* [2009] for a more complete discussion).

The mechanism of *Atreya et al.* (1978) relies upon solar ultraviolet light to convert ammonia to $N_2$. The process is initiated by the photolysis of gaseous ammonia, which produces the amino radical ($NH_2$). This species then undergoes dimerization to produce hydrazine ($N_2H_4$). The formation of hydrazine is crucial because this is where two nitrogen atoms become bonded together. Additional photochemical reactions result in the production of $N_2$ from hydrazine. The efficiency of this process depends strongly on the atmospheric temperature, which should be high enough but not too high (e.g., 150-250 K). Temperatures need to be high enough to keep appreciable amounts of ammonia and hydrazine in the gas phase. However, if temperatures are too high, then water vapor is also present, and hydroxyl radicals (OH) derived from water dissociation act as a scavenger of amino radicals. This would prevent the formation of the key intermediate hydrazine.

Shock chemistry can also convert atmospheric (or surface, see below) ammonia to $N_2$ (*McKay et al.*, 1988). This is where meteoroids generate shock waves while traversing the atmosphere. These shocks expand away from the source, and compress and heat the surrounding air to high pressures and temperatures (of order $10^3$ K). Affected air parcels reach chemical equilibrium at these conditions, then quickly cool which causes this state to be quenched. The shock-induced equilibrium favors the formation of $N_2$, and organic compounds if methane is present. In addition to requiring a sufficient flux of high-velocity (>4 km s$^{-1}$) impactors that transfer a substantial fraction of their energy into shocks (not so likely in the Kuiper belt [*Greenstreet et al.*, 2015] but plausible in Pluto's thought-to-be natal 20–30 AU region), this mechanism also requires the presence of ammonia gas. Hence, it is most relevant to syn- and post-instability accretion (Section 2), when the impact rate and speeds were highest and the atmosphere may have been warm enough to support $NH_3$ vapor.

Pluto could have experienced substantial accretional heating (see sections 2.4 and 2.5). If such heating enabled the differentiation of Pluto by ice melting, then it is reasonable to envision the mobilization of volatiles toward the surface. This could have led to the formation of an early atmosphere on Pluto, where ammonia photolysis might have occurred. However, modeling has yet to be performed to understand the chemistry of such an atmosphere and test whether a photochemical origin of $N_2$ is viable. Currently, it appears marginally plausible, and the low solar flux at Pluto is an additional challenge. $N_2$ production via atmospheric shocks seems less attractive because impact velocities, even early on, are expected to be relatively low onto Pluto (~1-to-3 km s$^{-1}$; see section 2.3).

An intriguing implication of forming $N_2$ from atmospheric $NH_3$ is that this scenario may provide a way to explain the apparent lack of CO on Pluto relative to $N_2$. It can be envisioned that the most volatile primordial species (e.g., CO) were outgassed earliest and lost from a warm atmosphere, and the present $N_2$ inventory was formed somewhat later (as proposed for Triton by *Lunine and Nolan*, 1992). This sequence of events is consistent with the much greater volatility of CO vs. ammonia. The question of Pluto's CO abundance, in particular its apparent low $CO/N_2$ ratio, is taken up in greater detail in Section 3.2.

*Sekine et al.* (2011) showed that cometary impacts on ammonia-bearing *ice* can lead to the production of $N_2$ as well. Partial production occurs if the impact velocity is >5.5 km s$^{-1}$, and the conversion efficiency reaches 100% at velocities greater than ~10 km s$^{-1}$, though this latter speed is too high to be relevant for Pluto. Impact-induced decomposition of ammonia could have played an important role on Titan during an early heavy bombardment. However, as for atmospheric shock chemistry, impact shock chemistry may not be a major contributor to $N_2$ production on Pluto because of inadequate impact velocities. If Titan's atmospheric $N_2$ is impact-derived, then it would have a different origin from Pluto's (*Sekine et al.*, 2011).

Ammonia can be decomposed to $N_2$ in planetary interiors (*Glein*, 2015). At equilibrium, $N_2$ formation is favored by higher temperatures, lower pressures, more oxidized systems, higher pH (if the fluid is water-rich), and higher concentrations of bulk N. It is commonly assumed that chemical equilibrium could be reached over geologic timescales, but this has not been tested for the ammonia-$N_2$ system. The nitrogen speciation could be kinetically controlled at lower temperatures, depending on the relationship between the rate of ammonia decomposition (which increases with temperature) and the residence time of fluids in the environment of interest. There are two general environments inside icy worlds where the decomposition of primordial ammonia to $N_2$ might occur. One is at the water-rock interface where internal heating can drive ocean water circulation through the seafloor (*Shock and McKinnon*, 1993). This type of environment has the whole ocean inventory of ammonia at its disposal, but the ammonia is subjected to elevated temperatures for only a limited duration while it is circulated through the rock. The second environment of interest is the deeper interior. This environment may contain ammonium-bearing minerals that formed during differentiation (e.g., *De Sanctis et al.*, 2015). Heating may cause these minerals to release ammonia, which could then undergo conversion to $N_2$ if the geochemical conditions are suitable.

The essential requirement for a geochemical origin of Pluto's $N_2$ from accreted ammonia is the occurrence of high temperatures in the interior. This requires Pluto to be a differentiated body with a rocky core. It is thought that Pluto is probably differentiated (see the chapter by Nimmo and McKinnon in this volume). Assuming that this requirement is met, seafloor hydrothermal circulation may be possible, as shown by the preliminary modeling of *Gabasova et al.* (2018). However, the observed composition of Pluto seems inconsistent with the idea of generating $N_2$ in seafloor hydrothermal systems. The aqueous $NH_3$-$N_2$ and $CH_4$-$CO_2$ systems behave similarly, so conditions that favor $N_2$ usually also favor $CO_2$ (*Glein et al.*, 2008). $CO_2$ production might be expected to accompany $N_2$ production if $N_2$ was formed in seafloor hydrothermal systems. Instead, the surface of Pluto is methane-rich and $CO_2$ has not been detected (*Grundy et al.*, 2016; cf. chapter by Cruikshank et al. in this volume). Metamorphism of ammoniated minerals in a rocky core could be a more promising source of $N_2$. *Bishop et al.* (2002) reported that mineral-bound ammonium is released at ~300°C. This temperature can easily be exceeded inside Pluto if a rocky core is present (*McKinnon et al.*, 1997; *Bierson et al.*, 2018; Fig. 6). However, questions remain regarding the post-differentiation inventory of ammoniated minerals in the core, whether the released ammonium would speciate to $N_2$, and how much $N_2$ formed in this way could be outgassed from the core and delivered to the surface.

*3.1.4. Organic matter as nitrogen source?* The idea that organic matter can serve as a source of $N_2$ on icy worlds is relatively new. There are two basic arguments. One is that these worlds could have started with a great deal of N-bearing organic matter if the current cometary data (from Halley, Wild 2, and 67P) are representative of pebbles and planetesimals that built bigger bodies in the outer solar system. The second argument is that organic matter can be "cooked" in the rocky cores of larger bodies if the body of interest is differentiated. This cooking process can provide sufficient thermal energy to break carbon-nitrogen bonds, which liberates nitrogen into the fluid phase. In a long-lived geologic system, volatilized nitrogen may come to chemical equilibrium that is determined by the temperature, pressure, and composition of the system. *Miller et al.* (2019) put these arguments in quantitative terms, and combined them with constraints on the ratios of $^{15}N/^{14}N$ and $^{36}Ar/^{14}N$ from the *Huygens* probe, to make the case that perhaps 50% of Titan's atmospheric $N_2$ was derived from organic N. This cooking process (organic pyrolysis) could also be relevant to the origin of methane on Titan, since methane is an abundant product (*Okumura and Mimura*, 2011). It is natural to wonder whether organic pyrolysis might be applicable to Pluto, whose surface composition resembles Titan's atmospheric composition. *Kamata et al.* (2019) suggested that methane and $N_2$ could be formed in this manner on Pluto, as a consequence of high interior temperatures (Fig. 6); *Stern et al.* (2018), while favoring an $NH_3$ source, earlier made a similar suggestion for the origin of Pluto's $N_2$. Below, we elaborate on some of the geochemical considerations.

Is an organic source of Pluto's $N_2$ large enough? A simple density mixing calculation can be performed to estimate the amount of organic N that could have been accreted by Pluto. We attempt to reproduce an uncompressed density for Pluto of 1.8 g cm$^{-3}$ (*McKinnon et al.*, 2017) in terms of a mixture of rock (3.0 g cm$^{-3}$), generic "water" (0.95 g cm$^{-3}$), and graphite (2.2 g cm$^{-3}$; see below). We further assume that the rock is ~30 wt. % $SiO_2$ (similar to the Murchison meteorite), and the initial ratio of $C_{org}/Si$ was ~5.5 as in comet 67P (*Bardyn et al.*, 2017). The latter assumptions imply that the mass ratio of graphite/rock should be ~0.33. The result of this calculation is that bulk Pluto could be composed of ~54 wt. % rock, ~28 wt. % water, and ~18 wt. % graphite (cf. section 3.3). The "rocky" core is predicted to be ~75% rock and ~25% graphite by mass. In this model, Pluto's present inventory of graphite is $1.9 \times 10^{23}$ mol. Assuming that the precursor to graphite was organic matter analogous to what was detected at comet 67P (with N/C ≈ 0.035; *Fray et al.*, 2017), then the initial inventory of organic N on Pluto would have been $67 \times 10^{20}$ mol. Therefore, the maximum theoretical yield of $N_2$ from organic N is estimated to be $34 \times 10^{20}$ mol. This is 11-85 times larger than *Glein and Waite*'s (2018) inventory based on *New Horizons* data. There is thus the potential to explain how Pluto got its $N_2$ inventory from an organic source, provided that 1-9% of accreted organic N was converted to $N_2$, delivered to the surface, and retained there.

We can gain insight into the fate of organic N in a rocky core on Pluto by determining what would be favored at chemical equilibrium. Here, we conduct an exercise in which equilibrium is calculated in the C-N-O-H system for a mixture of CI chondritic rock + cometary organic matter, based on *Lodders* (2003) and *Bardyn et al.* (2017). The adopted proportions are 100 C: 4 N: 165 H, which includes hydrogen atoms contributed by hydrated silicates (i.e., the organic mass fraction is roughly an order of magnitude larger than the several wt% in CI chondrites; cf. Sec. 3.3). A similar composition may have been produced on Pluto by aqueous alteration accompanying water-rock separation during the formation of a rocky core. The oxygen abundance is represented by the oxygen fugacity (oxygen partial pressure corrected for non-ideality), which provides a measure of the oxidation state of the system. The oxygen fugacity ($fO_2$) is treated as a free parameter, and is expressed relative to the value for the fayalite-magnetite-quartz (FMQ) mineral buffer

$$3Fe_2SiO_4, \text{fayalite} + O_2, \text{gas} = 2Fe_3O_4, \text{magnetite} + 3SiO_2, \text{quartz} \qquad , \qquad (5)$$

which is a commonly used geochemical point of reference (e.g., *Shock and McKinnon*, 1993). For simplicity, ideal gas calculations are performed (*McBride and Gordon*, 1996). We consider a pressure of 1900 bar, which may be the lowest pressure in the core (*McKinnon et al.*, 2017) and thus where ideal gas calculations will have the least error. In our example, the temperature is set to 500°C. This is close to the temperature for serpentine dehydration (e.g., *Glein et al.*, 2018), which could drive volatile outgassing from the core.

Figure 7 shows the computed speciation for the model system. The most relevant result is that $N_2$ is always the dominant form of nitrogen. It may seem counterintuitive that $N_2$ is favored even at strongly reducing conditions, where one may expect ammonia to predominate. However, this behavior is caused by the system being hydrogen-limited. There are not enough hydrogen atoms available (in this model) to prevent $N_2$ production. Most of the H inventory is used to make methane (water) at reducing (oxidizing) conditions owing to the larger abundance of C (O) compared with N. Hydrogen limitation also explains why graphite is the dominant form of carbon at lower oxygen fugacities. There remain important questions with regard to the accessibility of other hydrogen sources (hydrothermally-circulating ocean water, burial of free water in pores) to different depths in the hypothesized core. Yet, it has not escaped our attention that the predicted volatile assemblage at reducing conditions (Fig. 7) shows an intriguing gross similarity to Pluto's surface volatile composition (see the chapter by Cruikshank et al. in this volume). Future work will need to constrain geochemical conditions in the interior, as well as broaden the perspective to the possibility of multiple sources of nitrogen (as proposed for Titan; *Miller et al.*, 2019) in preparation for spacecraft *in situ* measurements of isotopically heavy nitrogen ($^{14}N^{15}N$) and primordial argon ($^{36}Ar$) on Pluto. These promise to further enlighten us on the origin of Pluto's $N_2$ (see *Glein and Waite*, 2018).

**3.2. The Missing CO Conundrum.**

The CO/$N_2$ ratio at Pluto's surface ($\sim 4\times 10^{-3}$; *Glein and Waite*, 2018) is orders of magnitude below that of known comets (e.g., CO/$N_2 \approx 35$ for comet 67P; *Rubin et al.*, 2019). Specifically, the hemispheric CO/$N_2$ ratio in Pluto's surface ice ranges between 2.5-and-$5\times 10^{-3}$ (*Owen et al.*, 1993; *Merlin*, 2015) while the atmospheric ratio is $5\times 10^{-4}$ (*Lellouch et al.*, 2017); these numbers are consistent with CO dissolved in solid solution in $N_2$ ice with their relative vapor pressures determined by Raoult's Law. While there is not yet a consensus solution to the "missing CO" problem, there is no shortage of possible solutions. One solution is that the low surface abundance of CO reflects low accreted or retained abundances of super-volatile species (CO, $N_2$, Ar), whereas the observed $N_2$ is a secondary product evolved from other N-bearing reservoirs, as detailed in sections 3.1.3. and 3.1.4. Alternative solutions include hydrothermal destruction of CO at the base of a subsurface ocean (*Shock and McKinnon*, 1993), hydrolysis of CO to formate (HCOO⁻) in the ocean itself (*Neveu et al.*, 2015), preferential burial of CO (relative to $N_2$) in Sputnik Planitia (*Glein and Waite*, 2018), and preferential sequestration of CO (relative to $N_2$) in subsurface clathrates (*Kamata et al.*, 2019). In this subsection we address these alternative possibilities.

In aqueous systems dissolved CO is a reactive and unstable species, and prefers to either be reduced to $CH_4$ or other organic species or oxidized to $CO_2$ and related species, depending on temperature, pH, and oxidation state (less so on pressure) (e.g., *Shock and McKinnon*, 1993). *Glein and Waite* (2018) looked specifically at metastable chemical speciation between aqueous CO, $CO_2$, $HCO_3^-$ (bicarbonate), $CO_3^{-2}$ (carbonate), HCOOH (formic acid), and $HCOO^-$ (formate) at conditions representative of the bottom of Pluto's putative ocean (0° C, 1900 bar) (i.e., not simply CO and formate). Hydrothermal processing (heating) was not assumed. Figure 8 shows the results. For all but the most acidic oceans, CO is very efficiently converted to other chemical species irrespective of hydrogen activity (meaning hydrogen molal concentration, corrected for non-ideality). That we even observe CO at all means that not *all* of Pluto's initial endowment of primordial CO need be aqueously processed (which seems plausible as volatile outgassing should have occurred before ice melting on the Pluto progenitors; Section 2.4), plus there is always the later input from Kuiper belt bombardment. Indeed, *Glein and Waite* (2018) estimate, based on *Singer and Stern* (2015), that Pluto's surface CO could have been completely supplied by comets over geologic time.

*Glein and Waite* (2018) also hypothesize that as volatile ices condensed to fill the Sputnik basin, over some tens of millions of years following the impact (*Bertrand and Forget*, 2016; *Bertrand et al.*, 2018), Rayleigh-type fractionation caused preferential deposition of CO while driving the atmospheric $CO/N_2$ ratio down, ultimately far down. Because convective transport with Sputnik continuously brings deeper ices to the surface (e.g., *McKinnon et al.*, 2016a), the implication is that the $N_2$-dominated optical surface would have to be controlled and maintained by the surface-atmosphere volatile cycle, while almost the entire ice sheet below the optical surface would necessarily be dominated by CO ice. While we do not view this as impossible *a priori*, it seems unlikely that an ultrathin surface veneer of $N_2$ ice can supply the $N_2$ vapor that appears to be condensing on the uplands to the east of Sputnik Planitia before returning as glacial flow into the basin (*Moore et al.*, 2016; *McKinnon et al.*, 2016a; *Umurhan et al.*, 2017; chapters by White et al. and Moore and Howard in this volume), and $N_2$ ice is ubiquitous across Pluto. Further work on such possible fractionation is warranted, however.

Finally, there is the possibility of CO sequestration by clathrates in Pluto's ice shell

(*Kamata et al.*, 2019). Clathrates are open-cage water-ice crystal structures in which cavities or sites are occupied by guest atoms or molecules (*Sloan and Koh*, 2008). The guest atoms or molecules are not bound to the ice crystal structure, but are "trapped" within, where they are free to rotate and vibrate, and thus capable of being detected spectrally with characteristic absorption bands (e.g., *Dartois et al.*, 2012). Clathrates are found on Earth in industrial and geological settings (e.g., $CH_4$ clathrate in permafrost or abyssal sediments, air clathrate in Antarctica), and have been proposed or invoked in a variety of Solar System settings over the decades (e.g., Mars, Ceres, Europa, Titan, Enceladus, comets, and Pluto). Owing to their instability at low pressures (except when very cold), they have never been directly detected beyond Earth to our knowledge (cf. *Luspay-Kuti et al.*, 2016). While the absence of evidence cannot be taken as evidence of absence, the reader should bear this in mind.

In Pluto's case, preferential freezing out of CO clathrate at the expense $N_2$ clathrate at the base of Plutos's floating ice shell is plausible, as $N_2$ clathrate is less stable (has a higher dissociation pressure than CO; *Sloan and Koh* [2008]), but whether this occurred or is occurring within Pluto today depends on the identities and concentrations of the gas species dissolved in the ocean. For example, it is easy to envisage a situation in which dissolved gases released by Pluto's core are either rich in $CH_4$ or $CO_2$, depending on oxidation state (see Fig. 7). Clathrates of either of these two species are much more stable than CO-clathrate, and could form preferentially ($CO_2$ clathrate if the ocean is acidic; Fig. 7 in *Glein and Waite* [2018]; *Bouquet et al.* [2019]). Either could provide the thermal insulation to stabilize Pluto's ocean as envisaged by *Kamata et al.* (2019), but would be irrelevant to the missing CO story. We note that $CO_2$ clathrate in particular is denser than water, and on its own would sink to the bottom of the ocean, where it would remain very effectively sequestered, and is one possible explanation for the lack of identified $CO_2$ ice at Pluto's surface.

### 3.3. Carbon-rich Pluto Models

The discussions above highlight the potential importance of Pluto's organic fraction—especially that trapped in a radiogenically heated core—for the evolution of its atmosphere and other volatiles. The rock/ice mass ratio of the Pluto system is often stated as being about 2/1 (*McKinnon et al.*, 2017), but this explicitly neglects the potential role of bulk carbonaceous matter, or implicitly assumes that CHON (carbon-hydrogen-oxygen-nitrogen-bearing organic matter) can be counted as low-density material similar to water ice (*McKinnon et al.*, 1997;

*Simonelli et al.*, 1989). Here we explicitly discuss the possibilities and some implications of truly carbon-rich compositions for Pluto and Charon.

Carbonaceous matter is an important cometary component and one likely important in the ancestral Kuiper belt (see, e.g., the review in *McKinnon et al.*, 2008). CI and CM carbonaceous chondrites, which can be considered as proxies for the rocky material accreted by Pluto, on their own contain both soluble organic material such as carboxylic and amino acids, nitrogen heterocycles, etc., and a dominant (>70% of the organic carbon) component of insoluble macromolecular organic compounds (e.g., *Sephton*, 2005). Soluble organic compounds can also be released through hydrothermal activity from the insoluble component (by thermogenesis) (e.g., *Yabuta et al.*, 2007). Pluto's icy component, likely originally similar in composition to that in comets (as discussed above), contained and contains a variety of hydrocarbons, nitriles and amines (*Mumma and Charnley*, 2011; *Altwegg et al.*, 2016).

Comets are notably rich in relatively non-volatile macromolecular organic matter (e.g., *Fray et al.*, 2016; *Bardyn et al.*, 2017) and surely Pluto was/is as well. Mass spectrometer measurements at comet 1P/Halley highlighted the importance of CHON particles (*Kissel and Krueger*, 1987), and the wealth of in situ *Rosetta* measurements at 67P/Churyumov-Gerasimenko (a Jupiter-family comet and thus one formed in the same region of the outer Solar System as Pluto) have only strengthened this inference (Fig. 9). The *Rosetta*-based elemental ratios derived for CHON in *Bardyn et al.* (2017), $C_{100}H_{100}O_{30}N_{3.5}$ (Fig. 9), compare favorably with the "classic" *Vega* Halley-based ratios of $\sim C_{100}H_{80}O_{20}N_4$ (*Kissel and Krueger*, 1987; *Alexander et al.*, 2017) derived from more limited, flyby data. Due to radiogenic and other heating the organic content of Pluto would have been altered/augmented by thermal and hydrothermal processing, as described above, as well as from the synthesis of hydrocarbons from inorganic species.

Post-*Rosetta* models of bulk cometary composition that match or come close to Pluto's density have been proposed, and contain a substantial organic/hydrocarbon component. *Davidsson et al.* (2016) propose in their "composition A" that 67P is 25 wt% metal+sulfides, 42 wt% rock/organics, and 32 wt% ice. For their assumed component densities, the overall grain (zero porosity) density is 1820 kg/m$^3$. *Fulle et al.* (2017), updating *Fulle et al.* (2016), argue for, by volume, 4 ± 1% Fe-sulfides (density 4600 kg/m$^3$), 22 ± 2% Mg,Fe-olivines and -pyroxenes (density 3200 kg/m$^3$ if crystalline), 54 ± 5% "hydrocarbons" (density 1200 kg/m$^3$), and 20 ± 8% ices (density 917 kg/m$^3$). This composition yields a comparatively lower primordial grain density (dust + ice) of 1720 ± 125 kg/m$^3$. Within uncertainties, both of these values are compatible with the uncompressed density of the Pluto system as a whole, ≈1800 kg m$^{-3}$ (*McKinnon et al.*, 2017).

Figure 10 shows representations of Pluto's possible internal structure based on these compositions. These structures are portrayed as if fully differentiated according to the densities of the principal compositional components. A notable difference is the characterization of the organic component. *Davidsson et al.* (2016) cite similarities between 67P's carbon component to the relatively refractory, insoluble organic material (IOM) found in carbonaceous chondrites, though they model this component with the density of graphite (H/C = 0; 2100 kg m$^{-3}$). *Fulle et al.* (2017), based on higher H/C values (e.g., *Fray et al.*, 2016), opt for a lower-density but macromolecular hydrocarbon. In this case the volume percent of the organic component is enormous compared with even the "extreme" carbon-rich models of the past (see Fig. 4 in *McKinnon et al.*, 1997). This comparison underscores the importance of organic maturation, which is controlled by interior temperatures where organic matter is/was present. Lower H/C would be associated with higher temperatures.

Organic-rich Pluto compositions naturally contain less rock than organic-poor ones, and thus would predict lower heat flows overall for Pluto. Radiogenic heat production in the models in Fig. 10a and 10b are, respectively, 0.77 and 0.79 times that of the Pluto models in *McKinnon et al.* (2017) (see also chapter by Nimmo and McKinnon in this volume). The cores in these models would also be comparatively cooler, as high H/C organics (e.g., benzene) likely convect readily, whereas graphite (if present) has a very high thermal conductivity. Indeed, graphitization of core organic material (section 3.1.4) in any scenario should markedly enhance core thermal conductivity.

Specific geological or geophysical evidence of massive organic layers within Pluto (or Charon) is lacking, either for or against, but should be sought. For example, impacts of large KBOs with either Pluto or Charon could potentially excavate into a near-surface carbonaceous layer (as in Fig. 10b). Despite the likely size of the Sputnik basin impactor (~200-km diameter or greater; *Johnson et al.*, 2016; *McKinnon et al.*, 2016b), the geology of Pluto is quite complex (chapter by White et al. in this volume) and still active, which complicates the identification of ancient basin ejecta. The large (~240-km across) basin Dorothy (Gale) on Charon is easier to interpret as Charon's northern terrain is relatively ancient and unmodified. No albedo or spectroscopic evidence of organic-rich ejecta is seen, which may impose useful lower limits on the thickness of Charon's ice shell (scaling Fig. 10b to Charon gives an ice shell thickness of 45 km). The fact remains, however, that Pluto, Charon, and KBOs generally likely inherited a large organic mass fraction during accretion (cf. Fig. 9c). The internal structures of large KBOs, and to an extent those of icy satellites as well, may be rather different than the paradigm that has reigned since the pioneering work of J. S. Lewis (e.g., *Lewis*, 1971).

### 3.4. Comparisons with Other Volatile-Rich Icy Worlds

The similarities in atmospheric and surface compositions among the triumvirate of Titan, Triton, and Pluto have long attracted attention. All three possess majority $N_2$ atmospheres with secondary $CH_4$, which suggests some commonality in the origin and/or evolution of their volatiles. Yet the different contexts of each world (Titan – a major giant planet satellite, Triton – a captured and tidally heated former KBO, and Pluto – a binary KBO dwarf planet) urges caution. It is possible that the dominant role for $N_2$ in all of these atmospheres stems more from the great volatility of nitrogen and its near-noble gas lack of chemical reactivity, and less from any cosmogonic commonality.

For Titan the results of the *Huygens* probe are telling. Both the low $^{14}N/^{15}N$ value for Titan's atmospheric nitrogen (≈168) and the very low ratio of primordial $^{36}Ar$ to $N_2$ (*Niemann et al.*, 2010) preclude an origin for Titan's nitrogen atmosphere from primordial molecular nitrogen trapped in the icy "satellitesimals" from which Titan formed. Otherwise, referring to the latter ratio, $^{36}Ar$ would have been trapped as well, according to the classic test proposed by *Owen* (1982). We note that *Rosetta* determined a coma production rate ratio for $^{36}Ar/N_2 \sim 5\times10^{-3}$ for comet 67P (*Rubin et al.*, 2018), which while itself low, is still orders of magnitude higher than that in Titan's atmosphere (~$2\times10^{-7}$). In any event, Titan's atmospheric $^{14}N/^{15}N$ and $^{36}Ar/N_2$ ratios instead point to the original molecular carrier of nitrogen into Titan being some combination of ammonia and organics (*Miller et al.*, 2019). Unfortunately, these types of in situ, isotopic data are not yet available for Triton or Pluto, and as discussed above, the colder conditions in the aKB may have been more conducive to the condensation and/or trapping of molecular nitrogen from the nebula there.

The differences between Triton and Pluto, bodies ostensibly accreted in the same region of the Solar System, are also instructive. Triton's surface is less rich in $CH_4$ ice compared with Pluto but $CO_2$ ice is distributed over a large areal fraction of Triton's surface (*Merlin et al.*, 2018, and references therein). Efficient hydrothermal conversion of carbon-bearing volatiles to $CO_2$, followed by degassing to the atmosphere during an early epoch of strong tidal heating was proposed for Triton by *Shock and McKinnon* (1993). Such a scenario is not applicable in full to Pluto, even in the aftermath of the Charon-forming collision. It is nonetheless puzzling why no $CO_2$ ice at all has been discovered on either Pluto or Charon, given that $CO_2$ is an important cometary volatile (*Mumma and Charnley*, 2011) and Charon's water-ice surface in particular is not covered with obscuring volatile ices or tholin-like organics. The $CO_2$-ice overtone lines near 2.0 μm are quite narrow, however, which may have precluded their detection in *New Horizons* LEISA spectra (*Grundy et al.*, 2016).

LEISA's low spectral resolution reduces its sensitivity to $CO_2$ ice, but Triton-like abundances would have been readily detected (W.M. Grundy, personal communication, 2020). Higher spectral resolution ground-based spectra of Pluto have also not seen it. Although higher spectral resolution is beneficial, ground-based observations have the disadvantage of much lower spatial resolution, but that is at least partly compensated by extremely high signal/noise. The lack of evidence for $CO_2$ in both LEISA and ground-based spectra is likely an important clue to the divergent evolutions of Triton and the Pluto system. We note that $CO_2$ has not been detected on any of the other Kuiper belt dwarf planets either, discussed next.

*3.4.1. Other dwarf planets in the Kuiper belt.* Whereas no other KBO is known to have an atmosphere, both volatile and involatile surface ices have been identified on a significant number of them. Water ice and methane ice, with their pronounced spectral absorptions, are the easiest to detect. Notable among these detections are methane ice on Eris and Quaoar, extensive methane ice on Makemake, and water ice on Haumea, Quaoar, Orcus, and Gonggong (*Brown*, 2012; *de Bergh et al.*, 2013; *Holler et al.*, 2017). Nitrogen ice, with its weak, narrow absorption feature at 2.15 μm, cannot be detected given the signal-to-noise of spectra obtained to date, but subtle band shifts in the $CH_4$-ice features on Eris, Makemake, and Quaoar are consistent with minor $CH_4$ dissolved in $N_2$-ice (as on Pluto and Triton), or vice versa ($N_2$ contamination of $CH_4$ ice) (e.g., *Tegler et al.*, 2008, 2010; *Lorenzi et al.*, 2015; *Barucci et al.*, 2015). All these ice detections or inferences have most commonly been interpreted in terms of long-term stability, with distance from the Sun (surface temperature and UV flux) and body size (surface gravity) having competing effects on loss or retention (*Schaller and Brown*, 2007; *Johnson et al.*, 2015). That $N_2$ ice is found or is probable on three of the four largest dwarf planet KBOs is, however, consistent with a common process or origin for surface nitrogen in the Kuiper belt. Some degree of endogenic control, especially for these larger, potentially geologically active KBOs, does not seem implausible.

The large dwarf Kuiper belt planet that does *not* display spectral evidence of $N_2$ ice, or any volatile ice, is Haumea. Its surface is dominated (spectrally) by $H_2O$-ice, a characteristic it shares with Pluto's large moon, Charon. Both rapidly-spinning Haumea and Charon are thought to be products of giant impacts (*Brown et al.*, 2007; *Leinhardt et al.*, 2010), but in Haumea's case it is the primary body. The surfaces of its two known satellites and KBO dynamical family are also water-ice rich. The outcomes of giant impacts depend on many variables (Section 2.3), but Haumea is apparently large enough (mean diameter ~ 1600 km; *Ortiz et al.*, 2017) to have differentiated and driven its volatile ices to the surface early in its history. It would appear that the giant impact that gave the Haumea system its unique character ejected to heliocentric orbit any volatile-ice-rich "crust" and atmosphere, while exposing its deeper water-ice mantle (hence the inference for differentiation) (see also *Carter et al.*, 2018). Perhaps such escaped fragments, or similar, could explain the CO-and-$N_2$-rich comet R2 discussed in section 3.1.2. But this then raises further questions. Did Pluto lose at least some volatile ices in the Charon-forming impact? Or is it a matter of timing, in that Charon formed early enough that Pluto could replenish or replace its surface volatile inventory, whereas Haumea could not because its giant impact came later (*Levison et al.*, 2008b; *Volk and Malhotra*, 2012)?

## 3.5. Synthesis and Unresolved Issues

We can envision the following for Pluto's $N_2$ and CO budgets: (1) Pluto started with cometary inventories of $N_2$ and CO; (2) subsurface aqueous chemistry led to the destruction of CO; (3) $N_2$ was outgassed or otherwise transported to the surface efficiently (e.g., cryovolcanically); (4) no substantial loss of $N_2$ (due to escape) has occurred at the surface, and it accumulated to form Sputnik Planitia; and (5) comets have delivered a small resupply of CO that mixes with surface $N_2$. This scenario is by no means the only possibility, but it is consistent with the evidence to date.

Future work should focus on quantitative evaluation of the thermochemical pathways available to Pluto's core, under a variety of compositional assumptions. These in turn would feed forward to models of CO, $N_2$, $CH_4$, and $CO_2$ clathrate formation, as Pluto's ocean cooled and its ice shell thickened. All of this would inform our cosmochemical and geochemical understanding of Pluto, as well as better constrain Pluto's likely thermal evolution and the characteristics of its subsurface ocean. It would also be valuable to run long-term climate models with not just $N_2$ and $CH_4$ ices, but with CO ice as well (see Section 4.2), to determine the degree to which the optical surface of Sputnik Planitia might mask a deeper, more CO-rich ice sheet.

In terms of better observational constraints, the coming era of "big glass" (telescopes) promises more definitive spectral detections of the icy constituents on large KBOs. The launch of the *James Webb Space Telescope* (JWST) should extend high signal-to-noise observations farther into the infrared, covering the stronger, fundamental vibration bands of $N_2$, CO, and $CO_2$. Detection of the water ice O-D stretch near 4.13 μm (*Clark et al.*, 2019) may allow determination of the D/H ratio (in water ice) for Charon and other bodies, with potentially profound implications for the provenance of the Pluto system with respect to other KBOs and comets (see, e.g., *Cleeves et al.*, 2014; *McKinnon et al.*, 2018a). Related would be determination of the $CH_3D/CH_4$ ratio in the atmosphere or on the surface, though proper interpretation would require understanding its potential fractionation with respect to D/H in water ice, the main, primordial hydrogen reservoir on Pluto-Charon.

Finally, the determination of Pluto's atmospheric $^{14}N/^{15}N$ and $^{36}Ar/N_2$ ratios would be extremely constraining for the provenance of Pluto's nitrogen. As *Glein and Waite* (2018) note, however, because of diffusive separation in Pluto's atmosphere it will be very difficult to make a definitive measurement of these ratios even with a future close flyby or orbiter mission (this caveat applies to measuring atmospheric deuterated methane as well). Perhaps the closest near-term measurement that might be made would be $HC^{14}N/HC^{15}N$ from ALMA (*Lellouch et al.*, 2017). In this case, however, proper interpretation would require a good understanding of the formation and isotopic fractionation of HCN in Pluto's atmosphere.

## 4. ACTIVE WORLDS: PLUTO AND CHARON THROUGH TIME

In this section we examine the behavior of the principal components of the Pluto system through time. Space constraints necessitate selectivity in our approach. We focus on Pluto, and specifically, notable aspects of its geological history in Sec. 4.1, followed by a concise summary of its atmospheric and climactic evolution in Sec. 4.2, as the latter is clearly of great importance for understanding Pluto's geology. Attention then turns to Charon in Sec. 4.3, with a focus on its orbital and tectonic history and how this history fits in with the overall formation and evolution of the Pluto system. Finally, in Sec. 4.4 we offer a précis of the persistent enigma of the small satellites, followed in Sec. 4.5 by an overview of the system and future work needed to advance understanding.

**4.1. Geological History of Pluto**

The geological terrains and history of Pluto are well covered by the chapters by White et al. and Moore and Howard in this volume and the references therein. In this section we tour through Pluto's geological history at a high level, starting in deep time and moving forward to present activity. We focus on impacts and internally driven processes, and highlight both reasonably substantiated conclusions and interpretations as well as less well-supported inferences and outright puzzles that may hopefully spur further work.

*4.1.1. The giant impact epoch.* Pluto's geological history begins of course with the giant impact that formed Charon. Charon's post-formation orbit would have been closer in and likely eccentric, at least initially (chapter by Canup et al. in this volume and Section 4.3). Correspondingly, Pluto would have been spinning much more rapidly and would have a grossly distorted oblate figure (see, e.g., figures in *Sekine et al.*, 2017). As tides raised on Pluto drove Charon's orbit to expand, Pluto would have despun and its global figure would have relaxed. But as the chapter by Nimmo and McKinnon in this volume discusses, no presently measurable geological evidence survives of this despinning epoch. That is, *New Horizons* did not detect a fossil bulge and the potentially colossal tectonic signatures of such a shape change (*Barr and Collins*, 2015) are not writ into Pluto's surface. This suggests that Pluto's icy lithosphere was then too thin or weak to withstand the shape change (in contrast to, e.g., that of Saturn's satellite Iapetus), which in turn suggests that Pluto's heat flow in this epoch was substantially higher than present-day estimates.

Models of Pluto's heat flow due to long-term radiogenic heat release (U, Th, $^{40}$K) generally depict a modest rise and fall over geologic time because the heat flow is moderated by the thermal buffering of the core: early heating is absorbed by raising core temperatures whereas later heat flow is maintained by the conductive release of this stored heat from the core (e.g., Fig. 6). Pluto, however, had access to other heat sources post-Charon formation. Noted in section 2.5, these include the heat released by the completion of differentiation, hydration of anhydrous core minerals, tidal dissipation, and the giant impact itself. Combined these may not only have sustained an early ocean but could have provided sufficient early heat that allowed Pluto's global figure to relax. Pluto's heat flow through time may actually have been a bit of a roller coaster, rising to an early peak then falling, only to rise again slowly due to long-term radiogenic heating over 1-2 billion years before entering a long decline to today (cf. *Bierson et al.*, 2020). The full details of such an evolution remain to be quantitatively assessed, however.

*4.1.2. Formation of Sputnik.* Near the base of Pluto's *known* geological record (since the non-encounter "far side" and polar south are so little known) lies the formation of the Sputnik basin. Formed by the impact of substantial KBO (some ~200 km or more across; *Johnson et al.*, 2016; *McKinnon et al.*, 2016b), such an impact is exceedingly unlikely in today's Kuiper belt environment (*Greenstreet et al.*, 2015). From Fig. 5 in that work, the chance of a single 1000-km scale basin forming on Pluto over the past 4-4.5 billion years is only ~1% (see also the chapter by Singer et al. in this volume). Because impacts are stochastic, this does not mean that Sputnik could not have formed more recently, only that it is much more likely to have formed very early, during the dynamically violent, planetary instability epoch or during Neptune's subsequent migration through the aKB (Section 2.2, and see Fig. 3), when impact rates were substantially higher. The work of *Greenstreet et al.* (2015) incorporates the time evolution (decline) of various Kuiper belt subpopulations through time, but did not attempt to assess the bombardment during the instability era owing to the uncertainty as to the details and timing of what actually occurred. This question could probably be fruitfully revisited.

What can be said is that the formation of the Sputnik basin probably occurred early in Pluto's history, which according to the most recent work on the giant planet instability, could have taken place within some 100 million years of the formation of the Solar System (Section 2.2). At that time, the Sun would have been only about 70% as luminous as today (*Gough*, 1981; *Siess et al.*, 2000; *Bahcall et al.*, 2001), but Pluto itself would have been much closer to the Sun, somewhere in the 20-30 AU range perhaps, at least initially. During such time the insolation Pluto (and Charon) received could have been of order twice today's values, and this high insolation could have been enjoyed for up to ~100 Myr. Once emplaced into the 3:2 resonance and once Neptune's orbital migration slowed to a crawl (e.g., Fig. 3), that is, once Pluto achieved its present orbit, insolation would have been only 70% of today's value due to the faint young Sun. All other things being equal, Pluto's atmosphere then would have been more tenuous and its surface-atmosphere volatile exchange hindered. The effects of such a warm epoch followed by a marked cool-down for the distribution of Pluto's volatile ice reservoirs, and the filling of Sputnik Planitia in particular, also remain to be assessed.

*4.1.3. Ancient terrains and tectonics.* Pluto's surface is marked by a wide range of impact crater densities, from the heavily cratered (if not saturated) highland of eastern Cthulhu to the uncratered ices (at *New Horizons* resolutions) of Sputnik Planitia itself (*Moore et al.*, 2016; chapter by Singer et al. in this volume). Broad regions contain sufficient numbers of large craters that they probably date back to 4 Gya or more even though they are not saturated with craters in the manner of the lunar highlands. These include western Cthulhu and the region to the northwest of Sputnik (Vega Terra and the bright plains to its north). These regions are notable because they are relatively flat, in contrast with much of the rest of Pluto. Topographic variance over this western expanse of the *New Horizons* encounter hemisphere is about ±1 km (*Schenk et al.*, 2018; Fig. 11). While many geologic processes can create level terrain (e.g., burial by sediments), the regional extent of this more-or-less even landscape is reminiscent of the subdued topography of the tidally heated icy satellites Europa and Triton. The subdued topography and high crater density of western Cthulhu and Vega Terra may be related to the inferred early warm epoch for Pluto discussed above in Sec. 4.1.1.

Foremost among the rugged ancient tectonic structures on Pluto is the great north-south ridge-and-trough system (NSRTS) that stretches from the northern polar region of Lowell Regio to the western border of Sputnik Planitia and then further south as far as *New Horizons* imaging allows (*Schenk et al.*, 2018; Fig. 12). Its full extent is unknown, but could extend well into Pluto's far side (sub-Charon) hemisphere (*Stern et al.*, 2020). What is known is that the NSFTS appears highly degraded and is crosscut by the western rim of the Sputnik basin. The orientations ("strikes") of its ridges, troughs, elevated plateaus, and elongate depressions also do not appear to be influenced by proximity to the SP basin. These observations together suggest that the SP impact occurred later in geologic time than the NSRTS. However, although the NSRTS appears older than Sputnik, it is conceivably stratigraphically younger. Its tectonic elements continue right up to the SP ice sheet, and it is not obvious how an ancient near-rim tectonic structure (even a deep-seated one) could have survived such a colossal impact. But nature can be surprising. See *Schenk et al.* (2018) for a full discussion.

The tectonic driver for the NSRTS is unknown, but it resembles, at least in its northern segment, major rift systems on certain icy satellites (e.g., Ithaca Chasma on Saturn's Tethys).

Early or first freezing of Pluto's ocean (not necessarily complete) could have been responsible. Equatorial crustal thickening is cited as a possible causative mechanism in the chapter by White in this volume, but how this would simultaneously account for both the depressed and elevated sections of the NSRTS is unclear. Regardless, the implication of this suggestion, if true, is that the NSRTS would have been aligned along a "paleo-equator" prior to a reorientation (true polar wander) of Pluto's ice shell.

Such a reorientation, *but due to Sputnik*, almost certainly affected Pluto's tectonics subsequent to the formation of the Sputnik basin (e.g., *Keane et al.*, 2016). The impact basin, unfilled post-formation, was likely close to isostatic, as the weakened, fractured and thinned ice shell adjusted to mechanical equilibrium with the underlying ocean (*Johnson et al.*, 2016). Because the compensating, uplifted ocean is deeper than the basin, the basin at first would have represented an overall negative gravity anomaly at spherical harmonic degree-2. Only later, when the basin filled with dense nitrogen-rich ices did it evolve to become the (inferred) positive gravity anomaly that drove the position of Sputnik to more closely align with the Pluto-Charon tidal axis (*Nimmo et al.*, 2016; chapter by Nimmo and McKinnon in this volume). From the perspective of Pluto today, the filling of the basin is inevitable, as (1) $N_2$ condensation is much enhanced in the less-insolated equatorial regions (due to Pluto's high obliquity) and at low elevations because of higher atmospheric surface pressures (see details on this atmospheric-topographic process in *Bertrand and Forget* [2016], *Bertrand et al.* [2018], and the chapter by Young et al. in this volume), and (2) because $N_2$ ice is glacially mobile at Pluto surface temperatures; if initially widely distributed it will, quite simply, flow downhill (*Umurhan et al.*, 2017; *Bertrand et al.*, 2018). From this perspective, ices could have filled the basin over some tens of millions of years (*Bertrand et al.*, 2018), but it should be borne in mind that the immediate post-Sputnik insolation and atmosphere may have been very different than today (as discussed above). Once emplaced in today's 3:2 resonant orbit early Pluto could, depending on ice albedo, have been colder on average and any surface-atmosphere volatile transport time scales considerably lengthened.

*4.1.4. Glaciated and mantled terrains.* Discussed in detail in the chapter by Moore and Howard in this volume and in Section 4.2.2 below, many of Pluto's surfaces display an array of erosional and in a few cases depositional signatures indicative of glacial action. Because glacial flow is observed today (from the eastern Tombaugh Regio uplands into the Sputnik basin), it is almost certain the topographic modifications seen elsewhere by *New Horizons* — incisions, channeling, and fluting — were due to past glacial action. What is not known is whether this is a recurring phenomenon, driven by orbital oscillations (akin to terrestrial Milankovich cycles), perhaps with high-pressure atmospheric excursions (e.g., *Stern et al.*, 2017), or whether it represents secular evolution of Pluto's volatile ice reservoir. For example, prior to the Sputnik impact, $N_2$ and $CH_4$ ices may have been more equitably distributed across Pluto's surface, and only as the Sun began to warm (over billions of years!) were they sufficiently mobilized to begin to cycle and flow as they do today. What is clear is that glacial $N_2$ must have covered much of Pluto's encounter hemisphere in the past, but is not there now. Remnant $N_2$ deposits are seen on certain crater floors and other low-lying areas. Alcyonia Lacus in particular records the loss of $N_2$ over time (see Fig. 5 in the chapter by Moore and Howard in this volume). So where is this nitrogen now? Is it underground as a liquid, in cracks and pores? Or did it escape to space, even though *New Horizons* determined that the nitrogen escape rate to space today is very limited (*Gladstone et al.*, 2016; *Young et al.*, 2018)? If present conditions are naively extrapolated in time, atmospheric escape is unlikely to have affected Pluto's nitrogen mass balance (Table 1). For Pluto's atmosphere, however, the present is not (quite) the key to the past, as the ultraviolet flux from the young Sun should have been far larger than today (*Zahnle and Walker*, 1982; *Claire et al.*, 2012), and moreover, for some of that early solar epoch Pluto orbited much closer to the Sun than today (Sec. 2).

    Even more enigmatic are the mantled terrains to the northeast of Sputnik, e.g., Hayabusa Terra. There a smooth, $CH_4$-ice mantled landscape alternates with impact craters and large, flat-floored depressions, the latter 10s of km across and up to 3 km deep. *Howard et al.* (2017b) suggest that the materials surrounding the depressions may be $CH_4$-ice rich, but note that such heights are not obviously consistent with the bearing strength of methane ice. Their preferred (if tentative) formation model invokes subsurface cryovolcanism or heat driving volatilization and explosive eruptions of the overlying volatile ice mantle, perhaps akin to maar formation on Earth (where silicate lavas react with groundwater or permafrost). A survey of craters superimposed on the washboard texture that embosses uplands to the northwest of Sputnik Planitia, which is thought to have formed as a consequence of nitrogen ice glaciation, suggests that such glaciation ended ~4 Gya (*White et al.*, 2019). But for both the mantled and glaciated terrains, the generally lower abundance of well-preserved craters implies that these terrains are not amongst the most

ancient on Pluto's near side (<4 Gya), though likely still old (see chapter by Singer et al. in this volume).

Potentially related to the glaciated and mantled terrains is the equatorial bladed terrain of Tartarus Dorsa. The bladed terrain comprises a giant deposit of methane ice that is virtually crater free and likely represents an enormous depositional episode of methane ice in Pluto's early- to mid-history (*Moore et al.*, 2018; chapter by White et al. in this volume). But as evidenced by their bladed decrescence (ablation) texture and the observation that underlying terrain appears to have been exhumed at their margins, the deposits now seem to be receding as a consequence of excursions in Pluto's climate causing sublimation to be favored over deposition (chapter by Moore and Howard in this volume). The lack of craters may be due to a combination of original deposition of the methane ice after the era of heaviest bombardment, and elimination of any craters that did form by the subsequent sublimation erosion.

*4.1.5. Middle-age and younger tectonics.* As was recognized going back to the initial post-*New-Horizons* geological assessments of Pluto (*Moore et al.*, 2016), Pluto's visible tectonics are overwhelmingly extensional. This led to a natural inference that ocean freezing and the resulting uplift and extension of the surface may be an important general driver (*Nimmo*, 2004). On its own, however, such extension might be expected to create a randomly oriented array of normal faults and graben, but as Section 3 in the chapter by White et al. in this volume makes clear, the orientations of Pluto's normal faults and grabens are anything but random. We have already noted the possibility of true polar wander stresses initiated by the formation of Sputnik basin and its possible subsequent evolution into a degree-2 gravity high. As a mass concentration (or mascon), Sputnik Planitia (SP) can drive its own regional tectonism as well. If its excess mass were sufficiently concentrated, it would raise a flexural arch around the basin and circumferential graben might have formed along the crest of the arch in a manner similar to that seen around lunar and martian impact basins (e.g., *Solomon and Head*, 1980). But Sputnik is a geographically large structure compared with Pluto's radius. It is so wide in angular extent (in terms of degrees on the surface) that in-plane membrane stresses are more important than flexural, bending stresses, which leads instead to *circumferential extension* as the lithosphere surrounding SP is "pushed" longitudinally outward (*Janes and Melosh*, 1990). In this way, the prominent sub-parallel graben and troughs of the Inanna, Dumuzi, and Virgil Fossae, which strike quasi-radially away from SP (*Keane et al.*, 2016), are more akin to the Valles Marineris canyon system on Mars (which strikes radially away from the center of Tharsis).

These prominent tectonic features are all the more remarkable because they cross-cut major craters and are youthful in appearance, with sharp, undegraded scarp crests (*Moore et al.*, 2016; chapter by White et al. in this volume). Yet the ostensible tectonic driver, SP, is ancient. This may imply reactivation of an older fracture pattern, or a more complex history and interplay of tectonic forces. For example, long-term build-up of background extensional stresses due to ocean freezing (e.g., *Hammond et al.*, 2016) could add to pre-existing loading stresses due to SP, and reach (or only reach) the extensional failure limit in Pluto's icy lithosphere later in geologic history.

The Mwindo Fossae extensional tectonic system in the far east of Hyabusa Terra is unusual in that it is isolated and converges to a nexus (*Moore et al.*, 2016; *McGovern et al.*, 2019). The fracture pattern is consistent with negative loading or tectonic uplift at the nexus, but there is nothing otherwise geologically unusual about the nexus region compared with the surrounding terrain, other than possibly being elevated by 0.5-1 km (*Schenk et al.*, 2018). As a fairly crisp-looking and undegraded set of fault scarps, the Mwindo Fossae are a further reminder that Pluto has been tectonically active late in its history, and that local thermal and dynamic perturbations to the ice shell remain possible.

Tartarus Dorsa (the bladed terrain region) actually consists of broad, elongate swells that are typically ~400 km long and ~100 km wide, and which tend to be separated from one another by troughs that include Sleipnir Fossa and others that form the southern components of Mwindo Fossae (see Fig. 9 in the chapter by White et al. in this volume). These swells form some of the highest standing terrain in the *New Horizons* close approach hemisphere, reaching 4.5 km above mean radius (*Moore et al.*, 2018; *Schenk et al.*, 2018). Despite some connection to the younger tectonic elements of the Mwindo Fossae to the north, the swells themselves appear to be much older basement upon which the methane-rich blades were emplaced (and now are in retreat). The origin of this striking basement topography is enigmatic, but the scale, amplitude, and elevation suggest the possibility of ancient compressive tectonics (*McGovern et al.*, 2019).

*4.1.6. Middle-age and younger cryovolcanism.* Wright Mons and its less-well-imaged partner to the south, Piccard Mons, are two large, quasi-annular massifs. Rising 4-to-5 km from their broad bases (~150 and ~250 km for Wright Mons and Piccard Mons, respectively), both possess deep central depressions, thus superficially resembling terrestrial volcanos (*Moore et al.*, 2016). Nothing about Wright Mons (which was well imaged) is consistent with impact or erosional or mass wasting processes. A predominantly tectonic origin is conceivable, that is, uplift or inflation of the subsurface, followed by collapse or deflation of the central region. But such an origin does not easily explain the characteristic hills or hummocks on the flanks of the edifice or in the surrounding terrain. One is left then, by elimination, with the suspicion that both edifices are actually constructional and thus true cryovolcanoes. If so they would be the first such clearly identified features found on an icy Solar System body (cf. *Moore and Pappalardo* [2011] for a discussion of the checkered history of such identifications).

Eruption of cryolavas from Pluto's interior can be assisted, at least in principle, by overpressure generated during ongoing freezing of a mixed ammonia-water ocean beneath an elastic ice lithosphere (e.g., *Manga and Wang*, 2007), though such freezing would need to be reconciled with any thermal stabilization due to clathrate formation at the base of the ice shell (Sec. 3.2; chapter by Nimmo and McKinnon in this volume). The shallow flanks of Wright Mons consist of closely packed, semi-regular hills ~8–10 km in diameter. The hills themselves are rubbly textured, reminiscent of the funiscular terrain between Enceladus' south polar sulci ("tiger stripes"), and could represent the individual constructional elements of the edifice (in contrast to the subaerial lava flows of basaltic shield volcanoes on the terrestrial planets). Few if any impact craters superpose Wright Mons, implying an upper limit on its age ($\lesssim$1 Gya; chapter by Singer et al. in this volume), but there is no evidence that either Wright or Piccard Mons are active today. While there is much we do not understand about the formation of these spectacular features, if we can claim any understanding at all, their mere existence attests to the reality of late endogenic activity on Pluto.

Evidence for even more recent cryovolcanic eruptions or effusions have been advanced by *Dalle Ore et al.* (2019) and *Cruikshank et al.* (2019, 2020). Fresh-appearing reddish surface materials and coatings are seen at or near the Virgil Fossae and Uncama Fossa, both to the west of SP and among Pluto's youngest extensional fracture systems. The reddish materials display the spectral signature of water ice along with an ammoniated compound, and it is hypothesized that the red chromophore (coloring agent) is organic and is a product of Pluto's internal chemical evolution, given the planet's likely incorporation of copious primordial organic matter (Section 3.3). The association of aqueous (i.e., low viscosity) fluid or even vapor-driven, "cryoclastic" eruptions with recent extensional faulting is self-consistent, as the latter promotes the former. How these fault systems tapped into a subsurface aqueous reservoir and whether that reservoir itself was, or was connected to, Pluto's putative ocean remain to be determined, as do the reasons that the (apparent) cryovolcanic expressions at Virgil Fossae and Uncama Fossa and those at Wright Mons are so different.

*4.1.7. The red layer.* Notable on the inner rimwalls of a number of the larger craters to the northwest of SP is a singular, stratigraphically exposed dark, red band or layer (*Moore et al.*, 2016; see chapter by Singer et al. in this volume). This layer is apparently also exposed in the faces of tilted mountain blocks within the Al-Idrisi Montes (*Moore et al.*, 2016; *White et al.*, 2017). Lying about one kilometer below Pluto's surface, this red layer must represent an important event in Pluto's geological history. It may represent the accumulation of dark, reddish, tholin-like haze particles (*Grundy et al.*, 2018), followed by a depositional hiatus. Or the accumulation may have been interrupted by another event, such as ejecta deposition from the Sputnik impact. If the latter is the case then it is an ancient feature of Pluto's crust. Or the red layer may represent a geologically more recent event, such as the regional eruption of reddish cryovolcanic material as described in Section 4.1.6 above. Whatever its origin, it deserves greater attention.

*4.1.8 Sputnik Planitia convection.* The spectacular cellular plains of Sputnik Planitia display the geometric organization and topographic signature of solid state convection in a kilometers deep $N_2$-ice sheet (*Moore et al.*, 2016; *McKinnon et al.*, 2016a). The van der Waals bonded molecular ices $N_2$, CO, and $CH_4$ are weak enough at Pluto temperatures that viscous flow is able to transport Pluto's nominal radiogenic heat flow (~few mW m$^{-2}$), provided the ice sheet exceeds a critical thickness, about 1 to 2 km for an ice sheet dominated by $N_2$ or $CH_4$ ice, respectively (see Fig. 3 in *McKinnon et al.* [2016a]) (see also the chapter by Umurhan et al. in this volume). Assuming a dominantly nitrogen ice sheet, *McKinnon et al.* (2016a) derived from numerical models a timescale for the renewal/replacement of the surface of a typical SP convection cell of ~500,000 years (± a factor of 2). A comparable timescale was derived by *Buhler and Ingersoll* (2018) in their study of sublimation pit formation on the surface of SP.

*Vilella and Deschamps* (2017) subsequently inferred, also based on numerical models, that SP's three-dimensional convection pattern would be more naturally explained if the convecting layer is heated from within or is cooled from above, rather than being heated from below. They derived an upper limit on the heat flow from the interior of Pluto of <1 mW m$^{-2}$, which is rather low for a body Pluto's size and inferred rock fraction (*McKinnon et al.*, 2017). Nor is there any obvious internal heat source for the ice sheet itself. Surface temperature fluctuations on an appropriate time scale are conceivable. The thermal skin depth appropriate to Pluto's 2.8 Myr orbital element variation, which drives its "Milankovich" cycles (see section 4.2), is $\sim(\kappa\tau/\pi)^{1/2}$ = 2 km (for the thermal diffusion coefficient $\kappa$ = 1.33×10$^{-7}$ m$^2$ s$^{-1}$ of N$_2$ ice [*Scott*, 1976] and period $\tau$ = 2.8 Myr). However, the annual mean surface temperature of the ice in SP (which is what counts, not the annual variation) only varies by ±0.5 K in the long-term climate model of *Bertrand et al.* (2018), which does not seem sufficient to trigger solid state convection in the ice sheet. But *Vilella and Deschamps* (2017) are correct in the sense that boundary conditions do count. As an example, if the base of the SP ice sheet is at or near the melting temperature for N$_2$ ice (63 K), then the 3-D geometry of the SP convection pattern can be recovered (*McKinnon et al.*, 2018b). Work on this topic is ongoing (e.g., *Wong et al.*, 2019).

**4.2. Atmosphere and Climate Evolution on Pluto**

As detailed in the chapters by Forget et al. and Young et al. in this volume, the climate of Pluto is a complex system in which the atmospheric dynamics are coupled with the N$_2$ cycle (sublimation and condensation processes induce surface pressure variations and control the winds) and the CH$_4$ and CO cycles (both partly control the radiative properties of the atmosphere, whereas CH$_4$ drives atmospheric photochemistry and haze formation). These cycles strongly depend on surface ice distribution and temperatures, themselves controlled by insolation changes (Fig. 13).

Pluto's climate is highly variable in time, with a surface pressure varying by a factor of 1000 over a present-day Pluto year, maybe even more in the past, according to different models (*Bertrand and Forget*, 2016; *Johnson et al.*, 2019). However, it is only marginally variable in space. Because Pluto's atmosphere is a very weak emitter in the thermal infrared and efficiently mixes trace gases, there are indeed only minor gradients of atmospheric temperature and composition across the globe, except in the lowest ~5 km near the surface where the air can be 10-20 K warmer over a dark volatile-free surface than over a N$_2$ ice-covered surface (see Section 2.1 in the chapter by Forget et al. in this volume).

There are two reasons why it is quite likely that the atmosphere and climate of Pluto have strongly varied in the past and will strongly vary in the future on timescales of millions of years.

First, the climate of Pluto depends on its Milankovitch orbital and rotation parameters, and in particular its obliquity (see the chapter by Young et al. in this volume; *Earle et al.*, 2017), which varies over a range of 23° (i.e., 115.5 ± 12.5°) over a period of ~3 Myr (*Dobrovolskis et al.*, 1997). Such large variations must have induced considerable climate changes, as is the case for the Earth and Mars (e.g., *Imbrie and Imbrie*, 1979; *Laskar et al.*, 2004). Second, Pluto is covered by geological landforms resulting from local accumulation or erosion of m-to-km thick layers of ice, such as flowing glaciers and ice mantles (*Howard et al.*, 2017a,b; *Moore et al.*, 2018; chapter by Moore and Howard in this volume). Their presence and their characteristics are difficult to reconcile with the present-day climate but are consistent with climate changes over periods of Myrs.

*4.2.1. Atmospheric response to the Milankovich cycles.* A robust and periodic solution for the Milankovitch parameters history of Pluto has been developed for the most recent ~100 Myrs, but cannot be mapped further back into the past due to the chaotic nature of the solutions (*Sussman and Wisdom*, 1988; *Earle and Binzel*, 2015). Several climate models have integrated these calculations to explore the most "recent" past climates of Pluto (*Earle et al.*, 2017; *Stern et al.*, 2017; *Bertrand et al.*, 2018, 2019). They reproduced and explained the formation of the major permanent volatile deposits in the mid-latitudes and equatorial regions, which receive less insolation and tend to be colder than the poles on average over several Myr, due to the relatively high obliquity of Pluto. During very high obliquity tilt periods (~104°), the surface pressure and atmospheric abundances of $CH_4$ and CO should have been minimal because the low-latitude permanent volatile ice reservoirs received less insolation. During so-called "moderate" obliquity periods (~127°), meaning those with obliquities farthest from 103°, these tendencies should have reversed (*Bertrand et al.*, 2018, 2019). The orbital changes in longitude of perihelion and eccentricity also impact the volatile cycles but the effects are of second and third order respectively.

Over the last 100 Myr, models suggest that the surface pressures remained in the range of 0.01 μbar to 1 mbar (*Bertrand et al.*, 2018, see their Fig. 16; *Johnson et al.*, 2019) and the $CH_4$ atmospheric mixing ratio in the range of 0.001–10%. This is enough for the atmosphere to have remained opaque to Lyman-alpha radiation and allowed for haze production during most of this time, as suggested as well by the thick layers of dark materials, likely settled haze particles, observed on Pluto's surface (*Grundy et al.*, 2018, *Bertrand et al.*, 2019, *Johnson et al.*, 2019).

Higher pressures up to ~100 mbar (close to the triple point allowing liquid $N_2$ or $CH_4$ on the surface) require extreme conditions with large and very dark volatile ice deposits covering Pluto's poles, and low soil thermal inertia (*Stern et al.*, 2017). These changes in surface pressure and trace gas abundances over time are expected to have impacted photochemistry, haze and cloud amounts and composition, and thermal structure (*Gao et al.*, 2017; *Young et al.*, 2018; *Johnson et al.*, 2019).

Pluto's atmospheric circulation has been shown to be very sensitive to Pluto's $N_2$ ice distribution (*Forget et al.*, 2017). Recent modeling results suggest that the current-day circulation regime is a retro-rotation (westward winds at all latitudes), maintained throughout a Pluto year and mostly controlled by cross-hemispheric transport of $N_2$, in particular within Sputnik Planitia (*Bertrand et al.*, 2020, chapter by Forget et al. in this volume). We can expect a similar circulation regime in Pluto's past because Sputnik Planitia likely remained the main reservoir of $N_2$ ice and forced cross-hemispheric transport of $N_2$, although this remains theoretical.

*4.2.2. Geological evidence of past climates.* Pluto's surface displays many geological features that reveal or suggest substantial changes in the "recent" past (possibly 100s of Myr old, but in many cases much more recent). The $N_2$-rich Sputnik Planitia ice sheet, and the surrounding terrains, exhibit numerous evidences of climate variation: active glacial flow on the edges of the sheet, icy dunes, fluvial features and ponds (which could have been shaped by liquid flows at the base of thick $N_2$ glaciers, now disappeared), deep sublimation pits, and erosion of water ice mountains (*Howard et al.*, 2017a; *White et al.*, 2017, 2019; *Bertrand et al.*, 2018; *Buhler and Ingersoll*, 2018; *Telfer et al.*, 2018). A variety of dissected terrains outside Sputnik Planitia are also thought to have been carved by ancient glaciers (*Howard et al.*, 2017b). The major $CH_4$-rich deposits include the massive Bladed Terrain at the equator and several ice mantles at mid-northern latitudes (*Howard et al.*, 2017a; *Moore et al.*, 2018; chapter by Moore and Howard in this volume). Climate models have been able to relate their latitudinal extension to the Milankovitch parameters history of Pluto, although it remains unclear which reservoir formed first and whether these $CH_4$ reservoirs also evolved over much longer timescales of several 100s of Myr (*Bertrand et al.*, 2019). The ~300-meter tall bladed texture of the equatorial deposits could have formed through condensation-sublimation of $CH_4$ ice over the last 10s of Myr (*Moores et al.*, 2017; *Moore et al.*, 2018). Finally, the mid-latitude ice mantles display subsurface layering up to several kilometers thick that could be the signatures of past climate processes (*Stern et al.*, 2017).

Several processes could have disrupted the past climates of Pluto: cryovolcanism (*Moore et al.*, 2016; *Cruikshank et al.*, 2020), tectonic activity (*Howard et al.*, 2017b), volatile escape (very marginal for $N_2$ ice but several 10s of meters of $CH_4$ ice lost over 4 billion years; *Gladstone et al.*, 2016; chapter by Strobel in this volume), darkening and contamination by haze sedimentation and by direct photolysis/radiolysis of the ices, and surface albedo feedbacks (*Earle et al.*, 2017; *Grundy et al.*, 2018; *Bertrand et al.*, 2020).

Beyond the timescale of ~100 Myr ago, Pluto's climate is relatively unknown due to the lack of constraints on the Milankovitch parameters and on surface conditions. Nevertheless, *Binzel et al.* (2017b) state that the presence of ancient craters at the equator demonstrates a certain stability of the Milankovitch cycles, which could extend back in time by hundreds of Myr (otherwise the craters would have been eroded away or completely buried by volatile ice). Early in Pluto history, and despite the lower insolation then, the impact flux may have allowed for warmer surface temperatures than those of today, at least transiently (following *Zahnle et al.*, 2014), which may have led to intervals of a thicker atmosphere if enough $N_2$ ice was already present and perhaps even liquid $N_2$ flowing directly on the surface. At some point of Pluto's history, the Sputnik Planitia impact should have rapidly trapped most of the $N_2$ ice inside the basin, thus limiting the available ice for sublimation and (for present-day $N_2$-ice albedos) the maximum surface pressures to only a few hundreds of microbars (*Bertrand and Forget*, 2016; *Bertrand et al.*, 2018).

### 4.3. Orbital and Tectonic Evolution of Charon

There is broad consensus that Charon formed as the result of a collision between Pluto and a similarly-sized protoplanet (e.g., *Canup*, 2005, 2011). However, the style of impact, the extent of differentiation of each colliding body, and the evolution and ultimate fate of debris from the collision are all debated (e.g., *Walsh and Levison*, 2015; *Desch and Neveu*, 2017; *Kenyon and Bromley*, 2019c; see Sec. 2). Post-impact, Charon would have orbited much closer to Pluto than its current semimajor axis (*Canup*, 2005, 2011). Outcomes of collisional models suggest a starting orbital distance for Charon of a few to greater than 10 $R_P$ (Pluto radii) and a substantial orbital eccentricity of 0.1 to 0.4 (*Cheng et al.*, 2014). Charon's current orbit at ~16 $R_P$, synchronous rotation, and circular orbit indicate that tides evolved the orbits of both Pluto and Charon.

The tidal evolution of Pluto and Charon depends sensitively on their interior structures, which controls the extent of deformation and dissipation that may occur (*Barr and Collins*, 2015). Measurements from *New Horizons* indicate that Pluto had (and may still have) an internal ocean (chapter by Nimmo and McKinnon in this volume). Charon is also thought to have possessed an ocean, with its tectonic features attributed to extensional stresses generated during ocean freezing (*Moore et al.*, 2016; *Desch and Neveu*, 2017; *Beyer et al.*, 2017; cf. *Malamud et al.* [2017] for a contrarian view). The presence of oceans can speed up the orbital evolution process and potentially generate large (100s of MPa) stresses within the icy shells of Pluto and Charon (*Barr and Collins*, 2015). In addition, diurnal tidal stresses caused by Charon's eccentric orbit would be greatly enhanced if it possessed an ocean, particularly when Charon orbited closer to Pluto (*Rhoden et al.*, 2015).

Charon's surface displays a variety of tectonic features (*Beyer et al.*, 2017; *Robbins et al.*, 2019). Within the encounter hemisphere the large canyon system dubbed Serenity Chasma dominates the geologic record. The canyon system trends roughly northeast-southwest. However, no tectonic patterns have yet been identified on Charon (including Serenity Chasma) that record or are consistent with despinning, outward migration, or an epoch of high eccentricity (*Beyer et al.*, 2017; *Rhoden et al.*, 2020). The lack of tidal fractures implies that either tidal stresses were never high enough to produce fractures or Charon's geologic record was reset after the epoch of tidally-driven fracture formation. The most likely potential explanations are either Charon never had an ocean, so tidal stress magnitudes were negligible and freezing stresses were unavailable, or that Charon's orbit circularized before the ocean froze out. In that case, there would be no eccentricity, recession, or despinning stresses to combine with the freezing stresses and generate a distinct pattern.

The rate of change of Charon's eccentricity (technically the eccentricity of the binary) from tidal dissipation in both bodies when in the dual synchronous state is given for small to moderate $e_C$ by (*Dobrovolskis et al.*, 1997):

$$\dot{e}_C \approx -\frac{21}{2}\frac{k_{2C}}{Q_C}\frac{m_P}{m_C}\frac{R_C^5}{a_C^5}ne_C - \frac{21}{2}\frac{k_{2P}}{Q_P}\frac{m_C}{m_P}\frac{R_P^5}{a_C^5}ne_C \quad , \tag{6}$$

where $a_C$ and $n$ are Charon's semimajor axis and mean motion, and $m$, $R$, $k_2$ and $Q$ are the mass, radius, second-degree potential Love number, and tidal dissipation factor for each body, with the subscripts P and C referring to Pluto and Charon, respectively. $k_2$ is a measure of the distortion a body may undergo in response to tides, and is smaller for differentiated solid bodies compared with uniform ones, but can be much larger for bodies with internal oceans (the maximum value is 3/2 for a uniform fluid body). $Q$ can be thought of as the effective quality factor of a planet or satellite, analogous to the $Q$ of a simple spring and dashpot, though many mechanisms can contribute to tidal dissipation in actual planets and satellites (see, e.g., *Sotin et al.*, 2009).

For Pluto-Charon then, the upper limit to the characteristic time for decay of its eccentricity $e_C/\dot{e}_C$ is $5\times10^4$ $Q$ yr; this upper limit assumes Charon's present-day semimajor axis and that both Pluto and Charon are solid bodies with ice-rock rigidities and equal $Q$s. If Pluto possessed an ocean early on, $e_C$ decay timescales would have been shorter, and shorter still when Charon was closer to Pluto (all other things being equal, dissipation in Pluto dominates that in Charon in equation 6). Even for a standard, or benchmark, $Q$ of 100 (*Murray and Dermott*, 1999), Charon's orbit likely circularized within 1 Myr of the generative giant impact.

Charon's orbital and spin evolution may have been quite complicated, however (*Cheng et al.*, 2014). After the giant impact, Pluto would have been spinning at much faster than the synchronous rate, only slowing as Charon's orbit evolved outward. In this case the coefficient of the second term in equation 6 is +57/8, and tides raised on Pluto would have acted to raise Charon's eccentricity, possibly to the point of orbital instability and escape. Obviously this did not happen, and the presence of Pluto's small satellites imposes an even stricter upper limit on Charon's eccentricity evolution. Most likely, $e_C$ increased until tidal distortion and dissipation within Charon increased sufficiently that $\dot{e}_C \approx 0$. Maintaining a substantial but finite eccentricity as tides raised on Pluto drove Charon's orbit out to dual-synchronous altitude is equivalent to tuning the relative $k/Q$s of the two bodies, as in *Cheng et al.* (2014). Ultimately, though, Pluto's spin slowed, the effect reversed, and $e_C$ decayed according to equation 6. The total time needed for Charon to evolve outward from an inner, post-giant impact orbit to the semimajor axis where both Pluto and Charon achieve spin-orbit synchronism is generally longer than the circularization timescale above, $\sim 10^5$ $Q_P$ yrs for realistic $k_{2P}$ values (*Dobrovolskis et al.*, 1997; *Cheng et al.*, 2014). Charon's orbital evolution thus may have been (was likely?) complete before capture of the Pluto-Charon system into the 3:2 mean-motion resonance by Neptune (Fig. 5 and Sec. 2.6).

Given that the formation of Charon's chasm system has been attributed by most workers to the volume expansion caused by a freezing ocean, we favor the interpretation that Charon's orbit circularized early in its evolution, the lack of tidal heating contributed to ocean freezing, and the freezing ocean generated most of the tectonic features we observe, removing evidence of past tidally-driven fracturing. This is especially true if the freezing of the ocean is related to the eruption of cryolavas that formed Vulcan Planitia and which buried much of the preexisting terrain (*Beyer et al.*, 2019). The mechanism by which Charon's observed fractures formed at their particular orientations, however, remains an open question. Further discussion of Charon's geology and geophysics can be found in the chapter by Spencer et al. in this volume.

**4.4. Puzzling Satellites**

An outstanding problem in our understanding of the Pluto-Charon system is the formation and evolution of the smaller moons: Styx, Nix, Kerberos, and, Hydra. Several studies have simulated their formation from the debris disk generated in the Charon-forming impact, with some success (*Walsh and Levison*, 2015; *Kenyon and Bromley*, 2019c). However, the large orbit distances of the moons, their survival throughout Charon's outward migration, and how they came to be near resonances with Charon are still challenging to explain (*Stern et al.*, 2018; chapter by Canup et al. in this volume).

The orbital and physical properties of the small satellites are described in the chapter by Porter et al. in this volume. From the point of view of this chapter, the most critical aspect is how these properties inform our understanding of the system as a whole. The orbits of the small satellites are relatively compact (35.9 to 54.5 $R_P$, where $R_P$ is Pluto's radius), near-circular, coplanar, and aligned with the Pluto-Charon orbital plane (*Weaver et al.*, 2016). These characteristics strongly point to an origin from a dissipative system of orbiting smaller particles, such as would be created in the Charon-forming giant impact (*Canup*, 2011). Both an *in situ* origin from an extended debris disk (*Kenyon and Bromley* [2014] and subsequent papers) and formation from an impact-generated proximal debris disk followed by resonant tidal evolution driven by Charon (*Stern et al.* [2006] and subsequent papers) have been proposed. The reader is directed to the chapter by Canup et al. in this volume for detailed discussion (see also *Peale and Canup*, 2015), but it suffices to say that no model satisfactorily explains the origin and dynamical characteristics of the small satellite system. Some models invoke collisional interactions among the satellites or with heliocentric (KBO) impactors (e.g., *Walsh and Levison*, 2015; Bromley and Kenyon, 2020). Given the contingent nature of such interactions, a first principles understanding of the origin and evolution of the small satellites may prove elusive.

The physical properties of the satellites themselves, as well as lack of detected small satellites beyond Hydra's orbit by *New Horizons*, are valuable clues nonetheless. The high albedos of the satellites ($\gtrsim 0.5$) and the clear prominence, in *New Horizons* near-infrared spectra, of water-ice on Nix, Hydra, and Kerberos, and the detection of the 2.2-μm absorption attributed to an ammonia-bearing species on Nix and Hydra (*Weaver et al.*, 2016; *Grundy et al.*, 2016; *Cook et al.*, 2018), all point to the satellites being predominantly ice. Such a composition is consistent with the giant impact model in which the small satellites form from the debris blasted off from the icy surface layers of one or both of the progenitor bodies (*Canup*, 2011). It is not consistent with any origin where the small satellites are derived (or even partly derived) from mixed rock-ice or primordial or later heliocentric KBO material, which would be rich in dark rocky and carbonaceous materials. Typical midsize KBO albedos are closer to 0.1 and reflect the latter compositions (see *Stern et al.,* 2018).

Strictly speaking, the albedos and spectral absorptions above refer to surfaces of the small satellites. Although it is difficult to understand how, for example, these ices could mask dark, compositional rocky interiors given the erosive cratering environments the satellites exist in (chapter by Canup et al. in this volume), it would nonetheless be more satisfying if the inferred iciness of the small satellites could be confirmed by their densities. Astrometry-based estimates of the satellite masses prior to the New Horizons encounter (*Brozović et al.*, 2015), when combined with volume estimates from *New Horizons* imagery, are not constraining (the uncertainties are too large; *McKinnon et al.* [2017]). Dynamical stability calculations over Gyr timescales by *Kenyon and Bromley* (2019a) imply, however, that the densities of the satellites must be under 2000 kg m$^{-3}$, and for Nix and Hydra (the largest), likely under 1600 kg m$^{-3}$. Such densities are consistent with ice, but they do not prove it. Given the likely structural disruptions from the impacts evident on their surfaces (*Weaver et al.*, 2016), we expect satellites of their size, if made of ice, to have densities under 1000 kg m$^{-3}$, more similar to the densities of the icy, inner satellites of Saturn (*Buratti et al.*, 2019). Further astrometric measurements of the Pluto system combined with numerical integrations should ultimately yield better density constraints.

*New Horizons* did not detect any new satellites at Pluto, a surprising result given the steady march of *HST* satellite discoveries in the years leading up to the encounter (*Stern et al.*, 2015; *Weaver et al.*, 2016). *Kenyon and Bromley* (2014) predicted that small satellites up to ~2-6 km across would be found beyond the orbit of outermost Hydra, based on their viscously spreading particle disk model for the origin of the small satellites. The *New Horizons* lower limit for detection was 1.7 km across for a Nix-like albedo of 0.5, out to an orbital radius of ~80,000 km (67 $R_P$) from Pluto, with less stringent limits at larger radii (*Weaver et al.*, 2016). While it is unfortunate that *New Horizons* data could not definitively test their hypothesis, it is commendable that the model of *Kenyon and Bromley* (2014) is testable.

The apparent emptiness of the Pluto system beyond Hydra, whose orbit is less than 1% of Pluto-Charon's Hill radius (referring to the gravitational sphere of influence of the Pluto system with respect to the Sun), is notable. This suggests an alternative explanation for the lack of more distant satellites: tidal stripping during Neptune's "wild days." Equal mass binaries are relatively uncommon amongst the dynamically excited (hot) KBO populations compared with the more distant cold classical KBOs (*Noll et al.*, 2020), and one explanation is that the former have been lost to collisions and dynamical effects (e.g., scattering by Neptune) whereas the latter have remained relatively dynamically undisturbed (see *Nesvorný and Vokrouhlický*, 2019). A question of some interest has been whether Pluto's satellite system could have survived the implantation of the system into the 3:2 MMR resonance with Neptune and any subsequent orbital migration (*Pires et al.*, 2015).

Capture into resonance is fundamentally agnostic as to whether a body is single, a binary, or a multiple system, as long as the binary or system is gravitationally bound (*Malhotra and Williams*, 1997). The critical issue is whether impacts, or tides during scattering encounters with Neptune, can cause the binary or system to become unbound or otherwise disturbed, especially if the orbital evolution of the Pluto satellite system was complete prior to the giant planet instability (as argued in Sec. 2.6). *Nesvorný and Vokrouhlický* (2019) as part of their study of binary stability also examined the stability of the KBO dwarf planet satellites during Neptune's migration and implantation/creation of the Kuiper belt. They found that all of Pluto's satellites of are expected to survive during the dynamical implantation of Pluto in the Kuiper belt. They also found that the low orbital eccentricities of Pluto's small moons (<0.01; chapter by Porter et al. in this volume) may have been excited during encounters of the Pluto system to Neptune, or by small impacts while the Pluto system was immersed in the massive planetesimal disk of the aKB.

It should also be said, however, that ~40% of the simulations in *Nesvorný and Vokrouhlický* (2019) resulted either in the loss of outermost Hydra or excitation of its eccentricity to > 0.1. This suggests the following speculation: the implantation may have been sufficiently destabilizing that all small satellites down to the Hydra's orbital distance escaped while the other small satellites were thrown into substantially perturbed orbits, orbits that led to collisions and reaccumulation into the satellite system we see today. Even if less likely, it highlights an interesting aspect of Pluto's dynamics in this early solar system epoch. The tidal effects of scattering encounters with Neptune can also perturb the orbital energies (semimajor axes) of the moons, potentially displacing one or more moons from mean-motion resonances with Charon (the four small moons are today close to but not in the 3:1, 4:1, 5:1, and 6:1 MMR with Charon). (*Nesvorný and Vokrouhlický*, 2019). Any such perturbations would only serve to complicate understanding of the small satellites' history.

**4.5. Synthesis and Unresolved Issues**

Prior to the *New Horizons* encounter, *Moore et al.* (2015) published a detailed look ahead at the geological processes potentially to be revealed on Pluto and on Charon. None were expected to be dull or quotidian. But Pluto and to a certain extent Charon exceeded all expectations by a wide margin. Pluto in fact turned out to be one of the more active solid bodies in the Solar System, rivaling Mars, with a wide array of geological, geophysical, atmospheric, and climatic processes on display, including some never before seen or seen as clearly. All the topics discussed in this section are either the subjects of ongoing research, or they need to be!

For example, the thermal and tectonic histories of Pluto and Charon need to be revisited, based on the likely initial states that evolved subsequent to the Charon-forming impact. The bombardment history of Pluto and Charon *prior* to its emplacement in the 3:2 mean-motion resonance can be modeled, because we now have specific scenarios for the formation of the Kuiper belt that have passed numerous tests. Our ideas about planetesimal and planet formation, including the dynamical instability that populated the modern Kuiper belt will no doubt continue to evolve, but interim implications for Pluto's composition and evolution can still be usefully drawn. The mere existence of Pluto is a key datum in our search for a better understanding of how the Solar System came to be, and better and deeper understanding of the Pluto system (including the small satellites) will provide additional context and clues.

Pluto's earliest post-giant-impact and post-Sputnik-basin-formation evolution deserve greater attention, in order to understand the evolution (if not the creation) of Pluto's volatile ice reservoirs and their effects on the planet's geology and geophysics. We do not yet know if the evidence of substantial volatile transport and glacial (or even fluvial) erosion writ into its surface reflects mainly a truly ancient ($\gtrsim 4$ Gya) geological era or whether this activity has continued, perhaps intermittently, into Pluto's middle age or even up to today. Ongoing glacial flow is seen of course, and famously so (*Moore et al.*, 2016). Pluto in this sense is even more active than Mars. But does the evidence cited in this chapter and in the chapter by Moore and Howard elsewhere in this volume for extensive $N_2$-ice cover in the geologic past imply a secular trend in which $N_2$ ice simply ended up in the Sputnik basin, or has there been substantial loss to space? The latter was the widely held assumption before the *New Horizons* encounter. Detailed evaluation of Pluto's likely atmospheric structure and evolution under the "faint young, but XUV active Sun" is needed. Transient warmer conditions due to large impacts, as well as potential extra-solar influences on the Pluto system, such as effects of nearby supernovae and passage through dense molecular clouds (see *Stern*, 1986; *Stern and Shull*, 1988), might also be fruitfully considered.

Numerous important geological features and processes on Pluto and Charon remain unexplained. What was the cause of the great north-south rift system on Pluto? What accounts for the deep broad depressions and pits in the methane-ice-rich mantled terrains to the northeast of Sputnik Planitia? How were the putative cryovolcanic edifices Wright and Piccard Mons formed? Is there something unique about their location on Pluto? Why don't we see similar structures on Charon, or on Triton, or on icy satellites generally? What does the dark red layer seen in many crater rimwalls and on the exposed faces of many faulted mountain blocks signify? And why do mountain blocks preferentially congregate at the western margin of SP? A partial answer to the latter at least is discussed in Sec. 2.6.2 of the chapter by White et al. in this volume: specifically, the coincidence of the low-viscosity, dense nitrogen ice (or liquid) intruding into water ice crust that has been weakened and fractured by the NSRTS and other tectonic systems, circumstances that are not replicated on the other sides of Sputnik to anywhere near the same extent.

The evidence that Pluto possesses an ocean is circumstantial, but a self-consistent story based on the position of and the tectonics surrounding SP is reasonably convincing (chapter by Nimmo and McKinnon in this volume). Do the sharpness and high stratigraphic position of the most recent extensional faults on Pluto imply active tectonism today? If so, does the evidence for fluid or gas-driven cryovolanism outlined in sec. 4.1.6 also imply ongoing cryovolcanism today? And how is any of this, or Charon's tectonics and plains cryovolcanism, related to ocean freezing?

Finally, we judge that convective overturn is occurring today in the $N_2$-ice sheet within SP. Beyond the inferences for Pluto's heat flow, volatile ice rheology, and the maintenance of a vigorous surface-atmosphere volatile cycle, the ability to study solid state convection in the raw (as it were) is unprecedented. Solid-state convection occurs on Earth (i.e., plate tectonics) and is inferred to occur or have occurred on many other Solar System bodies both rocky and icy. But the details are always hidden from view, beneath the lithosphere of a given world. For the SP ice sheet there is no lithosphere, and the physical structure of convective flow is directly exposed. Pluto thus provides a natural laboratory to study one of the most important processes in geophysical fluid dynamics.

Obviously, we would like to learn more about Pluto and Charon, to see their non-encounter, "farside" hemispheres and terrains that that were in polar darkness in 2015 (*Stern et al.*, 2020). High-resolution remote sensing as well as geophysical measurements would be extraordinarily valuable. Such observations could be made by a future mission to Pluto, logically an orbiter, but given the very long lead time for such a mission, research might focus on a deeper understanding of what *New Horizons* data imply for the Pluto system and planetary formation and evolution in general. Adaptive optics imaging from the coming generation of large, Earth-based telescopes may match or exceed *HST* in terms of resolution, however. At the very least this should allow monitoring of the evolution of albedo patterns and thus surface-atmosphere interactions on Pluto in the coming decades. There are many years of work ahead.

Turning to the small satellites, given their importance to understanding the origin of the Pluto system, further efforts should be made to constrain their masses and thus densities. Efforts to incorporate additional years of *HST* astrometric observations as well as *New Horizons* imaging are underway (*Jacobson et al.*, 2019). Future searches for small trans-Hydran satellites using the next generation of space-based telescopes would also be valuable as a definitive test of any extended debris disk origin hypothesis (*Kenyon and Bromley*, 2019b; Bromley and Kenyon, 2020).

## 5. SUMMARY

The *New Horizons* encounter with the Pluto system was no mere box-checking exercise. By flying by the last of the classical planets and the first known Kuiper belt planet, and for the first time exploring *in situ* major bodies in the Solar System's "3$^{rd}$ zone," a paradigm shift was initiated in our understanding of the possibilities for planetary evolution and expression in modest-scale worlds far from their parent stars. The Pluto-Charon system provides a broad variety of constraints on planetary formation, structure, composition, chemistry, and evolution:

*Origin.* The emerging view of planetesimal formation via gravitational instability in the protoplanetary gas-and-particle disk aligns with the requirements imposed by Charon-forming giant impact models. Initial planetesimals (50-100 km scale) form between ~20-30 AU. Accretion timing appears consistent with subsequent slow and/or stalled pebble accretion followed by hierarchical coagulation after nebular gas dispersal (~few Myr). Partially differentiated proto-Pluto precursors (the probable initial condition) imply slow and/or "pebbly" (impact heat gets radiated) and late (little $^{26}$Al) accretion in the trans-Neptunian planetesimal disk. The dynamic environment (number density, velocity dispersion) in the post-gas planetesimal disk is favorable for Charon formation, and the subsequent giant planet instability and Neptune's migration emplaces Pluto-Charon in its present 3:2 mean-motion resonance with Neptune.

*Interior.* Partially differentiated precursors plus the Charon-forming impact should have driven both Pluto and Charon toward full ice-from-rock differentiation, and concomitantly toward early interior ocean formation. The latter is consistent with the general absence of compressional tectonics on both bodies. While evidence for an ocean on Pluto (and former ocean within Charon) remains circumstantial, evidence continues to accrue from detailed tectonic modeling of the Sputnik basin as a mascon and geologically young eruptions of $NH_3$-bearing cyrofluids and clastics (presumably ultimately sourced from a deep, possibly pressurized ocean) on Pluto. Preservation of Pluto's ocean and maintenance of an uplifted ocean beneath Sputnik (the hypothesized source of the mascon, along with the ice sheet) would have been strongly aided by clathrate formation within or at the base of Pluto's floating ice shell.

*Composition and Chemistry.* Pluto's low surface CO/$N_2$ has been variously explained by CO burial in the Sputnik Planitia $N_2$-ice sheet, destruction by aqueous chemistry in the ocean, or preferential sequestration in subsurface clathrate. There is no fundamental problem explaining Pluto's global nitrogen abundance, though the ultimate provenance of this nitrogen remains to be determined. Within Pluto's likely organic-rich, chemically reducing core, thermochemistry favors the production of metastable organics, graphite, $CH_4$ and $N_2$ — potentially explaining the bulk composition of Pluto's atmosphere and that of similar, sizeable icy worlds.

*Tectonics and Heat Flow.* Contradictory estimates for Pluto's lithospheric heat flow exist, but the preponderance of evidence is for a low, close to steady state radiogenic value (a few mW/m$^2$) throughout most of Pluto's history (see chapter by Nimmo and McKinnon in this volume for details). The significant exception may be the *lack of* evidence for the collapse of Pluto's post-formation rotational bulge, which requires a sufficient combination of higher heat flow and lithosphere weakness. The block tectonics of Charon's Oz Terra bear no clear geometric relation to eccentricity tidal stresses. Strong eccentricity tides are not a given for Charon during its post-formation tidal evolution away from Pluto, however, which puts the onus on ocean freezing and, possibly, tidal bulge collapse to explain the extensive disruption of Charon's ancient crust. A possible explanation is that Charon's tidal evolution was sufficiently rapid, and subsequent geologic activity, including the eruption of ammoniated cryolavas that formed Vulcan Planitia, has obscured most evidence of the tidal evolution epoch.

*Atmosphere and Climate.* The variations of orbital and rotation parameters of Pluto over the last millions of years have led to substantial insolation changes, thus triggering volatile transport and extensive resurfacing, including glaciers and ice mantle formation, as well as more than 1000-fold annual variations of surface pressure and CO and $CH_4$ atmospheric abundances. On Pluto the global nitrogen ice distribution and the induced nitrogen condensation-sublimation flows strongly control the atmospheric circulation. GCMs predict a general retrograde atmospheric circulation for current-day Pluto that could have been in place in Pluto's past as well, and could account for many of the geological features and longitudinal asymmetries in ice distribution observed on Pluto.

*Sputnik.* Much of Pluto's geophysical, geological, and atmospheric behavior has been and is controlled or strongly influenced by Sputnik, which raises the question of how other dwarf planets in the Kuiper Belt (and Triton) behave in the absence of (or in the presence of more than one) giant impact basin. Dynamical arguments suggest that there were once ~1000-4000 Pluto-mass bodies in the trans-Neptunian planetesimal disk. Simulations show that the current scattered disk comprises ~0.5–1% of the original planetesimals in the aKB (see *Morbidelli and Nesvorný*, 2020). Thus up to several dozen Pluto-class dwarf planets may still be out there, in the Kuiper Belt's scattered disk and its extended (detached) component.

## 6. CODA

So what has *New Horizons'* exploration of the Pluto system taught us? It has taught us once again that Nature's imagination exceeds our own. It has reinforced the emerging paradigm that planetary-level behavior is not the sole province of terrestrial-composition (rock+metal) planets or even relatively large worlds (Mars-scale and beyond). As one moves farther from the Sun, as long as solid bodies can partake compositionally of increasingly geologically mobile and volatile materials (carbonaceous matter, all manner of ices), all of the characteristic expressions of internal and insolation-driven geological activity found on the active terrestrial planets (Mars and Earth especially) can reappear in new robes. Some are similar, some are novel; all are fascinating. While the most active icy satellites (Europa, Enceladus) characteristically derive their activity from resonant tidal heating, Pluto is proof that tidal heating is not absolutely necessary, within limits. The differences between Pluto and Charon do illustrate that size matters; however, the fuzzy boundary between worlds that enjoy early activity before sliding into senescence and those that remain active after 4.6 billion years occurs at a much smaller size scale than previously thought.

Our understanding of the Pluto system, and of the Kuiper belt in which it resides, are set for much further improvement. Sections 2.6, 3.5, and 4.5 in this chapter, and other chapters in this volume, detail on-going, critical, important, or hoped-for progress in new Earth-based astronomical observations, continued analyses of *New Horizons* and other data, experimental measurements of the relevant properties of planetary ices and their geochemical/petrological interactions with rock and carbonaceous matter, and ever improving numerical simulations of geological, geophysical, geochemical, atmospheric and dynamical processes. But the single most important advance in the decades ahead will come from continued exploration of the Kuiper belt by spacecraft. A return to Pluto, with an orbiter mission, could obviously address the majority of the science questions laid out in this volume, and would be a great leap forward. But equally important would be further reconnaissance of other KBOs, of any size or dynamical class, including Centaurs, which are derived from the Kuiper belt's scattered disk. One only needs to consider the advances in planetary science that resulted from the July 2015 *New Horizons* encounter with the Pluto system, and equally, those that resulted from the subsequent New Year's 2019 encounter with the small cold classical KBO Arrokoth. Future telescopic surveys should allow the planning of a flyby mission with multiple KBO encounters more-or-less along the spacecraft's trajectory. *New Horizons*, like the Pioneers (10 and 11) through the asteroid belt to Jupiter and Saturn before it, was a pathfinder mission, and a highly capable one at that. But the in-depth exploration of the Kuiper belt has only begun, and its scientific riches beckon. *Carpe Tertium Zona!*

***Acknowledgments.*** We thank reviewers F. Nimmo, W. Grundy, A. Morbidelli, D. Nesvorný, and Z. Leinhardt, uber-editor S. A. Stern, and O. White and J. Moore for their comments and suggestions regarding the manuscript. All authors also thank the *New Horizons* project for supporting this research and its many components over the years. C. R. G. was partly supported by the NASA Astrobiology Institute through its JPL-led team entitled Habitability of Hydrocarbon Worlds: Titan and Beyond. T. B. was supported for this work by an appointment to the National Aeronautics and Space Administration (NASA) Post-doctoral Program at the Ames Research Center administered by Universities Space Research Association (USRA) through a contract with NASA.

# REFERENCES

Abod C. P., Simon J. B., Li R., Armitage P. J., Youdin A. N., and Kretke K. A. (2019) The mass and size distribution of planetesimals formed by the streaming instability. II. The effect of the radial gas pressure gradient. *Astrophys. J., 883*, 192, DOI: 10.3847/1538-4357/ab40a3.

Alexander C. M. O'D., Cody G. D., De Gregorio B. T., Nittler L. R., and Stroud R. M. (2017) The nature, origin and modification of insoluble organic matter in chondrites, the major source of Earth's C and N. *Chem. Erde*, *77*, 227–256.

Alexander R., Pascucci I., Andrews S., Armitage P., and Cieza L. (2014) The dispersal of protoplanetary disks. In *Protostars and Planets VI* (H. Beuther et al., eds.), pp. 475–496. Univ. of Arizona, Tucson, DOI: 10.2458/azu_uapress_9780816531240-ch021.

Altwegg K., Balsiger H., Bar-Nun A., et al. (2016) Prebiotic chemicals — amino acid and phosphorus — in the coma of Comet 67P/Churyumov-Gerasimenko. *Sci. Adv.*, *2*, e1600285.

Arakawa S., Hyodo R., and Genda H. (2019) Early formation of moons around large trans-Neptunian objects via giant impacts. *Nature Astron.*, *3*, 802–807, DOI: 10.1038/s41550-019-0797-9.

Atreya S. K., Donahue T. M., Kuhn W. R. (1978) Evolution of a nitrogen atmosphere on Titan. *Science*, *201*, 611-613.

Atreya S. K., Lorenz R. D., and Waite J. H. (2009) Volatile origin and cycles: Nitrogen and methane. In *Titan from Cassini-Huygens* (R. H. Brown et al., eds.), pp. 77–99. Springer, New York.

Bagenal F., Horányi M., McComas D. J., et al. (2016) Pluto's interaction with its space environment: solar wind, energetic particles, and dust. *Science*, *351*, aad9045.

Bahcall J. N., Pinsonneault M H., and Basu S. (2001) Solar models: Current epoch and time dependences, neutrinos, and helioseismological properties. *Astrophys. J.*, *555*, 990–1012.

Bannister M. T., Gladman B. J., Kavelaars J. J., et al. (2018) OSSOS. VII. 800+ trans-Neptunian objects — the complete data release. *Astrophys. J. Suppl.*, *236*, 18, DOI: 10.3847/1538-4365/aab77a.

Bardyn A., Baklouti D., Cottin H., et al. (2017) Carbon-rich dust in comet 67P/Churyumov-Gerasimenko measured by COSIMA/Rosetta. *Mon. Not. R. Astron. Soc.*, *469*, S712–S722, DOI: 10.1093/mnras/stw2640.

Barr A. C. and Collins (2015) Tectonic activity on Pluto after the Charon-forming impact. *Icarus*, *246*, 146–155, DOI: 10.1016/j.icarus.2014.03.042.

Barr, A. C. and Schwamb M. E. (2016) Interpreting the densities of the Kuiper belt's dwarf planets. *Mon. Not. Royal Astron. Soc.*, *460*, 1542–1548, DOI:10.1093/mnras/stw1052.

Barucci M. A., Dalle Ore C. M., Perna1 D., Cruikshank D. P., Doressoundiram A., and Alvarez-Candal A. (2015) (50000) Quaoar: Surface composition variability. *Astron. Astrophys.*, *584*, A107, DOI: 10.1051/0004-6361/201526119.

Bertrand T. and Forget F. (2016) Observed glacier and volatile distribution on Pluto from atmosphere-topography processes. *Nature*, *540*, 86–89.

Bertrand T., Forget F., Umurhan, O. M., et al. (2018) The nitrogen cycles on Pluto over seasonal and astronomical timescales. *Icarus*, *309*, 277–296, DOI: 10.1016/j.icarus.2018.03.012.

Bertrand. T., Forget. F., Umurhan, O. M., Moore, J. M., Young, L. A., Protopapa, S., Grundy, W. M., Schmitt, B., Dhingra, R. D. and Binzel R. P. (2019) The $CH_4$ cycles on Pluto over

seasonal and astronomical timescales. Icarus, 329, 148–165, DOI: 10.1016/j.icarus.2019.02.007.

Bertrand T., Forget F., White O., and Schmitt B. (2020) Pluto's beating heart regulates the atmospheric circulation: Results from high resolution and multi-year numerical climate simulations. *J. Geophys. Res. Planets*, *125*, e06120, DOI: 10.1029/2019JE006120.

Beuther H., Klessen R. S., Dullemond C. P., and Henning T. K., eds. (2014) *Protostars and Planets VI*. Univ. of Arizona, Tucson. 914 pp.

Beyer R.A., Nimmo F., McKinnon W.B., et al. (2017) Charon tectonics. *Icarus*, *287*, 161–174,O DOI: 10.1016/j.icarus.2018.12.032.

Beyer R.A., Spencer J. R., McKinnon W.B., et al. (2019) The nature and origin of Charon's smooth plains. *Icarus*, *323*, 16–32. DOI: 10.1016/j.icarus.2016.12.018.

Bieler A., Altwegg K., Balsiger H., et al. (2015) Abundant molecular oxygen in the coma of comet 67P/Churyumov-Gerasimenko. *Nature*, *526*, 678–81.

Bierson C. J. and Nimmo F. (2019) Using the density of Kuiper Belt Objects to constrain their composition and formation history. *Icarus*, *326*, 10–17, DOI: 10.1016/j.icarus.2019.01.017.

Bierson C. J., Nimmo F., McKinnon W. B. (2018) Implications of the observed Pluto-Charon density contrast. *Icarus*, *309*, 207–219.

Bierson C. J., Nimmo F., Stern S. A. (2020) Evidence for a hot start and early ocean on Pluto. *Nature Geosci.*, *13*, 468–472, DOI: 10.1038/s41561-020- 020-0595-0.

Binzel R. P., Earle A. M., Buie M. W., Young L. A., Stern S. A., Olkin C. B., Ennico K., Moore J. M., Grundy W., Weaver H.A., Lisse C. M., Lauer T. R. (2017a) Climate zones on Pluto and Charon. *Icarus*, *287*, 30–36IOI: 10.1016/j.icarus.2016.07.023.

Binzel R. P., Olkin C. B., Young L. A. (2017b) Editorial. *Icarus*, *287*, 1, DOI: 10.1016/j.icarus.2017.02.001.

Bishop J. L. Banin A., Mancinelli R. L., Klovstad M. R. (2002) Detection of soluble and fixed $NH_4^+$ in clay minerals by DTA and IR reflectance spectroscopy: A potential tool for planetary surface exploration. *Planet. Space Sci.*, *50*, 11–19.

Biver N., Bockelée-Morvan D., Paubert G., Moreno R., Crovisier J., Boissier J., Bertrand E., Boussier H., Kugel F., McKay A., Dello Russo N., and DiSanti M.A. (2018) The extraordinary composition of the blue comet C/2016 R2 (PanSTARRS). *Astron. Astrophys.*, *619*, A127.

Bouquet A., Mousis O., Glein C. R., Danger G., and Waite J. H. (2019) The role of clathrate formation in Europa's ocean composition. *Astrophys. J.*, *885*, 14, DOI: 10.3847/1538-4357/ab40b0.

Bromley B. C. and Kenyon S. J. (2020) A Pluto–Charon concerto: An impact on Charon as the origin of the small satellites. *Astron. J.*, *160*, 85.

Brown, M. E. (2012) The compositions of Kuiper belt objects. *Annu. Rev. Earth Planet. Sci.*, *40*, 467–494.

Brown M. E., Barkume K. M., Ragozzine D., and Shlerchaller E. L. (2007) A collisional family of icy objects in the Kuiper belt. *Nature*, *446*, 294–296.

Brozović M., Showalter M., Jacobson R. A., and Buie M. W. (2015) The orbits and masses of satellites of Pluto. *Icarus*, *46*, 317–29, DOI: 10.1016/j.icarus.2014.03.015.

Buhler P. B. and Ingersoll A. P. (2018) Sublimation pit distribution indicates convection cell surface velocities of ∼10 cm per year in Sputnik Planitia, Pluto. *Icarus*, *300*, 327-340.

Buratti B. J., Thomas P. C., Roussos E., et al. (2019) Close Cassini flybys of Saturn's ring moons Pan, Daphnis, Atlas, Pandora, and Epimetheus. *Science*, *364*, eaat2349, DOI: 10.1126/science.aat2349.

Canup, R. M. (2005) A giant impact origin of Pluto–Charon. *Science*, *307*, 546–550.

Canup, R. M. (2011) On a giant impact origin of Charon, Nix, and Hydra. *Astron. J.*, *141*, 35–44.

Carter, P. J., Z. Lienhardt Z. M., Elliott T., Stewart S. T., and Walker M. J. (2018) Collisional stripping of planetary crusts. *Earth Planet. Sci. Lett.*, *484*, 276–286, DOI: 10.1016/j.epsl.2017.12.012.

Castillo-Rogez J., Johnson T. V., Lee M. H., Turner N. J., Matson D. L., and Lunine J. (2009) $^{26}$Al decay: Heat production and a revised age for Iapetus. *Icarus*, *204*, 658–662.

Cheng W. H., Lee M. H., and Peale S. J. (2014) Complete tidal evolution of Pluto-Charon. *Icarus*, *233*, 242–258.

Claire M. W., Sheets J., Cohen M., Ribas I., Meadows V. S., and Catling D. C. (2012) The evolution of solar flux from 0.1 nm to 160 μm: Quantitative estimates for planetary studies. *Astrophys. J.*, *757*, 95, DOI: 10.1088/0004-637X/757/1/95.

Clark R. N., Brown R. H., Cruikshank D. P., and Swayze G. A. (2019). Isotopic ratios of Saturn's rings and satellites: Implications for the origin of water and Phoebe. *Icarus*, *321*, 791–802.

Cleeves L. I., Bergin E. A., Alexander C. M. O'D., Du F., Graninger D., Öberg K. I., and Harries T. J. (2014) The ancient heritage of water ice in the solar system. *Science*, *345*, 1590–1593.

Cook J. A., Dalle Ore C. M., Protopapa S., et al. (2018) Composition of Pluto's small satellites: analysis of *New Horizons* spectral images. *Icarus*, *315*, 30–45.

Cook J. A., Dalle Ore C. M., Protopapa S., et al. (2019) The distribution of $H_2O$, $CH_3OH$, and hydrocarbon-ices on Pluto: analysis of *New Horizons* spectral images. *Icarus*, *331*, 148–169.

Cruikshank D. P., Umurhan O. M., Beyer R. A., et al. (2019) Recent cryovolcanism in Virgil Fossae on Pluto. *Icarus*, *330*, 155-168, DOI: 10.1016/j.icarus.2019.04.023.

Cruikshank D. P., Dalle Ore C. M., Scipioni F., et al. (2020) Cryovolcanic flooding in Viking Terra on Pluto. *Icarus*, 113786.

Dalle Ore C. M., Protopapa S., Cook J. A., et al. (2018) Ices on Charon: Distribution of $H_2O$ and $NH_3$ from New Horizons LEISA observations. *Icarus*, *300*, 21–32.

Dalle Ore C. M., Cruikshank D. P., Protopapa S., Scipioni F., McKinnon W. B., Cook J. C., Grundy W. M., Stern S. A., Moore J. M., Verbiscer A., Parker A. H., Singer K. H., Umurhan O. M., Weaver H. A., Olkin C. B., Young L. A., Ennico K., and the New Horizons Surface Composition Theme Team (2019) Detection of ammonia on Pluto's surface in a region of geologically recent tectonism. *Sci. Adv.*, *5*, eaav5731, DOI: 10.1126/sciadv.aav5731.

Dartois D. E., Bouzit M., and B. Schmitt B. (2012) Clathrate hydrates: FTIR spectroscopy for astrophysical remote detection. In *European Conference on Laboratory Astrophysics – ECLA* (C. Stehlé, C. Joblin and L. d'Hendecourt, eds.), pp. 219–224. EAS Publications Series, 58.

Davidsson B. J. R., Sierks H., Güttler C., et al. (2016) The primordial nucleus of comet 67P/Churyumov-Gerasimenko. *Astron. Astrophys.*, *592*, A63. DOI: 10.1051/0 0 04-6361/201526968.

de Bergh C., Schaller E. L., Brown M. E., Brunetto R., Cruikshank D. P., Schmitt B. (2013) The ices on transneptunian objects and Centaurs. In *The Science of Solar System Ices* ( M. Gudipati M. and J.Castillo-Rogez, eds.), pp. 107–146. Springer, New York.

De Sanctis M. C., Ammannito E., Raponi A., et al. (2015) Ammoniated phyllosilicates with a likely outer Solar System origin on (1) Ceres. *Nature, 528*, 241-244.
64

Desch S. J. (2007) Mass distribution and planet formation in the solar nebula. *Astrophys. J.*, *671*, 878–893, DOI: 10.1086/522825.

Desch S. J., Cook J. C., Doggett T. C., Porter S. B. (2009) Thermal evolution of Kuiper belt objects, with implications for cryovolcanism. *Icarus, 202*, 694-714.

Desch, S.J. and Neveu M. (2017) Differentiation and cryovolcanism on Charon: a view before and after New Horizons. *Icarus*, *287*, 175–186, DOI: 10.1016/j.icarus.2016.11.037.

Desch S. J., Kalyaan A., and Alexander C. M. O'D. (2018) The effect of Jupiter's formation on the distribution of refractory elements and inclusions in meteorites. *Astrophys. J. Suppl.*, *238*, 11, DOI: 10.3847/1538-4365/aad95f.

Dobrovolskis A. R., Peale S. J. and Harris A. W. (1997) Dynamics of the Pluto-Charon binary. In *Pluto and Charon* (S. A. Stern and D.J. Tholen, eds.), pp. 159–190. Univ. of Arizona, Tucson.

Durham W. B., Kirby S. H., and Stern L. A. (1993) Flow of ices in the ammonia-water system. *J. Geophys. Res.*, *98*, 17,667-17,682.

Earle A. M. and Binzel R. P. (2015) Pluto's insolation history: Latitudinal variations and effects on atmospheric pressure. *Icarus*, *250*, 405-412.

Earle A. M., Binzel R. P., Young L. A., Stern S. A., Ennico K., Grundy W., Olkin C. B., Weaver H. A., and New Horizons Geology and Geophysics Imaging Team (2017) Long-term surface temperature modeling of Pluto. *Icarus*, *287*, 37-46.

Emsenhuber A., Jutzi M., and Benz W. (2018) SPH calculations of Mars-scale collisions: The role of the equation of state, material rheologies, and numerical effects. *Icarus*, *301*, 247–257, DOI:

Forget F., Bertrand T., Vangvichith M., Leconte J., Millour E., and Lellouch E. (2017) A post-new horizons global climate model of Pluto including the $N_2$, $CH_4$ and CO cycles. *Icarus*, *287*, 54–71.

Fraser W. C., Brown M. E., Morbidelli A., Parker A., and Batygin K. (2014) The absolute magnitude distribution of Kuiper Belt objects. *Astrophys. J.*, *782*, 100.

Fray, N., Bardyn A., Cottin H., et al. (2016) High-molecular-weight organic matter in the particles of comet 67P/Churyumov–Gerasimenko. *Nature*, *538*, 72–74, DOI: 10.1038/nature19320.

Fray N. Bardyn A., Cottin H., et al. (2017) Nitrogen-to-carbon atomic ratio measured by COSIMA in the particles of comet 67P/Churyumov–Gerasimenko. *Mon. Not. R. Astron. Soc.*, *469*, S506–S516.

Fulle, M., Della Corte V., Rotundi A., et al. (2016) Comet 67P/Churyumov-Gerasimenko preserved the pebbles that formed planetesimals. *Mon. Not. R. Astron. Soc.*, *462*, S132–S137, DOI: 10.1093/mnras/stw2299.

Fulle, M., Della Corte V., Rotundi A., et al. (2017) The dust-to-ices ratio in comets and Kuiper belt objects. *Mon. Not. R. Astron. Soc.*, *469*, S45–S49, DOI: 10.1093/mnras/stw983.

Füri E. and Marty B. (2015) Nitrogen isotope variations in the solar system. *Nature Geosci.*, *8*, 515–522.

Gabasova L., Tobie G., and Choblet G. (2018) Compaction-driven evolution of Pluto's rocky core: Implications for water-rock interactions. In *Lunar and Planetary Science XLIX*, Abstract #2512. Lunar and Planetary Institute, Houston.

Gao P., Fan S., Wong M. L., et al. (2017) Constraints on the microphysics of Pluto's photochemical haze from New Horizons observations. *Icarus*, *287*, 116–123.

Gladman B. and Chan C. (2006) Production of the extended scattered disk by rogue planets. *Astrophys. J. Lett.*, *643*, L135–L138, DOI: 10.1086/505214.

Gladstone G. R. and Young L. A. (2019) New Horizons observations of the atmosphere of Pluto. *Annu. Rev. Earth Planet. Sci.*, 57, 119–140. DOI: 0.1146/annurev-earth-053018-060128.

Gladstone G. R. and 32 colleagues (2016) The atmosphere of Pluto as observed by New Horizons. *Science, 351*, aad8866.

Glein C. R. et al. (2008) The oxidation state of hydrothermal systems on early Enceladus. *Icarus*, *197*, 157-163.

Glein C. R. (2015) Noble gases, nitrogen, and methane from the deep interior to the atmosphere of Titan. *Icarus*, *250*, 570-586.

Glein C. R. and Waite J. H. (2018) Primordial N$_2$ provides a cosmochemical explanation for the existence of Sputnik Planitia, Pluto. *Icarus*, *313*, 79–92, DOI: 10.1016/j.icarus.2018.05.007.

Glein C. R., Postberg F., Vance S. D. (2018) The geochemistry of Enceladus: Composition and controls. In *Enceladus and the Icy Moons of Saturn* (P. M. Schenk, R.N. Clark, C.J.A. Howett, A.J. Verbiscer, and J.H. Waite, eds.), pp. 39-56. Univ. of Arizona, Tucson.

Goldreich P. and Ward W. R. (1973) The formation of planetesimals. *Astrophys. J.*, *183*, 1051–1062.

Goldreich P. G., Lithwick Y., and Sari R. (2004) Planet formation by coagulation: A focus on Uranus and Neptune. *Annu. Rev. Astron. Astrophys.*, *42*, 549–601, DOI: 10.1146/annurev.astro.42.053102.134004.

Gomes R. S. (2003) The origin of the Kuiper Belt high-inclination population. *Icarus*, *161*, 404–418.

Gomes R. S., Gallardo T., Fernández, J. A., and Brunini A. (2005) On the origin of the high-perihelion scattered disk: The role of the Kozai mechanism and mean motion resonances. *Celest. Mech. Dyn. Astr.*, *91*, 109-129.

Gough D.O. (1981) Solar interior structure and luminosity variations. *Solar Phys.*, *74*, 21–34.

Greenstreet S., Gladman B., and McKinnon W. B. (2015) Impact and cratering rates onto Pluto. *Icarus*, *258*, 267–288.

Grundy W. M., Young L. A., Stansberry J. A., et al. (2010) Near-infrared spectral monitoring of Triton with IRTF/SpeX II: Spatial distribution and evolution of ices. *Icarus*, *205*, 594–604.

Grundy W. M., Binzel R. P., Buratti B. J., et al., (2016) Surface compositions across Pluto and Charon. *Science*, *351*, aad9189, DOI: 10.1126/science.aad9189.

Grundy W. M., Bertrand T., Binzel R. P., et al. (2018) Pluto's haze as a surface material. *Icarus*, *314*, 232–245.

Grundy W. N., Bird M. K., Britt D. T., et al., et al. (2020a) Color, composition, and thermal environment of Kuiper belt object (486958) Arrokoth. *Science*, *367*, eaay3999. DOI:10.1126/science.aay3999.

Grundy W. M., et al. (2020b) Editorial. *Icarus*, *xxx*, 1.

Haisch K. E., Lada E. A., and Lada C. J. (2001) Disk frequencies and lifetimes in young clusters. *Astrophys. J.*, *553*, L153–L156.

Hammond N. P., Barr A. C., Parmentier E. M. (2016) Recent tectonic activity on Pluto driven by phase changes in the ice shell. *Geophys. Res. Lett.*, *43*, 6775–6782. DOI: 10.1002/2016GL069220.

Hersant F., Gautier D., and Lunine J. I. (2004) Enrichment in volatiles in the giant planets of the Solar System. *Planet. Space Sci.*, *52*, 623–641.
67

Holler B. J., Young L. A., Bus S. J., and Protopapa S. (2017) Methanol ice on Kuiper Belt objects 2007 OR$_{10}$ and Salacia: Implications for formation and dynamical evolution. *EPSC Abs.*, *11*, EPSC2017-330.

Howard A. D. and 14 colleagues (2017a) Present and past glaciation on Pluto. *Icarus*, *287*, 287–300.

Howard A. D. and 16 colleagues (2017b) Pluto: Pits and mantles on uplands north and east of Sputnik Planitia. *Icarus*, *293*, 218–230.

Imbrie J. and Imbrie K. P. (1979) *Ice Ages: Solving the Mystery*. Harvard Univ. Press, Cambridge, MA. 224 pp.

Izidoro A., Raymond S. N., Morbidelli A., Hersant F., and Pierens A. (2015) Gas giant planets as dynamical barriers to inward-migrating super-Earths. *Astrophys. J. Lett.*, *800*, L22, DOI: 10.1088/2041-8205/800/2/L22.

Jacobson R. A., Brozović M., Showalter M., Verbiscer, Buie M., and Helfenstein P. (2019) The orbits and masses of Pluto's satellites. *Pluto System After New Horizons 2019*, Laurel, MD, Abstract #2133.

Janes D. M. and Melosh H. J. (1990) Tectonics of planetary loading—a general model and results. *J. Geophys. Res.*, *95*, 21345–21355.

Johansen A. and Lambrechts M. (2017) Forming planets via pebble accretion. *Annu. Rev. Earth. Planet. Sci.*, *45*, 359–387, DOI: 10.1146/annurev-earth-063016-020226.

Johansen A., Oishi J. S., Mac Low M.-M., Klahr H., Henning T., and Youdin A. (2007) Rapid planetesimal formation in turbulent circumstellar disks. *Nature*, *448*, 1022–1025, DOI: 10.1038/nature06086.

Johansen A., Blum J., Tanaka H., Ormel C., Bizzarro M., and Rickman H. (2014) The multifaceted planetesimal formation process. In *Protostars and Planets VI* (H. Beuther et al., eds.), pp. 547–570. Univ. of Arizona, Tucson.

Johansen A., Mac Low M.-M., Lacerda P., and Bizzaro M. (2015) Growth of asteroids, planetary embryos, and Kuiper belt objects by chondrule accretion. *Sci. Adv.*, *1*, e1500109, DOI: 10.1126/sciadv.1500109.

Johnson B. C., Bowling T. J., Trowbridge A. J., and Freed A. M. (2016) Formation of the Sputnik Planum basin and the thickness of Pluto's subsurface ocean. *Geophys. Res. Lett.*, *43*, 10,068–10,077, DOI: 10.1002/2016GL070694, 2016.

Johnson P., Young L. A., Protopapa S., et al. (2019) Pluto's minimum surface pressure and implications for haze production. In *Pluto System After New Horizons*, Abstract #7025. Lunar and Planetary Institute, Houston.

Johnson R. E., Oza A., Young L. A., Volkov A. N., and Schmidt C. (2015) Volatile loss and classification of Kuiper Belt objects. *Astrophys. J.*, *809*, 43, DOI: 10.1088/0004-637X/809/1/43.

Kamata S., Nimmo F., Sekine Y., Kuramoto K., Noguchi N., Kimura J., and Tani A. (2019) Pluto's ocean is capped and insulated by gas hydrates. *Nat. Geosci.*, *12*, 407-410.

Kammer J. A., Stern S. A., Young L. A., et al. (2017) *New Horizons* upper limits on $O_2$ in Pluto's present day atmosphere. *Astron J.*, *154*, 55.

Keane J. T., Matsuyama I., Kamata S., Steckloff J. K. (2016) Reorientation and faulting of Pluto due to volatile loading within Sputnik Planitia. *Nature*, *540*, 90–93.

Kenyon S. C. and Bromley B. C. (2012) Coagulation calculations of icy planet formation at 15–150 AU: A correlation between the maximum radius and the slope of the size distribution for trans-Neptunian objects. *Astron. J.*, *143*, 63, DOI: 10.1088/0004-6256/143/3/63.

Kenyon S. J. and Bromley B. C. (2014) The formation of Pluto's low-mass satellites. *Astrophys. J., 147,* 8.

Kenyon S. J. and Bromley B. C. (2019a) A Pluto–Charon sonata: Dynamical limits on the masses of the small satellites. *Astrophys. J., 158*, 69.

Kenyon S. C. and Bromley B. C. (2019b) A Pluto-Charon sonata: The dynamical architecture of the circumbinary satellite system. *Astron. J.*, *157*, 79.

Kenyon S. C. and Bromley B. C. (2019c) A Pluto-Charon sonata. III. Growth of Charon from a circum-Pluto ring of debris. *Astron. J.*, *158*, 142.

Kissel J. and Krueger F. R. (1987) The organic component in dust from comet Halley as measured by the PUMA mass spectrometer on board Vega 1. *Nature*, *326*, 755–760.

Kouchi A. and Sirono S. (2001) Crystallization heat of impure amorphous $H_2O$ ice. *Geophys. Res. Lett.*, *28*, 827–830.

Krasnopolsky V. (2016) Isotopic ratio of nitrogen on Titan: Photochemical interpretation. *Planet. Space Sci.*, *134*, 61–63.

Kruijer T. S., Burkhardt C., Budde G., and Kleine T. (2017) Age of Jupiter inferred from the distinct genetics and formation times of meteorites. *Proc. Natl. Acad. Sci. USA*, *114*, 6712–6716.

Laskar J., Correia A. C. M., and Gastineau M. (2004) Long term evolution and chaotic diffusion of the insolation quantities of Mars. *Icarus*, *170*, 343–364.

Lawler S. M., Shankman C., Kavelaars J. J., et al. (2018) OSSOS. VIII. The transition between two size distribution slopes in the scattering sisk. *Astron. J.*, *155*, 197, DOI: 10.3847/1538-3881/aab8ff.

Lawler S. M., Pike R. E., Kaib N., et al. (2019) OSSOS. XIII. Fossilized resonant dropouts tentatively confirm Neptune's migration was grainy and slow. *Astron. J.*, *157*, 6, DOI: 10.3847/1538-3881/ab1c4c.

Leinhardt Z., Kraus R. A., and Stewart S. T. (2010) The formation of the collisional family around the dwarf planet Haumea. *Astrophys. J.*, *714*, 1789–1799.

Lellouch E., Gurwell M., Butler B., et al. (2017) Detection of CO and HCN in Pluto's atmosphere with ALMA. *Icarus*, *286*, 289–307, DOI: 10.1016/j.icarus.2016.10.013.

Levison H. F., Morbidelli A., VanLaerhoven C., Gomes R., and Tsiganis K. (2008a) Origin of the structure of the Kuiper belt during a dynamical instability in the orbits of Uranus and Neptune. *Icarus*, *196*, 258–273.

Levison, H. F., Morbidelli A., Vokrouhlický D., and Bottke W. F. (2008b) On a scattered- disk origin for the 2003 EL$_{61}$ collisional family–An example of the importance of collisions on the dynamics of small bodies. *Astron. J.*, *136*, 1079–1088.

Lewis J. S. (1971) Satellites of the outer planets: Thermal models. *Science*, *172*, 1127–1128, DOI: 10.1126/science.172.3988.1127.

Lodders K. (2003) Solar system abundances and condensation temperatures of the elements. *Astrophys. J.*, *591*, 1220-1247.

Lorenzi V., Pinilla-Alonso N., and Licandro J. (2015) Rotationally resolved spectroscopy of dwarf planet (136472) Makemake. *Astron. Astrophys.*, *577*, A86, DOI: 10.1051/0004-6361/201425575.

Lunine J. I. and Nolan M. C. (1992) A massive early atmosphere on Triton. *Icarus*, *100*, 221–234.

Luspay-Kuti A., Mousis O., Hässig M., et al. (2016) The presence of clathrates in comet 67P/Churyumov-Gerasimenko. *Sci. Adv.*, *2*, e1501781, DOI: 10.1126/sciadv.1501781.

Malamud U., Perets H. B., and Schubert G. (2017) The contraction/expansion history of Charon with implications for its planetary-scale tectonic belt. *Mon. Not. R. Astron. Soc.*, *468*, 1056–1069, DOI: 10.1093/mnras/stx546.

Malhotra R. (1993) The origin of Pluto's peculiar orbit. *Nature*, *365*, 819–821.

Malhotra R. and Williams J. G. (1997) Pluto's heliocentric orbit. In *Pluto and Charon* (S.A. Stern and D.J. Tholen, eds.) pp. 127-157. Univ. Arizona Press, Tucson.

Mandt K. E., Mousis O., Lunine J., and Gautier D. (2014) Protosolar ammonia as the unique source of Titan's nitrogen. *Astrophys. J. Lett.*, *788*, L24, DOI: 10.1088/2041-8205/788/2/L24.

Manga M. and Wang, C. Y. (2007) Pressurized oceans and the eruption of liquid water on Europa and Enceladus. *Geophys. Res. Lett.*, *34*, L07202, DOI: 10.1029/2007GL029297.

McBride B. J. and Gordon S. (1996) Computer program for calculation of complex chemical equilibrium compositions and applications: II. Users manual and program description. *NASA Reference Publication 1311*. NASA, Cleveland. 177 pp.

McGovern P. J., White O. L., and Schenk P. M. (2019) Tectonism across Pluto: Mapping and interpretations. *Pluto System After New Horizons 2019*, Laurel, MD, Abstract #7063.

McKay A. J., Disanti M. A., Kelley M. S. P., et al. (2019) The peculiar volatile composition of CO-dominated comet C/2016 R2 (PanSTARRS). *Astron. J.*, *158*, 128, DOI: 10.3847/1538-3881/ab32e4.

McKay C. P., Scattergood T. W., Pollack, J. B., Borucki, W. J., and van Ghyseghem H. T. (1988) High-temperature shock formation of $N_2$ and organics on primordial Titan. *Nature*, *332*, 520-522.

McKinnon, W. B. (1989) On the origin of the Pluto–Charon binary. *Astrophys. J.*, *344*, L41–L44.

McKinnon W. B. (2002) On the initial thermal evolution of Kuiper Belt Objects. *Proc. Asteroids, Comets, Meteors (ACM 2002)*, ESA SP-500, 29-38.

McKinnon W. B. and Mueller S. (1988) Pluto's structure and composition suggest origin in the solar, not a planetary, nebula. *Nature*, *335*, 240–243.

McKinnon W.B., Stern S. A., Weaver H. A., et al. (2017) Origin of the Pluto-Charon system: Constraints from the New Horizons flyby. *Icarus*, *287*, 2–11, DOI: 10.1016/j.icarus.2016.11.019.

McKinnon W. B., Simonelli D. P., and Schubert G. (1997) Composition, internal structure, and thermal evolution of Pluto and Charon. In *Pluto and Charon* (S. A. Stern and D.J. Tholen, eds.), pp. 295–343. Univ. of Arizona, Tucson.

McKinnon W. B., Prialnik D., Stern S.A., and Coradini A. (2008) Structure and evolution of Kuiper belt objects and dwarf planets. In *The Solar System beyond Neptune* (M. A. Barucci, H. Boehnhardt, D. Cruikshank, and A. Morbidelli, eds.), pp. 213–241. Univ. of Arizona, Tucson.

McKinnon W.B., Nimmo F., Wong T., et al. (2016a) Convection in a volatile nitrogen-ice-rich layer drives Pluto's geological vigour. *Nature*, *534*, 82–85.

McKinnon W. B., Schenk P. M., Moore J. M., et al. (2016b) An impact basin origin for Sputnik "Planitia" and surrounding terrains, Pluto. *Geological Society of America Abstracts with Programs, 48*, DOI: 10.1130/abs/2016AM-285142.

McKinnon W. B., Lunine J. I., Mousis O., Waite J. H., and Zolotov M. Yu. (2018a). The mysterious origin of Enceladus: A compositional perspective. In *Enceladus and the Icy Moons of Saturn* (P.M. Schenk, R.N. Clark, C.J.A. Howett, A.J. Verbiscer, and J.H. Waite, eds.), pp. 17-38. Univ. Arizona, Tucson, DOI: 10.2458/azu_uapress_9780816537075-ch002.

McKinnon W. B., Schenk P. M., and Bland M. T. (2018b) Pluto's heat flow: A mystery wrapped in an ocean inside an ice shell. In *Lunar and Planetary Science XLIX*, Abstract #2715. Lunar and Planetary Institute, Houston.

McKinnon W. B., Richardson D., Marohnic J., et al. (2020) The solar nebula origin of (486958) Arrokoth, a primordial contact binary in the Kuiper Belt. *Science*, *367*, eaay6620, DOI: 10.1126/science.aay6620.

Melosh H. J. and Ivanov B. (2018) Slow impacts on strong targets bring on the heat. *Geophys. Res. Lett.*, *45*, 2597-2599.

Merlin, F., (2015) New constraints on the surface of Pluto. *Astron. Astrophys.*, *582*, A39, DOI: 10.1051/0004-6361/201526721.

Merlin F., Lellouch E., Quirico E., and Schmidtt B. (2018) Triton's surface ices: Distribution, temperature and mixing state from VLT/SINFONI observations. *Icarus*, *314*, 274–292.

Miller K. E., Glein C. R., and Waite J. H. (2019) Contributions from accreted organics to Titan's atmosphere: New insights from cometary and chondritic data. *Astrophys. J.*, *871*, 59.

Moore J. M. and McKinnon W. B. (2021) Geologically diverse Pluto and Charon: Implications for the dwarf planets of the Kuiper belt. *Annu. Rev. Earth. Planet. Sci.*, *48*, in press.

Moore J. M. and Pappalardo R. T. (2011) Titan: An exogenic world? *Icarus*, *212*, 790–806.

Moore, J. M., Howard A. D., Schenk P M., McKinnon W. B., Pappalardo R. T., Ewing R. C., Bierhaus E. B., Bray V. J., Spencer J. R., Binzel R. P., Buratti B., Grundy W. M., Olkin C. B., Reitsma H. J., Reuter D. C., Stern S. A., Weaver H., Young L. A., and Beyer R. A. (2015). Geology before Pluto: Pre-encounter considerations. *Icarus*, *246*, 65–81, DOI: 10.1016/ j.icarus.2014.04.028.

Moore J. M. and 40 colleagues (2016) The geology of Pluto and Charon through the eyes of New Horizons. *Science*, *351*, 1284-1293.

Moore J. M. and 25 colleagues (2018) Bladed Terrain on Pluto: Possible origins and evolution. *Icarus*, *300*, 129-144.

Moores,J. E., Smith C. L., Toigo A. D. and Guzewich, S. D. (2017) Penitentes as the origin of the bladed terrain of Tartarus Dorsa on Pluto. *Nature*, *541*, 188-190.

Morbidelli A. and Rickman H. (2015) Comets as collisional fragments of a primordial planetesimal disk. *Astron. Astrophys.*, *583*, A43, DOI: 10.1051/0004-6361/201526116.

Morbidelli A. and Nesvorný D. (2020) Kuiper belt: Formation and evolution. In *The Trans-Neptunian Solar System* (D. Prialnik, M. A. Barucci, L. A. Young, eds.), pp. 25–59. Elsevier, DOI: 10.1016/B978-0-12-816490-7.00002-3.

Morbidelli A., Bottke W. F, Nesvorný D., and Levison H. F. (2009) Asteroids were born big. *Icarus*, *204*, 558–573.

Mousis O., Guilbert-Lepoutre A., Lunine J. I., Cochran A. L., Waite J. H., Petit J.-M., and Rousselot P. (2012) The dual origin of the nitrogen deficiency in comets: Selective volatile trapping in the nebula and postaccretion radiogenic heating. *Astrophys. J.*, *757*, 146

Mumma M. J. and Charnley S. B. (2011) The chemical composition of comets — Emerging taxonomies and natal heritage. *Annu. Rev. Astron Astrophys.*, *49*, 471–524.

Murray C. D. and Dermott S. F. (1999) *Solar System Dynamics*. Cambridge Univ., Cambridge, U.K . 603 pp.

Murray-Clay R. A. and Chiang E. I. (2016) Brownian motion in planetary migration. *Astrophys. J.*, *651*, 1194–1208, DOI: 10.1086/507514.

Nesvorný D. (2018) Dynamical evolution of the early Solar System. *Annu. Rev. Astron. Astrophys.*, *56*, 137–174, DOI: 10.1146/annurev-astro-081817-052028.

Nesvorný D. and Morbidelli A. (2012) Statistical study of the early solar system's instability with four, five, and six giant planets. *Astron. J.*, *14*, 117.

Nesvorný D. and Vokrouhlický, D. (2016) Neptune's orbital migration was grainy, not smooth. *Astrophys. J.*, *825*, 94.

Nesvorný D. and Vokrouhlický, D. (2019) Binary survival in the outer solar system. *Icarus*, *331*, 49–61.

Nesvorný D., Youdin A. N., and Richardson D. C. (2010) Formation of Kuiper belt binaries by gravitational collapse. *Astron. J.*, *140*, 785-793, DOI: 10.1088/0004-6256/140/3/785.

Nesvorný, D., Vokrouhlický D., Bottke W. F., and Levison H. F. (2018) Evidence for very early migration of the Solar System planets from the Patroclus-Menoetius binary Jupiter Trojan. *Nature Astron.*, *2*, 878-882.

Nesvorný D., Li R., Youdin A. N., Simon J. B., and Grundy W. M. (2019) Trans-Neptunian binaries as evidence for planetesimal formation by the streaming instability. *Nature Astron.*, *3*, DOI: 10.1038/s41550-019-0806-z.

Neveu, M., Desch S. J., Shock E. L., and Glein C. R. (2015) Prerequisites for explosive cryovolcanism on dwarf planet-class Kuiper belt objects. *Icarus*, *246*, 48–64.

Niemann H. B., Atreya S. K., Demick J. E., Gautier D., Haberman J. A., Harpold D. N., Kasprzak W. T., Lunine J. I., Owen T. C., and Raulin F. (2010) Composition of Titan's lower atmosphere and simple surface volatiles as measured by the Cassini-Huygens probe gas chromatograph mass spectrometer experiment. *J. Geophys. Res.*, *115*, E12006, DOI: 10.1029/2010JE003659.

Nimmo F. (2004) Stresses generated in cooling viscoelastic ice shells: Application to Europa. *J. Geophys. Res.*, *109*, E12001.

Nimmo F., Hamilton D. P., McKinnon W. B., et al. (2016) Reorientation of Sputnik Planitia implies a subsurface ocean on Pluto. *Nature*, *540*, 94-96.

Nogueira E., Brasser R., and Gomes R. (2011) Reassessing the origin of Triton. *Icarus*, *214*, 113–130.

Noll K. S., Grundy W. M., Nesvorný D., and Thirouin A. (2020) The Trans-neptunian binaries (2018). In *The Trans-Neptunian Solar System* (D. Prialnic, A. Barucci, L.A. Young, eds.), pp. 205–224. Elsevier, Amsterdam, DOI: 10.1016/B978-0-12-816490-7.00009-6.

Öberg K. I. and Wordsworth R. (2019) Jupiter's composition suggests its core assembled exterior to the $N_2$ snowline. *Astron. J.*, *158*, 194, DOI: 10.3847/1538-3881/ab46a8.

Okumura F. and Mimura K. (2011) Gradual and stepwise pyrolyses of insoluble organic matter from the Murchison meteorite revealing chemical structure and isotopic distribution. *Geochim. Cosmochim. Acta*, *75*, 7063-7080.

Ortiz J. L., Santos-Sanz P., Sicardy B., et al. (2017) The size, shape, density and ring of the dwarf planet Haumea from a stellar occultation. *Nature*, *550*, 219–223.

Owen T. C. (1982) The composition and origin of Titan's atmosphere. *Planet. Space Sci.*, *30*, 833–838.

Owen T. C., Roush T. L., Cruikshank D. P., Elliot J. L., Young L. A., de Bergh C., Schmitt B., Geballe T. R., Brown R. H., and Bartholomew M. J (1993) Surface ices and atmospheric composition of Pluto. *Science*, *261*, 745–48.

Palme H., Lodders K., and Jones A. (2014) Solar system abundances of the elements. In *Treatise on Geochemistry, volume 2: Planets, Asteriods, Comets and the Solar System* (H. H. Holland and K. K. Turekian, eds.), pp. 15–36. Elsevier, Amsterdam.

Peale S. J. and Canup R. M. (2015) The origin of natural satellites. In *Treatise on Geophysics: Second Edition, volume 10: Physics of Terrestrial Planets and Moons* (G. Schubert, ed.), *Vol. 10*, pp. 559–604. Elsevier, Amsterdam.

Pires P., Winter S. M. G., and Gomes R. S. (2015) The evolution of a Pluto-like system during the migration of the ice giants. *Icarus, 246*, 330-338, DOI: 10.1016/j. icarus.2014.04.029.

Poch O., Istiqomah I., Quirico E., et al. (2019) Ammonium salts are a reservoir of nitrogen on a cometary nucleus and possibly on some asteroids. *Science*, *367*, eaaw7462, DOI: 10.1126/science.aaw7462.

Postberg F., Kempf S., Schmidt J., Brilliantov N., Beinsen A., Abel B., Buck U., and Srama R. (2009) Sodium salts in E-ring ice grains from an ocean below the surface of Enceladus. *Nature*, *459*, 1098–1101.

Protopapa S., Grundy W. M., Reuter D. C., et al. (2017) Pluto's global surface composition through pixel-by-pixel Hapke modeling of New Horizons Ralph/LEISA data. *Icarus*, *287*, 218–228.

Quarles B. and Kaib N. (2019) Instabilities in the early solar system due to a self-gravitating disk. *Astron. J.*, *157*, 67, DOI: 10.3847/1538-3881/aafa71.

Rhoden A.R., Henning, W., Hurford, T.A., and Hamilton, D.P. (2015) The interior and orbital evolution of Charon as preserved in its geologic record. *Icarus*, *246*, 11–20.

Rhoden A. R., Skjetne H. L., Henning W. G., et al. (2020) Charon: A brief history of tides. *J. Geophys. Res. Planets*, *125*, e06449. DOI: 10.1029/2020JE006449.

Robbins S. J. and 28 colleagues (2019) Geologic landforms and chronostratigraphic history of Charon as revealed by a hemispheric geologic map. *J. Geophys. Res. Planets*, *124*, 155–174.

Robeiro R. de S., Morbidelli A., Raymond Sean N., Izidoro A., Gomes R., and Neto E. V. (2020) Dynamical evidence for an early giant planet instability. *Icarus*, *339*, 113605, DOI: 10.1016/j.icarus.2019.113605.

Robuchon, G. and Nimmo, F. (2011) Thermal evolution of Pluto and implications for surface tectonics and a subsurface ocean. *Icarus*, *216*, 426–439, DOI: 10.1016/j.icarus.2011.08.015.

Rubin M., Altwegg K., Balsiger H., et al. (2015) Molecular nitrogen in Comet 67P/Churyumov-Gerasimenko indicates a low formation temperature. *Science*, *384*, 232–235.

Rubin M., Altwegg K., Balsiger H., et al. (2018) Krypton isotopes and noble gas abundances in the coma of comet 67P/Churyumov-Gerasimenko. *Sci. Adv.*, *4*, eaar6297, DOI: 10.1126/sciadv.aar6297.

Rubin M., Altwegg K., Balsiger H., et al. (2019) Elemental and molecular abundances in comet 67P/Churyumov-Gerasimenko. *Mon. Not. R. Astron. Soc.*, *489*, 594-607.

Schaller E. L. and Brown M. E. (2007) Volatile loss and retention on Kuiper belt objects. *Astrophys. J.*, *659*, L61–L64.

Schenk P. M., R. A. Beyer R. A., McKinnon W. B., et al. (2018) Basins, fractures and volcanoes: Global cartography and topography of Pluto from New Horizons *Icarus*, *314*, 400–433, DOI: 10.1016/j.icarus.2018.06.008.

Schubert G., Stevenson D. J., and Ellsworth K. (1981) Internal structures of the Galilean satellites. *Icarus*, *47*, 46–59.

Scott E. R. D., Krot A. N., and Sanders I. S. (2018) Isotopic dichotomy among meteorites and its bearing on the protoplanetary disk. *Astrophys. J.*, *854*, 164.

Scott T. A. (1976) Solid and liquid nitrogen. *Phys. Rep. (Phys. Lett. C)*, *27*, 89–157.

Sephton M. (2005) Organic matter in carbonaceous meteorites: Past, present, and future research. *Phil. Trans. R. Soc. A*, *363*, 2729–2742, DOI: 10.1098/rsta.2005.1670.

Sekine Y., Genda H., Sugita S., Kadono T., and Matsui T. (2011) Replacement and late formation of atmospheric $N_2$ on undifferentiated Titan by impacts. *Nat. Geosci.*, *4*, 359-362.

Sekine Y., Genda H., Kamata S., and Funatsu T. (2017) The Charon-forming giant impact as a source of Pluto's dark equatorial regions. *Nat. Astron.*, *1*, 0031, DOI: 10.1038/s41550-016-0031.

Shannon A. and Dawson B. (2018) Limits on the number of primordial Scattered disc objects at Pluto mass and higher from the absence of their dynamical signatures on the present-day trans-Neptunian Populations. *Mon. Not. R. Astron. Soc.*, *480*, 1870–1882, DOI: 10.1093/mnras/sty1930.

Shakura N. I. and Sunyaev R. A. (1973) Black holes in binary systems. Observational appearance. *Astron. Astrophys.*, *24*, 337–355.

Shinnaka Y., Kawakita H., Emmanuël J., Decock A., Hutsemékers D., Manfroid J., and Arai A. (2016) Nitrogen isotopic ratios of $NH_2$ in comets: Implication for $^{15}N$-fractionation in cometary ammonia. *Mon. Not. R. Astron. Soc.*, *462*, S195–S209.

Sheppard S., Trujillo C. A., Tholen D. J., and Kaib N. (2019) A new high perihelion trans-Plutonian inner Oort cloud object: 2015 TG387. *Astron. J.*, *157*, 139, DOI: 10.3847/1538-3881/ab0895.

Shock E. L. and McKinnon W. B. (1993) Hydrothermal processing of cometary volatiles—Applications to Triton. *Icarus*, *106*, 464-477.

Siess L., Dufour E., and Forestini M. (2000) An internet server for pre-main sequence tracks of low- and intermediate-mass stars. *Astron. Astrophys.*, *358*, 593–599.

Simon J. B., Armitage P. J., Li R., and Youdin A N. (2016) The mass and size distribution of planetesimals formed by the streaming instability. I. The role of self-gravity. *Astrophys. J.*, *822*, 55, DOI: 10.3847/0004-637X/822/1/55.

Simonelli D. P., Pollack J. B., McKay C. P., Reynolds R. T., and Summers A. L. (1989) The carbon budget in the outer solar nebula. *Icarus*, *82*, 1–35.

Singer K. N. and Stern S. A. (2015) On the provenance of Pluto's nitrogen ($N_2$). *Astrophys. J. Lett.*, *808*, L50, DOI: 10.1088/2041-8205/808/2/L50.

Sloan E. D. and Koh C. A. (2008) *Clathrate Hydrates, 3rd ed.* CRC Press, Boca Raton, Fla.

Solomon S. C. and Head J. W. (1980) Lunar mascon basin: Lava filling, tectonics, and evolution of the lithosphere. *Rev. Geophys. Space Phys.*, *18*, 107–141.

Sotin C., Tobie G., Wahr J., McKinnon W. B. (2009) Tides and tidal heating on Europa. In *Europa* (R.P. Pappalardo, W.B. McKinnon, and D.J. Tholen, eds.) pp. 85-117. Univ. Arizona, Tucson.

Spencer J. R., Stren S. A., Moore J. M., et al. (2020) The geology and geophysics of Kuiper Belt object (486958) Arrokoth. *Science*, *367*, eaay3999. DOI:10.1126/science.aay3999.

Squyres S. W., Reynolds R. T., Summers A. L., and Shung F. (1988) Accretional heating of the satellites of Saturn and Uranus. *J. Geophys. Res.*, *93*, 8779-8794.

Stern S. A. (1986) The effects of mechanical interaction between the interstellar medium and comets. *Icarus*, *68*, 276–283.

Stern S. A. (1991) On the number of planets in the solar system: Evidence of a substantial population of 1000 km bodies. *Icarus*, *90*, 271–281.

Stern S. A. and Shull, J. M. (1988) The influence of supernovae and passing stars on comets in the Oort cloud. *Nature, 332*, 407–411.

Stern S. A. and Tholen D. J., eds. (1997) *Pluto and Charon*. Univ. of Arizona, Tucson. 728 pp.

Stern S. A. and Trafton L. M. (2008) On the atmospheres of objects in the Kuiper belt. In *The Solar System beyond Neptune* (M. A. Barucci, H. Boehnhardt, D. Cruikshank, and A. Morbidelli, eds.), pp. 365–380. Univ. of Arizona, Tucson.

Stern S. A., McKinnon W. B., and Lunine J. I. (1997) On the origin of Pluto, Charon, and the Pluto-Charon binary. In *Pluto and Charon* (S.A. Stern and D.J. Tholen, eds.), pp. 605-663. Univ. Arizona Press, Tucson.

Stern, S. A., Weaver, H. A., Steffl, A. J., et al. (2006) A giant impact origin for Pluto's small moons and satellite multiplicity in the Kuiper belt. *Nature, 439*, 946-948.

Stern S. A., Bagenal F., Ennico K., et al. (2015) The Pluto system: initial results from its exploration by New Horizons. *Science*, *350*, aad1815, DOI: 10.1126/science.aad1815.

Stern S. A., Binzel R. P., Earle A. M., et al. (2017) Past epochs of significantly higher pressure atmospheres on Pluto. *Icarus*, *287*, 47–53.

Stern S. A., Grundy W. M., McKinnon W. B., Weaver H. A., and Young L. A. (2018) The Pluto system after New Horizons. *Annu. Rev. Astron. Astrophys.*, 56, 357–392. DOI: 0.1146/annurevastro-081817-051935.

Stern S. A., et al. (2019) Initial results from the New Horizons exploration of 2014 MU$_{69}$, a small Kuiper Belt object. *Science*, *364*, eaaw9771, DOI: 10.1126/science.aaw9771.

Stern S. A., White O. L., McGovern P. J., et al. (2020) Pluto's far side. *Icarus*, DOI: 10.1016/j.icarus.2020.113805.

Stevenson D. J., Harris A. W., and Lunine J. I. (1986) Origins of satellites. In *Satellites* (J.A. Burns and M.S. Matthews, eds.), pp. 39–88. Univ. of Arizona, Tucson.

Sussman, G. J. and Wisdom, J. (1988) Numerical evidence that the motion of Pluto is chaotic. *Science*, *241*, 433-437, DOI:10.1126/science.242.4684.433.

Tegler S. C., Grundy W. M., Vilas F., Romanishin W., Cornelison D. M., Consolmagno G. J. (2008) Evidence of N$_2$-ice on the surface of the icy dwarf Planet 136472 (2005 FY9). *Icarus*, *195*, 844–850.

Tegler S. C., Cornelison D. M., Grundy W.M., et al. (2010) Methane and nitrogen abundances on Pluto and Eris. *Astrophys. J.*, *725*, 1296–1305.

Telfer M. W., and 19 colleagues (2018) Dunes on Pluto. *Science*, *360*, 992-997.

Trujillo C., Trilling D., Gerdes D., et al. (2019) Deep Ecliptic Exploration Project (DEEP) observing strategy. *EPSC Abs.*, *13*, EPSC-DPS2019-2070.

Tsiganis K., Gomes R., Morbidelli A., and Levison H. F. (2005) Origin of the orbital architecture of the giant planets of the Solar System. *Nature, 435*: 459–461.

Umurhan O. M., Howard A. D., Moore J. M., et al. (2017) Modeling glacial flow on and onto Pluto's Sputnik Planitia. *Icarus*, *287*, 301–319.

Vilella K. and Deschamps F. (2017) Thermal convection as a possible mechanism for the origin of polygonal structures on Pluto's surface. *J. Geophys. Res. Planets*, *122*, 1056-1076.

Volk K and Malhotra R. (2012). The effect of orbital evolution on the Haumea (2003 EL$_{61}$) collisional family. *Icarus*, *221*, 106–115, DOI: 10.1016/j.icarus.2012.06.047.

Walsh K. J. and Levison H. F. (2015) Formation and evolution of Pluto's small satellites. *Astron. J.*, *150*, 11.

Wang H., Weiss B. P., Bai X.-N., Downey B. G., Wang J., Wang J., Suavet C., Fu R. R., and Zucolotto M. E. (2017) Lifetime of the solar nebula constrained by meteorite paleomagnetism. *Science*, *355*, 623–627, DOI: 10.1126/science.aaf5043.

Weaver H. A., Buie M. W., Buratti B. J., et al. (2016) The small satellites of Pluto as observed by New Horizons. *Science*, *351*, aae0030.

White O. L., Moore J. M., McKinnon W. B., et al. (2017). Geological mapping of Sputnik Planitia on Pluto. *Icarus*, *287*, 261-286.

White O. L., Moore J. M., Howard A. D., et al. (2019). Washboard and fluted terrains on Pluto as evidence for ancient glaciation. *Nature Astron.*, *3*, 62–68.

Williams J. P and Cieza L. A. (2011) Protoplanetary disks and their evolution. *Ann. Rev. Astron. Astrophys.*, 49, 67-117, DOI: 10.1146/annurev-astro-081710-102548.

Wong T., Hansen U., Weisehöfer T., and McKinnon W. (2019) Formation of cellular structures on Sputnik Planitia from convection. *American Geophysical Union, Fall Meeting 2019*, abstract #P42C-07.

Yabuta H., Williams L. B., Cody G. D., Alexander C. M. O'D., and Pizzarello S. (2007) The insoluble carbonaceous material of CM chondrites: A possible source of discrete organic compounds under hydrothermal conditions. *Meteoritics & Planet. Sci.*, *42*, 37–48.

Youdin A. N. and Goodman J. (2005) Streaming instabilities in protoplanetary disks. *Astrophys. J.*, *620*, 459-469 (2005), DOI: 10.1086/426895.

Young L. A., Kammer J. A., Steffl A. J., et al. (2018) Structure and composition of Pluto's atmosphere from the New Horizons solar ultraviolet occultation. *Icarus*, *300*,174–199. DOI: 10.1016/j.icarus.2017.09.006.

Zahnle K. J. and Walker J. C. G. (1982) The evolution of solar ultraviolet luminosity. *Rev. Geophys.*, *20*, 280–292, DOI: 10.1029/RG020i002p00280.

Zahnle K. J., Korycansky D. G., and Nixon C. A. (2014). Transient climate effects of large impacts on Titan. *Icarus*, *229*, 378–391, DOI: 10.1016/j.icarus.2013.11.006.

Table 1. Nitrogen Inventory on Pluto

| Reservoir | Moles of $N_2$ |
|---|---|
| Atmosphere | $1\times10^{15}$ |
| Escape[a] | $5\times10^{16}$ |
| Photochemistry | $2\times10^{18}$ |
| Surface (Sputnik Planitia) | $(0.4\text{-}3)\times10^{20}$ |
| Subsurface liquid $N_2$ | Comparable?[b] |

[a]Assumes past similar to present (*Glein and Waite*, 2018).
[b]E.g., if outer 5 km of Pluto's ice shell were a 5% porous "nitrogenifer."

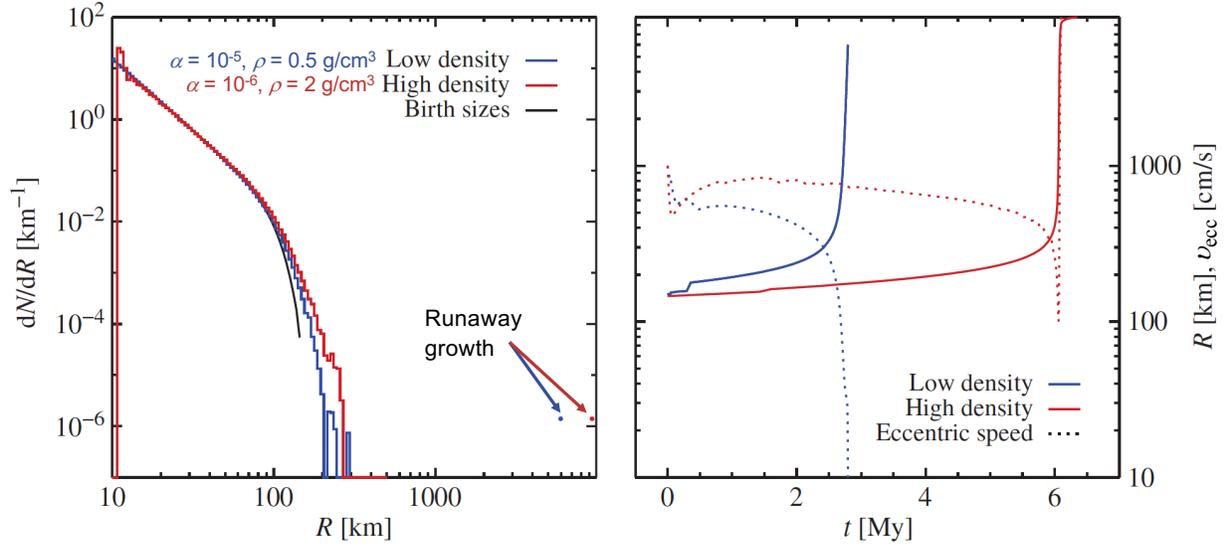

**Fig. 1.** See Plate TBD for color version. Planetesimal growth in the original trans-Neptunian protoplanetary disk (25 AU). Two models are shown: a low-density model where the solid density is set to $\rho = 0.5$ g/cm$^3$, similar to comets, and a high-density model where the internal density is set to $\rho = 2$ g/cm$^3$, similar to Pluto. The initial planetesimal size distribution is based on a streaming instability model. Both values of the turbulent stirring $\alpha$ (*Shakura and Sunyaev*, 1973) are "low" when compared with the $10^{-4}$ commonly used in disk viscous evolution models, and are thought to reflect the mild turbulence caused by streaming and Kelvin-Helmholtz instabilities in what would otherwise be a dead, laminar midplane. The right panel shows the size of the largest body as a function of time as well as its speed relative to a circular orbit ($v_{ecc}$). Runaway growth is facilitated by a steep decline in orbital eccentricity because the high pebble accretion rate damps the eccentricity. Modified from *Johansen et al.* (2015).

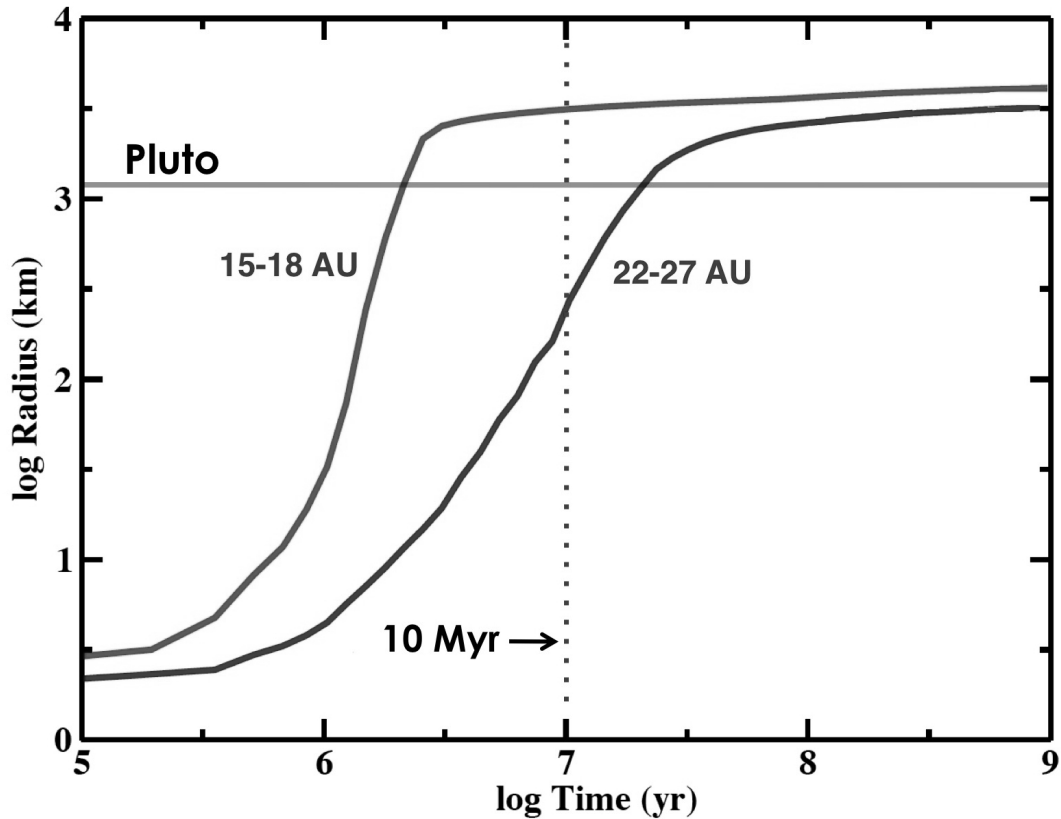

**Fig. 2.** Selected results from a numerical, multi-annulus hierarchical coagulation accretion simulation. Shown is the time evolution of the size of the largest accreted object in two trans-Neptunian ranges. Starting conditions place most of the mass in 1-km planetesimals; the initial cumulative size distribution of planetesimals is nearly flat, with a surface density distribution proportional to $a^{-3/2}$. The calculation includes gas drag for the removal of smaller bodies from the calculation, but no explicit pebble accretion; fragmentation is included but no long-range dynamical stirring by proto-Neptune. After a short period of slow growth, objects rapidly grow from ~10 km to ~1000 km (runaways due to substantial gravitational focusing) and then grow more slowly to ~3000–5000 km as the largest planetesimals ("oligarchs") stir the smaller planetesimals to larger and larger orbital velocities. Modified from *Kenyon and Bromley* (2012).

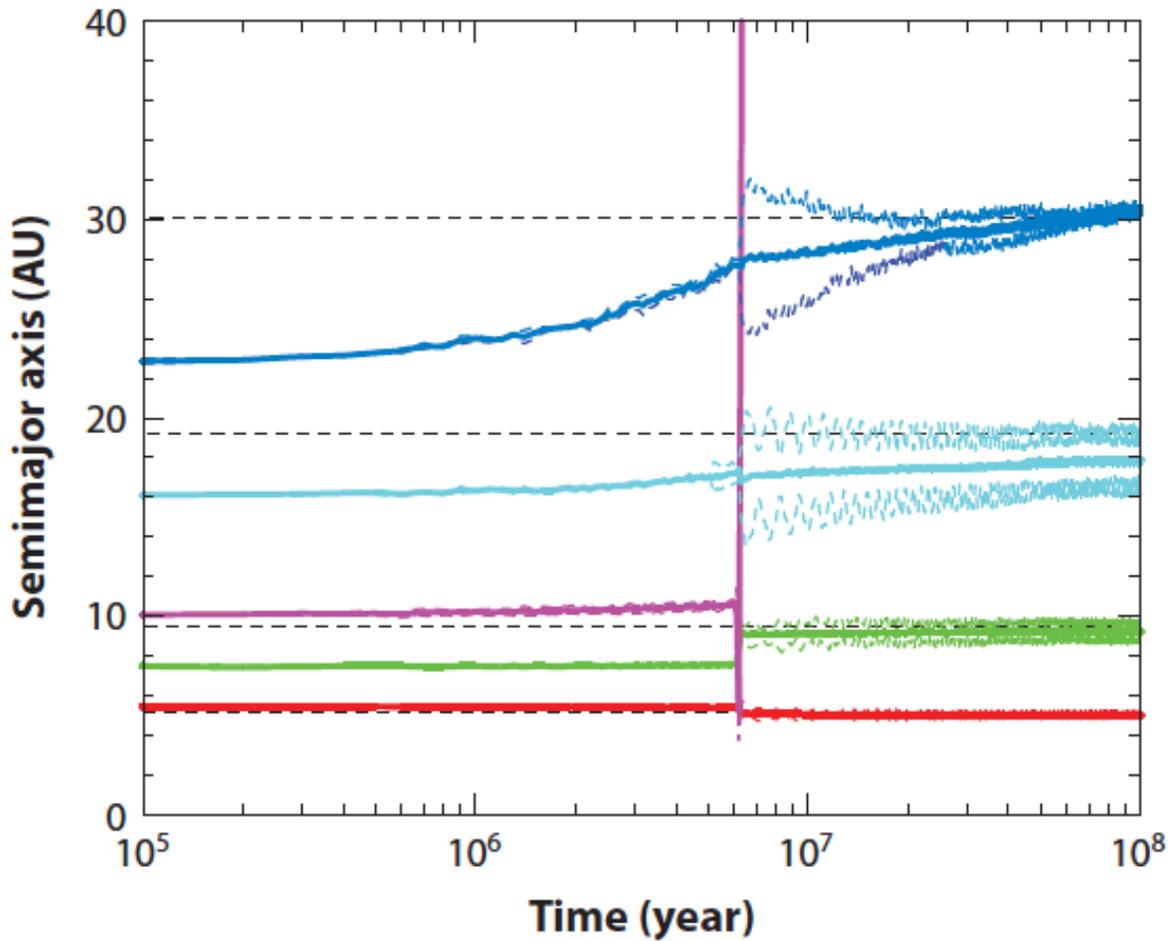

**Fig. 3.** See plate TBD for color version. A possible orbital history of the giant planets. Five planets were started in the (3:2, 4:3, 2:1, 3:2) mean-motion resonant chain along with a 20 $M_\oplus$ planetesimal disk between 23 AU and 30 AU. The semimajor axes (solid lines) and perihelion and aphelion distances (dashed lines) of each planet's orbit are indicated. The horizontal dashed lines show the semimajor axes of planets in the present Solar System. The final orbits obtained in the model are a good match to those in the present Solar System. Adapted from the statistical study of *Nesvorný and Morbidelli* (2012).

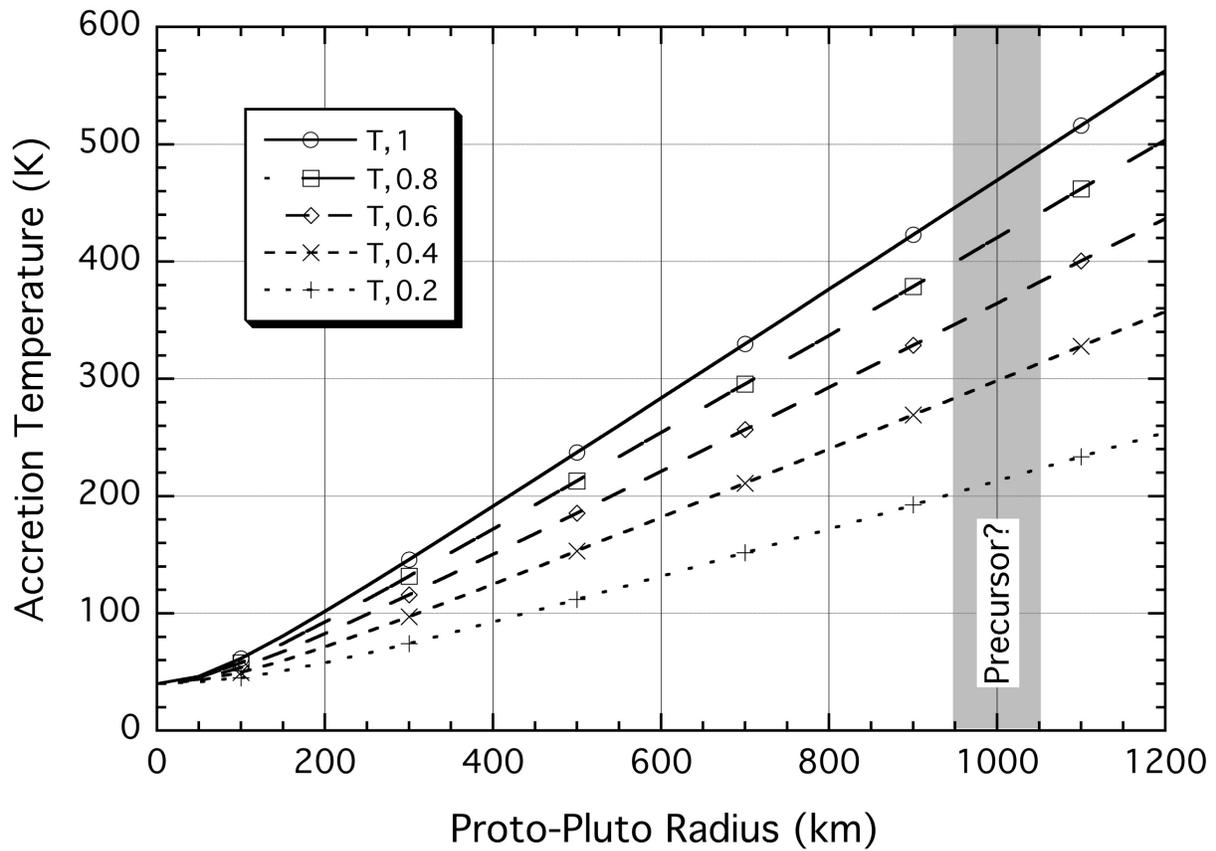

**Fig. 4.** Model accretion temperatures for proto-Pluto bodies, prior to the Charon-forming giant impact. The curve labeled T,1 assumes essentially complete retention of impact energy, the curve T,0.8 corresponds to retention of about 80% of the impact energy, etc. The approach kinetic energy (the second term in equation 2) is neglected for all five curves. Bodies with roughly half to two-thirds of the mass of the Pluto system correspond to the size range labeled "Precursor?"

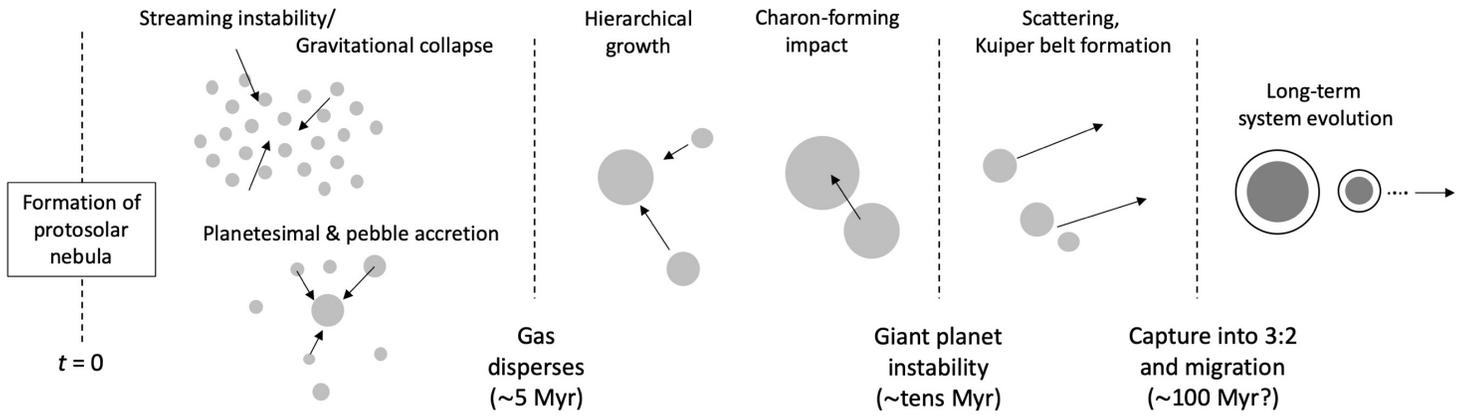

**Fig. 5.** Stages in the formation and evolution of the Pluto system, in the context of the formation of the Kuiper belt (see text).

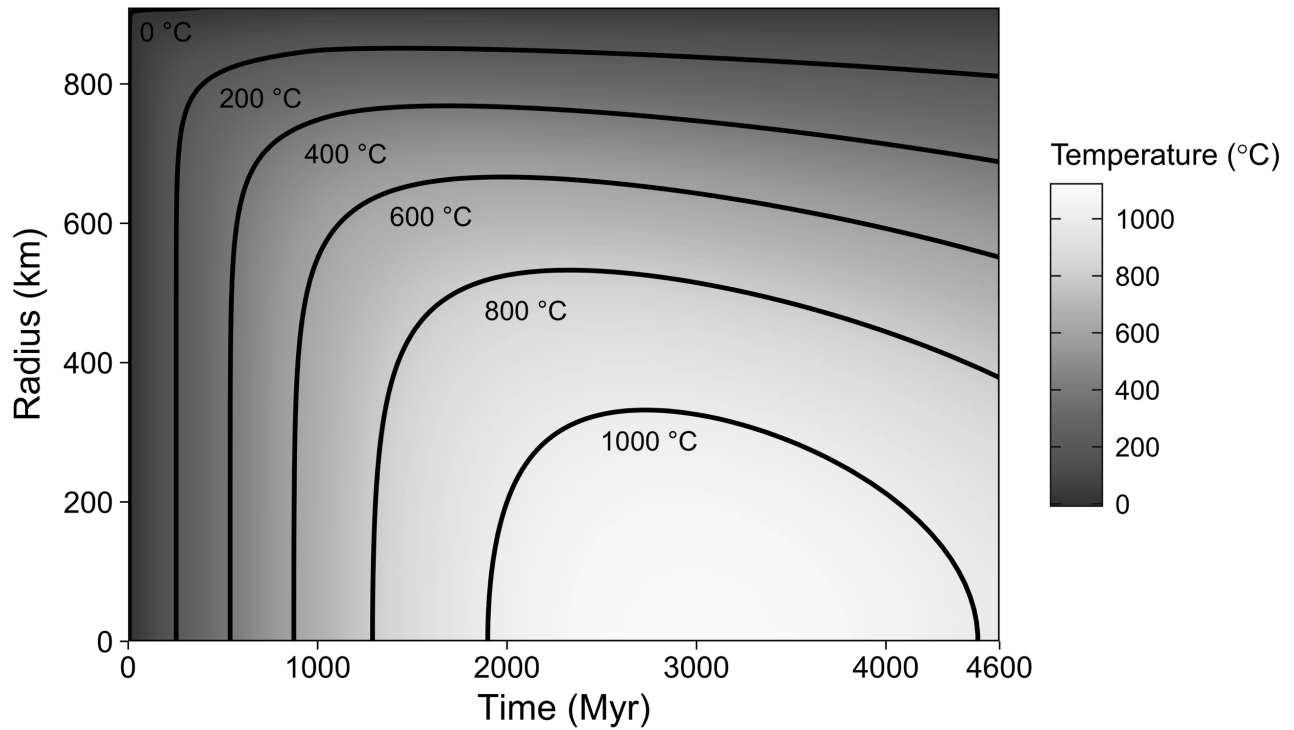

**Fig. 6.** Example of the possible thermal evolution of a rocky core inside Pluto. Temperatures were calculated for a carbonaceous chondritic radiogenic heating rate and heat conduction out of the core. The water-rock interface is at a radius of ~910 km, and its temperature was set to that of the overlying ocean. From *Kamata et al.* (2019).

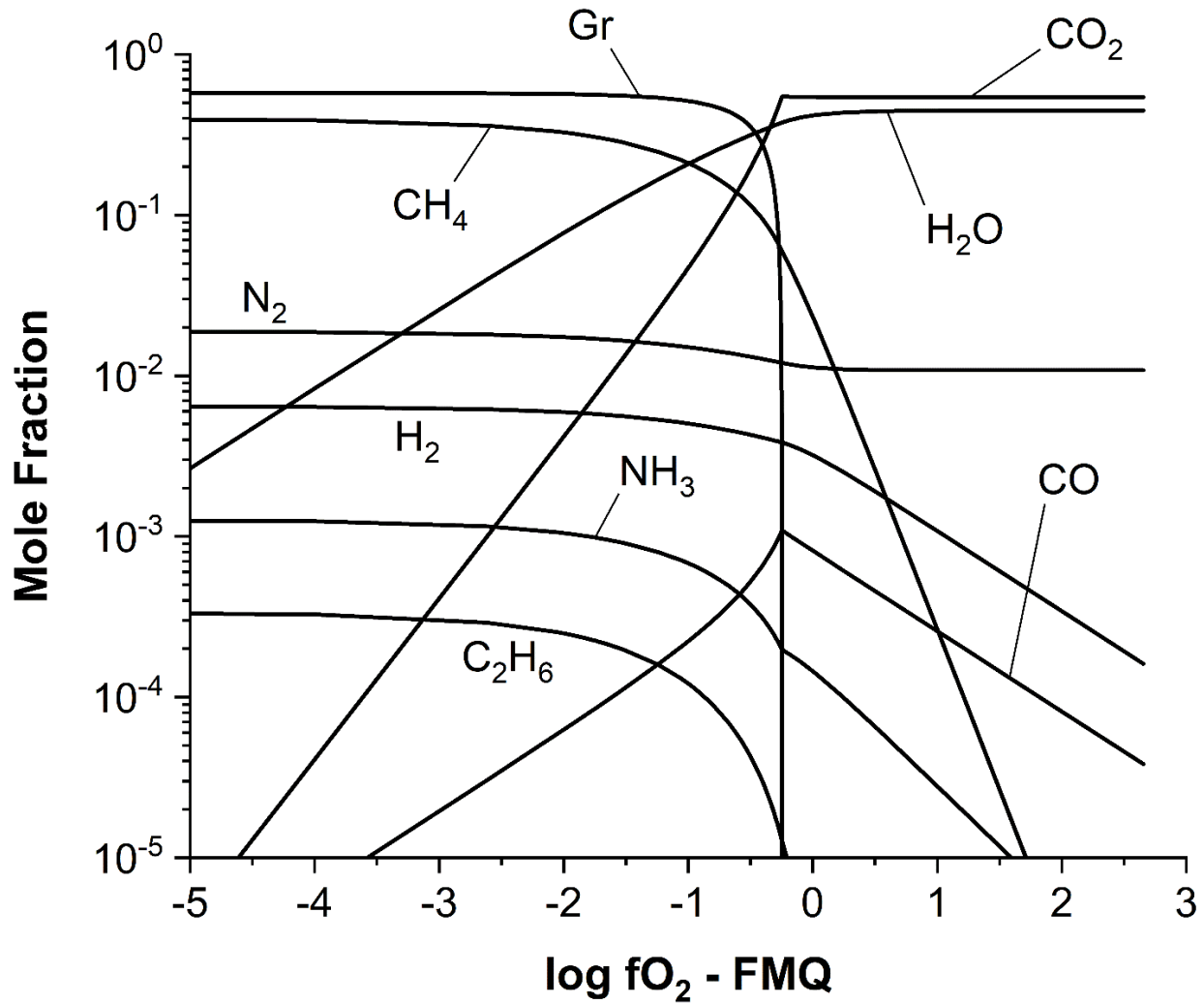

**Fig. 7.** Equilibrium state illustrating what could happen in an initial core on Pluto that is heated to 500°C and 1900 bar. Oxygen fugacity ($fO_2$) is plotted relative to the fayalite-magnetite-quartz (FMQ) mineral buffer (equation 5). Note that the standard unit for fugacity is bars, but the difference in log units is independent of the pressure unit chosen. The initial core was assumed to be composed of a mixture of hydrated silicates and accreted organic matter. The system is more reduced toward the left side of the plot and more oxidized toward the right side of the plot. Note that the "plunge" of graphite (Gr) corresponds to its disappearance in oxidized systems.

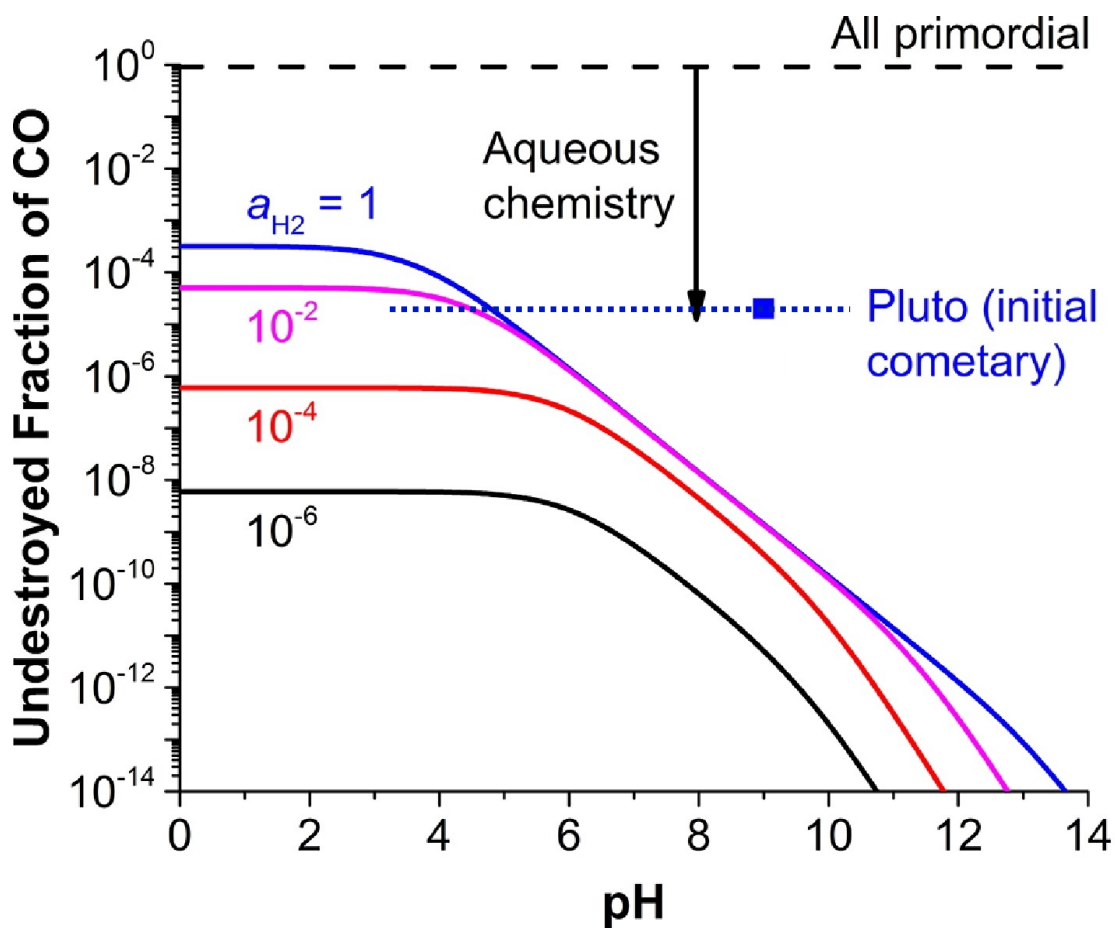

**Fig. 8.** Fraction of initial CO that remains in aqueous solution at metastable equilibrium with formate and carbonate species at 0 °C and 1900 bar, as a function of the pH and $H_2$ activity $a_{H_2}$ (a measure of the oxidation state). The symbol labeled "Pluto" shows how much CO destruction would need to occur for this model to reproduce the estimated surface ratio of $CO/N_2$ from an initial cometary ratio. The blue square denotes a pH value derived for Enceladus' ocean, for comparison (from *Postberg et al.*, 2009). Figure modified from *Glein and Waite* (2018).

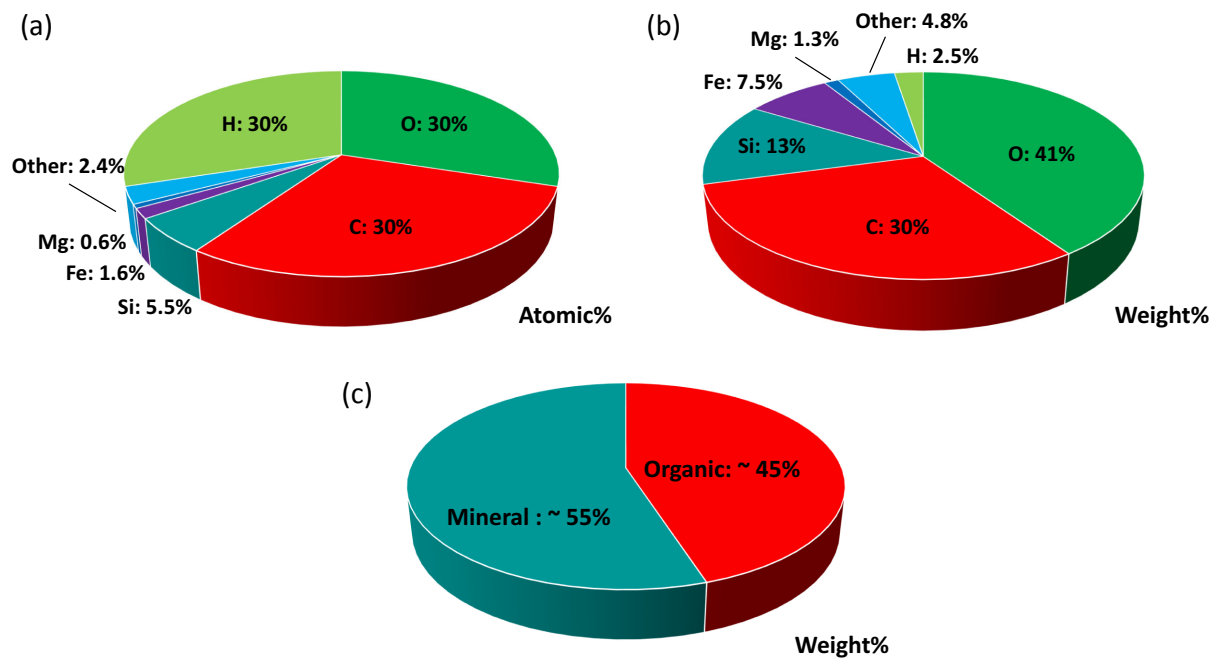

**Fig. 9.** See plate TBD for color version. Averaged composition of comet 67P's dust particles as deduced from COmetary Secondary Ion Mass Analyzer (COSIMA)/*Rosetta* mass spectrometer measurements and supplemental hypotheses detailed in *Bardyn et al.* (2017). The averaged composition is given **(a)** by atomic fraction and **(b)** by atomic mass fraction, and **(c)** is the mineral and organic content estimated in mass fraction. See *Bardyn et al.* (2017) for details.

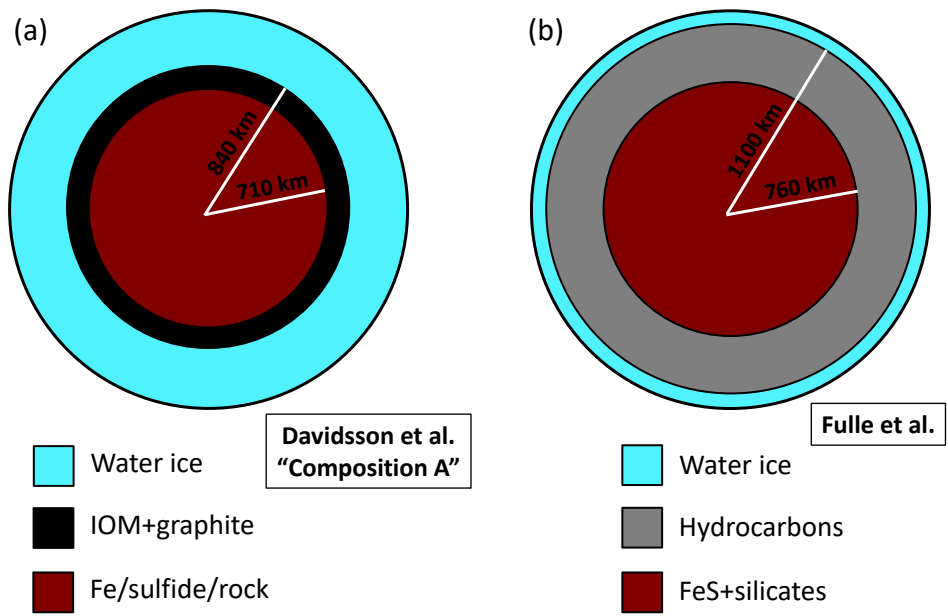

**Fig. 10.** See plate TBD for color version. Alternative Pluto internal models based on proposed cometary compositions. Rock/sulfide, carbonaceous matter, and ice are depicted as if separated under the influence of gravity (differentiated). **(a)** "Composition A" (*Davidsson et al.*, 2016) proposes a 16.5 wt% insoluble organic matter (IOM) like contribution modeled as graphite/amorphous carbon, equivalent to a deep 130-km-thick layer. **(b)** The model based on *Fulle et al.* (2017) is more radical, more than half high-H/C macromelecular organics (1200 kg m$^{-3}$) by volume. Thermally, model (a) could permit a water ocean at depth, but the total thickness of the ice shell in model (b), ~85 km, is too thin for the temperature profile at depth to reach ice melting for likely present-day heat flows.

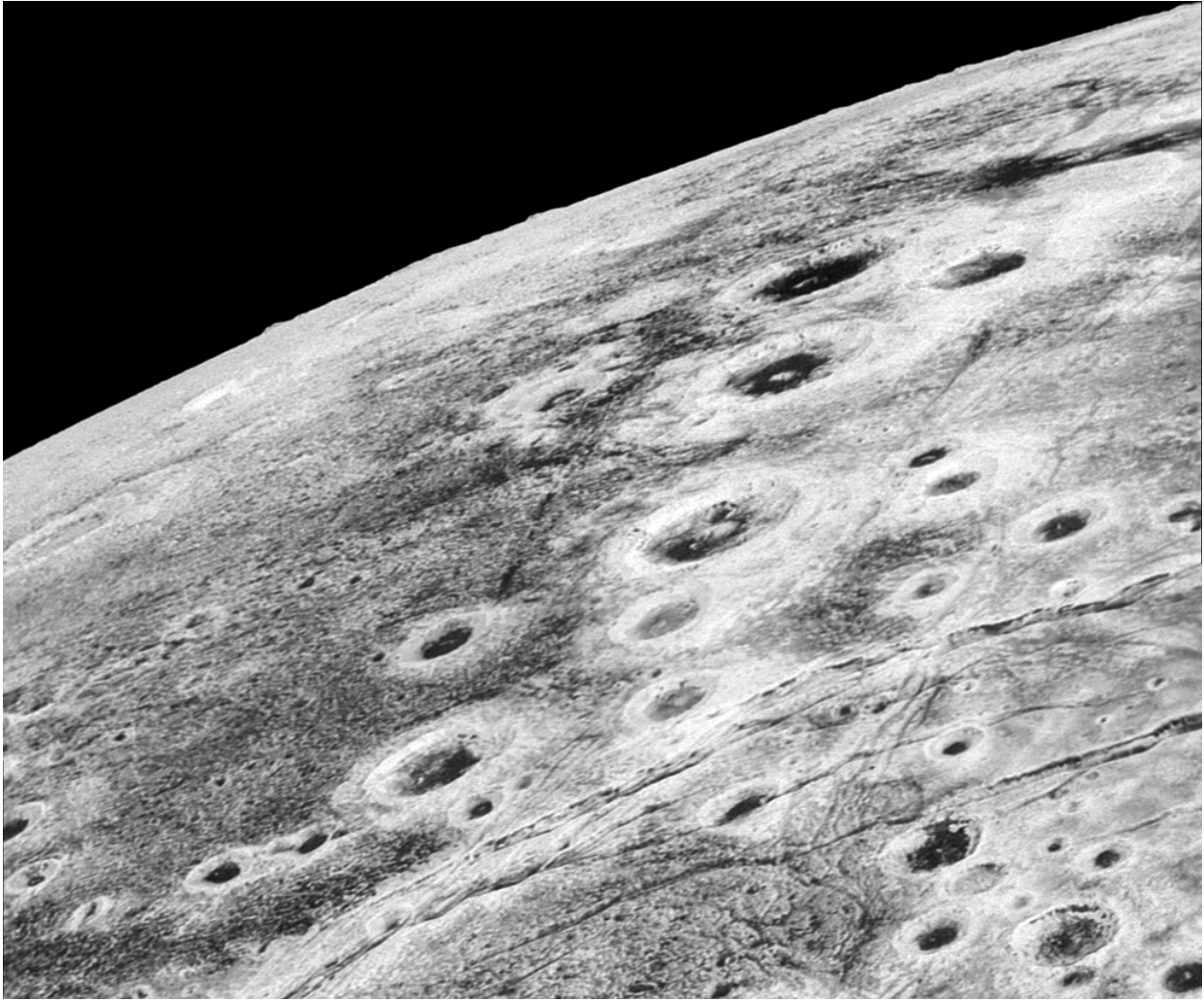

**Fig. 11.** The view across Vega Terra to the southwest. Several prominent ~20-km diameter craters occupy the middle distance, and the younger Djanggawul Fossae crosscut in the foreground. Vega Terra and the horizon beyond are unusually flat for Pluto (*Schenk et el.*, 2018).

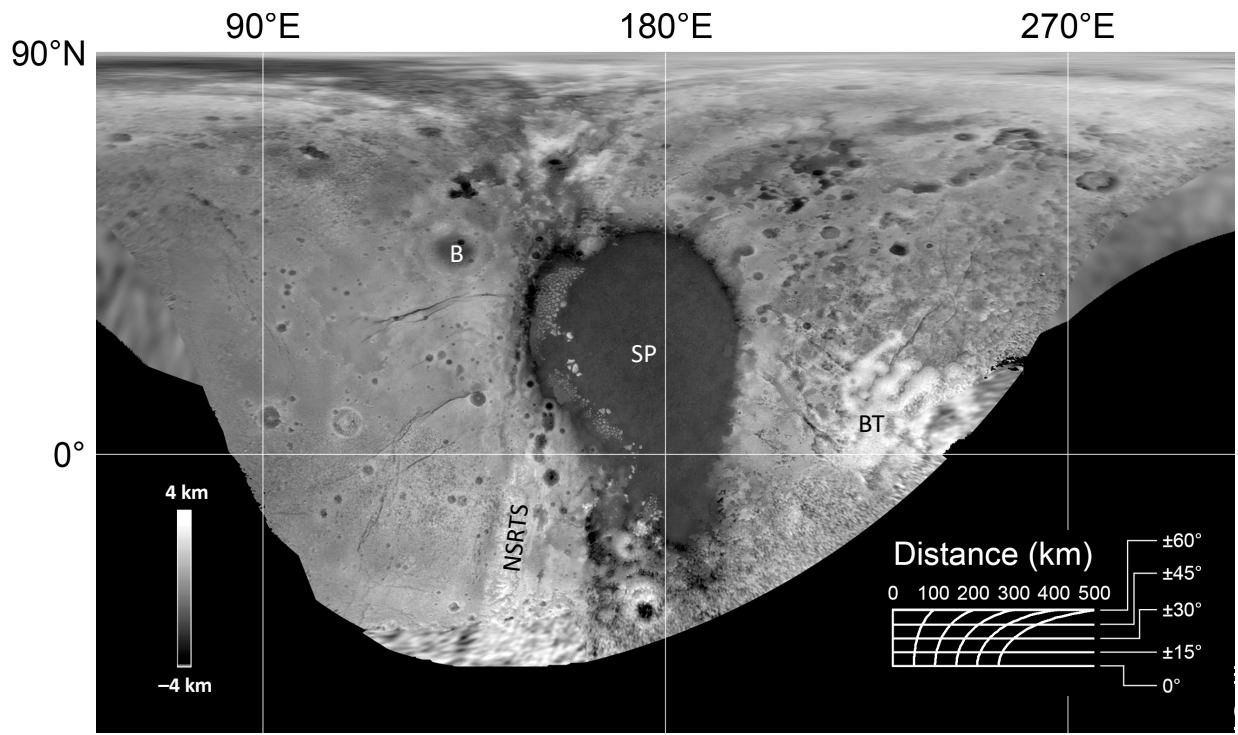

**Fig. 12.** Best hemispheric topography of Pluto. From combined LORRI-LORRI, MVIC-MVIC and LORRI-MVIC stereo, with a vertical resolution of 100-400 m (modified from *Schenk et al.*, 2018). Cylindrical map projection centered on 180° longitude. Dark areas were unilluminated or do not have resolvable stereogrammetric data from the 2015 encounter. Notable features are Sputnik Planitia (SP) at center, the high-standing Bladed terrain at right (BT), the great north-south ridge-and-trough system left of center (NSRTS), the Burney multiring basin (B), and the fairly flat plains to the west. North is up.

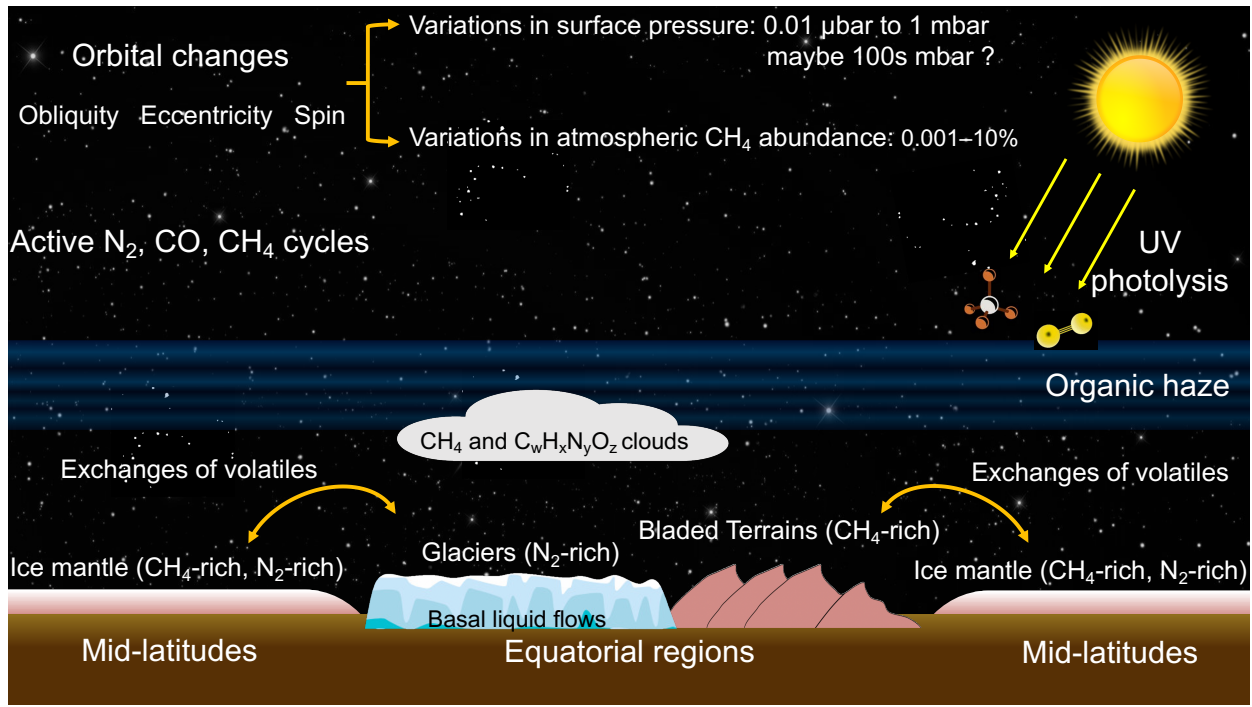

**Fig. 13.** See plate TBD for color version. Schematic view of the main dynamic surface-atmospheric processes on Pluto.